\documentclass[twocolumn]{aastex631}

\usepackage{graphicx}
\usepackage{natbib}
\usepackage{color}
\usepackage{appendix}
\usepackage{hyperref}

\definecolor{red}{rgb}{1.0,0.0,0.0}

\newcommand{\Rstar}{\mbox{$R_{\star}$}}
\newcommand{\Mstar}{\mbox{$M_{\star}$}}
\newcommand{\Msun}{\mbox{$M_{\rm \sun}$}}

\newcommand{\Rsun}{\mbox{$R_{\sun}$}}
\newcommand{\degree}{\mbox{$^{o}$}}
\begin{document}

\title{SCExAO/CHARIS Near-Infrared Scattered-Light Imaging and Integral Field Spectropolarimetry of the AB Aurigae Protoplanetary System}
\author{Erica Dykes}
\affiliation{Department of Physics and Astronomy, University of Texas-San Antonio, San Antonio, TX, USA}
\author{Thayne Currie}
\affiliation{Department of Physics and Astronomy, University of Texas-San Antonio, San Antonio, TX, USA}
\affiliation{Subaru Telescope, National Astronomical Observatory of Japan, 
650 North A`oh$\bar{o}$k$\bar{u}$ Place, Hilo, HI  96720, USA}
\author{Kellen Lawson}
\affiliation{NASA-Goddard Space Flight Center, Greenbelt, MD, USA}
\author{Miles Lucas}
\affiliation{Institute for Astronomy, University of Hawaii-Manoa, Honolulu, HI, USA}
\author{Tomoyuki Kudo}
 \affiliation{Subaru Telescope, National Astronomical Observatory of Japan, 
 650 North A`oh$\bar{o}$k$\bar{u}$ Place, Hilo, HI  96720, USA}
 \author{Minghan Chen}
  \affiliation{Department of Physics, University of California, Santa Barbara, Santa Barbara, California, USA}
\author{Olivier Guyon}
 \affiliation{Subaru Telescope, National Astronomical Observatory of Japan, 
 650 North A`oh$\bar{o}$k$\bar{u}$ Place, Hilo, HI  96720, USA}
 \affil{Steward Observatory,  Department of Astronomy, The University of Arizona, Tucson, AZ 85721, USA}
 \affil{College of Optical Sciences, University of Arizona, Tucson, AZ 85721, USA}
 \affil{Astrobiology Center, National Institutes of Natural Sciences, 2-21-1 Osawa, Mitaka, Tokyo 181-8588, Japan}
 \author{Tyler D. Groff}
 \affiliation{NASA-Goddard Space Flight Center, Greenbelt, MD, USA}
 \author{Julien Lozi}
 \affiliation{Subaru Telescope, National Astronomical Observatory of Japan, 
 650 North A`oh$\bar{o}$k$\bar{u}$ Place, Hilo, HI  96720, USA}
 \author{Jeffrey Chilcote}
 \affiliation{Department of Physics and Astronomy, University of Notre Dame, South Bend, IN, USA}
 \author{Timothy D. Brandt}
 \affiliation{Department of Physics, University of California, Santa Barbara, Santa Barbara, California, USA}
\author{Sebastien Vievard}
 \affiliation{Subaru Telescope, National Astronomical Observatory of Japan, 
650 North A`oh$\bar{o}$k$\bar{u}$ Place, Hilo, HI  96720, USA}
  \author{Nour Skaf}
\affiliation{Department of Astronomy \& Astrophysics, University of California, Santa Cruz, CA 95064, USA}
 \author{Vincent Deo}
 \affiliation{Subaru Telescope, National Astronomical Observatory of Japan, 
 650 North A`oh$\bar{o}$k$\bar{u}$ Place, Hilo, HI  96720, USA}
 \author{Mona El Morsy}
\affiliation{Department of Physics and Astronomy, University of Texas-San Antonio, San Antonio, TX, USA}
\author{Danielle Bovie}
\affiliation{Department of Physics and Astronomy, University of Texas-San Antonio, San Antonio, TX, USA}
\author{Taichi Uyama}
\affiliation{Department of Physics and Astronomy, California State University-Northridge, Northridge, CA USA}
\author{Carol Grady}
\affiliation{Eureka Scientific, 2452 Delmer Street Suite 100, Oakland, CA, USA}
\author{Michael Sitko}
\affiliation{Space Sciences Institute, 4765 Walnut St, Suite B, Boulder, CO 80301}
\author{Jun Hashimoto}
\affil{Astrobiology Center, National Institutes of Natural Sciences, 2-21-1 Osawa, Mitaka, Tokyo 181-8588, Japan}
 \affiliation{National Astronomical Observatory of Japan, 2-21-2, Osawa, Mitaka, Tokyo 181-8588, Japan}
 \affiliation{Department of Astronomy, Graduate School of Science, The University of Tokyo, 7-3-1, Hongo, Bunkyo-ku, Tokyo, 113-0033, Japan}
 \author{Frantz Martinache}
 \affiliation{Universit\'{e} C\^{o}te d'Azur, Observatoire de la C\^{o}te d'Azur, CNRS, Laboratoire Lagrange, France}
  \author{Nemanja Jovanovic}
\affiliation{Department of Astronomy, California Institute of Technology, 1200 East California Boulevard, Pasadena, CA 91125}
 \author{Motohide Tamura}
  \affil{Astrobiology Center, National Institutes of Natural Sciences, 2-21-1 Osawa, Mitaka, Tokyo 181-8588, Japan}
 \affiliation{National Astronomical Observatory of Japan, 2-21-2, Osawa, Mitaka, Tokyo 181-8588, Japan}
 \affiliation{Department of Astronomy, Graduate School of Science, The University of Tokyo, 7-3-1, Hongo, Bunkyo-ku, Tokyo, 113-0033, Japan}
 \author{N. Jeremy Kasdin}
 \affiliation{Department of Mechanical Engineering, Princeton University, Princeton, NJ USA}

\shortauthors{Dykes et al.}

\begin{abstract}
We analyze near-infrared integral field spectropolarimetry of the AB Aurigae protoplanetary disk and protoplanet (AB Aur b), obtained with 
SCExAO/CHARIS in 22 wavelength channels covering the J, H, and K passbands ($\lambda_{\rm o}$ = 1.1--2.4 $\mu m$) over angular separations of $\rho$ $\approx$ 0\farcs{}13 to 1\farcs{}1 ($\sim$20--175 au).  Our images resolve spiral structures in the disk in each CHARIS channel.  At the longest wavelengths, the data may reveal an extension of the western spiral seen in previous polarimetric data at $\rho$ $<$ 0\farcs{}3 out to larger distances clockwise from the protoplanet AB Aur b, coincident with the ALMA-detected $CO$ gas spiral. While AB Aur b is detectable in complementary total intensity data, it is a non-detection in polarized light at $\lambda$ $>$ 1.3 $\mu $m.  While the observed disk color is extremely red across $JHK$, the disk has a blue intrinsic scattering color consistent with small dust grains.  The disk's polarization spectrum is redder than  AB Aur b's total intensity spectrum. The polarization fraction peaks at $\sim$ 0.6 along the major disk axis.   Radiative transfer modeling of the CHARIS data shows that small, porous dust grains with a porosity of $p$ = 0.6--0.8 better reproduce the scattered-light appearance of the disk than more compact spheres ($p$ = 0.3), especially the polarization fraction.   This work demonstrates the utility of integral field spectropolarimetry to characterize structures in protoplanetary disks and elucidate the properties of the disks' dust.
\end{abstract}
\section{Introduction}
Most pre-main sequence stars in nearby 1--5 Myr-old star-forming regions are surrounded by gas and dust-rich \textit{protoplanetary} disks that form the building blocks of planets \citep[e.g.][]{Lada2006,Luhman2010}.   Optical/near-infrared (IR) scattered-light imaging of protoplanetary disks reveals numerous structures potentially connected to the formation of massive planets in these disks, including asymmetries, clumps, rings, gaps, and spirals \citep[e.g.][]{Muto2012,Benisty2017,Avenhaus2018,Garufi2018}.   Disks with these resolved structures are especially common around 1.5--3 $M_{\rm \odot}$ Herbig AeBe stars \citep[e.g.][]{Benisty2022,Rich2022}, the evolutionary precursors of 20-200 Myr-old stars around which we have directly imaged a dozen 10-100 au-separation, fully-formed superjovian planets \citep[e.g.][]{Marois2008a,Lagrange2010,Carson2013,Chauvin2017,Currie2023,Tobin2024}.  


\textit{Polarimetric differential imaging} (PDI) has proven to be a powerful technique to detect these protoplanetary disks in scattered light and constrain their properties \citep[e.g.][]{Quanz2011,Garufi2018,Benisty2022}.  Techniques used to achieve total intensity imaging detections of protoplanetary disks often distort disk signals \citep[e.g.][]{Rich2019,Currie2019a,Currie2023b,Benisty2022}.  However, PDI well suppresses the stellar halo without distorting the disk, exploiting the fact that the star is unlikely to be significantly polarized, while small dust grains in protoplanetary disks produce a detectable polarization signal \citep[e.g.][]{Avenhaus2018}.   
High-fidelity disk detections in polarized light also help to distinguish between bona fide protoplanets embedded in disks (usually visible in total intensity) and disk features \citep[e.g.][]{Thalmann2016, Currie2019a, Keppler2018,Currie2022a,Lawson2022}.    

The young Herbig AeBe star AB Aurigae is surrounded by a highly-structured protoplanetary disk that is the site of active planet formation.   The disk contains numerous spiral arms extending out to 200--500 au \citep[e.g.][]{Grady1999,Fukagawa2004,Oppenheimer2008}.  Millimeter continuum imaging resolves a $\sim$ 170 au-scale ring of pebble-sized dust \citep{Tang2012,Tang2017,Francisvandermarel2020}.  Previous broadband near-IR polarized intensity imaging reveals spirals on 20--100 au scales, which may be coincident with CO gas spirals also seen in the millimeter \citep{Hashimoto2011} and may be consistent with a spiral density wave expected from an unseen planet at $\sim$ 25 au \citep{Boccaletti2020}.

Most recently, optical and near-IR high-contrast imaging with the \textit{Subaru Coronagraphic Extreme Adaptive Optics} Project (SCExAO) and the \textit{Hubble Space Telescope} (HST) identify AB Aurigae b, an embedded protoplanet 0\farcs{}6 ($\approx$ 100 au) from the star \citep{Currie2022a} at a position consistent with that predicted for a planet driving one of the CO gas spirals \citep[see also][]{Currie2024}.  AB Aur b's optical/near-IR spectral energy distribution appears inconsistent with scattered starlight.  Its detection in $H_{\alpha}$ may identify accretion or an additional scattered light component to its emission.   While it appears as a clump-like signal brighter and distinguishable from surrounding disk emission in total intensity across the major near-IR passbands, preliminary near-IR polarized intensity imaging presented in \citet{Currie2022a} only detected uniform emission from the disk, not a spatially-concentrated clump.   The morphology and location of AB Aur b in the disk is consistent with that expected for a planet being formed by rapid gravitational collapse \citep[see also][]{Boss1997,Speedie2024}.

Further investigations of AB Aurigae over a large wavelength range in both total and polarized light may provide a fuller picture of the disk's structures, better constrain the disk's scattering properties, and investigate polarized emission from AB Aur b and other potential planetary signals.    Previous analysis of AB Aur's disk properties focused on broadband imaging at 2.0 $\mu m$ in both total and polarized intensity with HST to constrain the polarization function and 2--4 $\mu m$ total intensity imaging with the \textit{Large Binocular Telescope} (LBT) to investigate water ice abundances \citep{Perrin2009,Betti2022}.   The HST data favor a disk with a polarization fraction peaking at $\sim$0.6; the LBT data are more difficult to interpret as different lines of evidence (e.g. different disk colors) favor different conclusions about the presence/absence of icy grains.   In both cases, residual speckles limit the level of detail at which the data can be interpreted.  Neither study clearly resolves the disk spiral structure seen in other work with better spatial resolution  or higher-quality AO correction  \citep[][]{Currie2022a,Boccaletti2020}.  

In this paper, we analyze near-IR integral field spectropolarimetry (IFSP) of AB Aurigae obtained with the Subaru Telescope using SCExAO coupled to the Coronagraphic High-Angular Resolution Imaging Spectrograph (CHARIS).   These data were briefly introduced in \citet{Currie2022a} to demonstrate that AB Aur b lacked a strong polarization signal. Here, we analyze these data in more detail, exploring how polarimetric scattered-light images of AB Aur's protoplanetary disk illuminate our understanding of the disk's morphology and scattering properties.   We combine these data with the total intensity data presented in \citet{Currie2022a} to investigate the polarization fraction of AB Aur disk's across the major near-IR passbands. Our analyses provide protoplanetary disk spectra in polarized light across the $J$, $H$, and $K$ passbands and insights into the dust properties of AB Aur's disk.

\begin{deluxetable*}{llllllllll}
     \tablewidth{0pt}
    \tablecaption{AB Aurigae Observing Log}
    \tablehead{\colhead{UT Date} & \colhead{Instrument} &  \colhead{coronagraph} & \colhead{Seeing (\arcsec{})} &{Passband} & \colhead{$\lambda$ ($\mu m$)$^{e}$} 
    & \colhead{$t_{\rm exp}$} & \colhead{$N_{\rm exp}$} & \colhead{$\Delta$PA ($^{o}$)} & \colhead{Observing} \\
    {} & {} & {} & {} & {} & {} & {} & {} & {} & \colhead{Strategy}  }
    \startdata
     20201004 & SCExAO/CHARIS & Lyot & 0.6-0.8 & $JHK$ & 1.16--2.39& 60.49 & 73 & 80.0 & PDI\\
    \enddata
    \label{table:obslog}
    \end{deluxetable*}

    \begin{figure*}[ht]
  \centering
   \includegraphics[width=0.495\textwidth,clip]{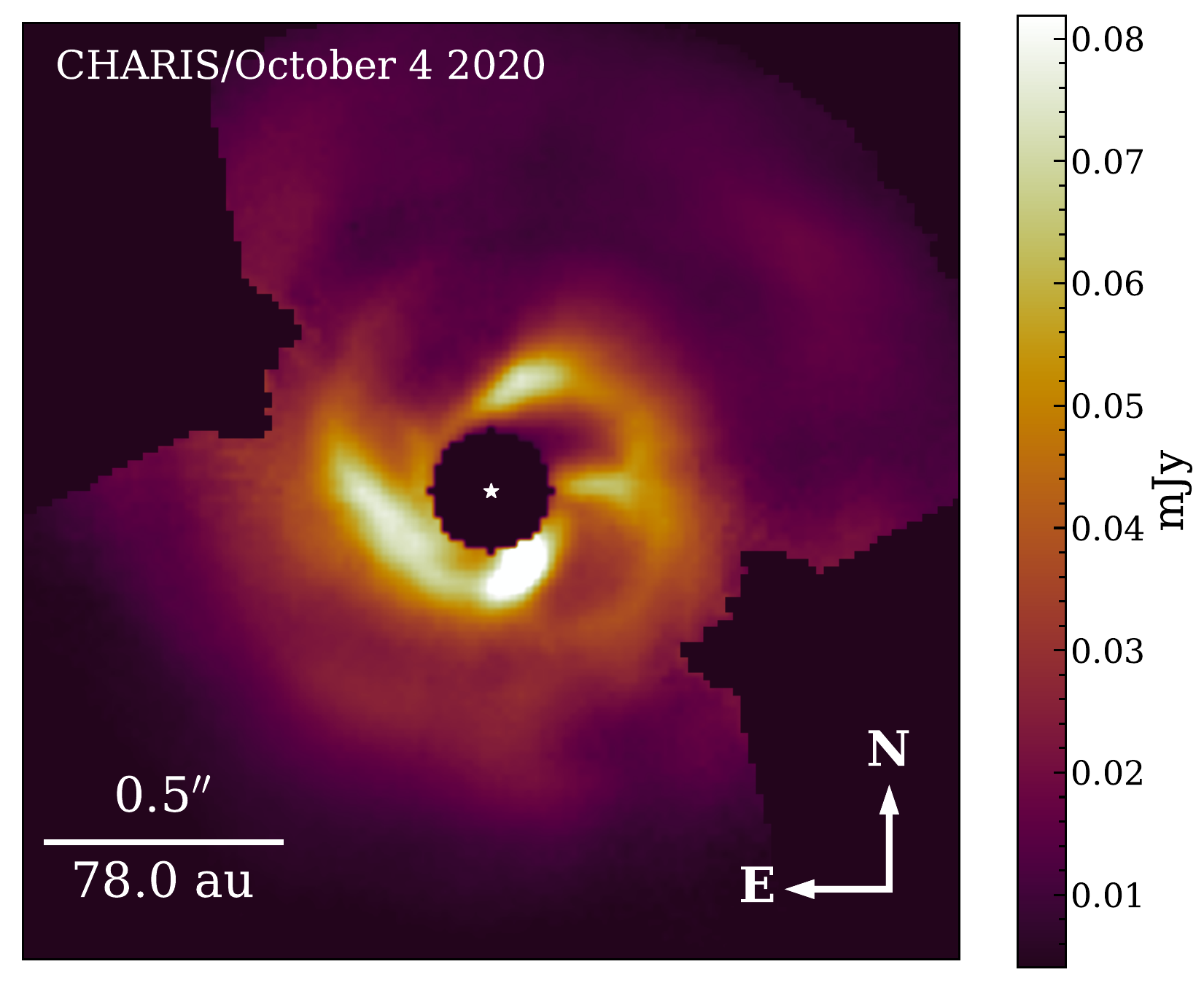}
   \includegraphics[width=0.495\textwidth,clip]{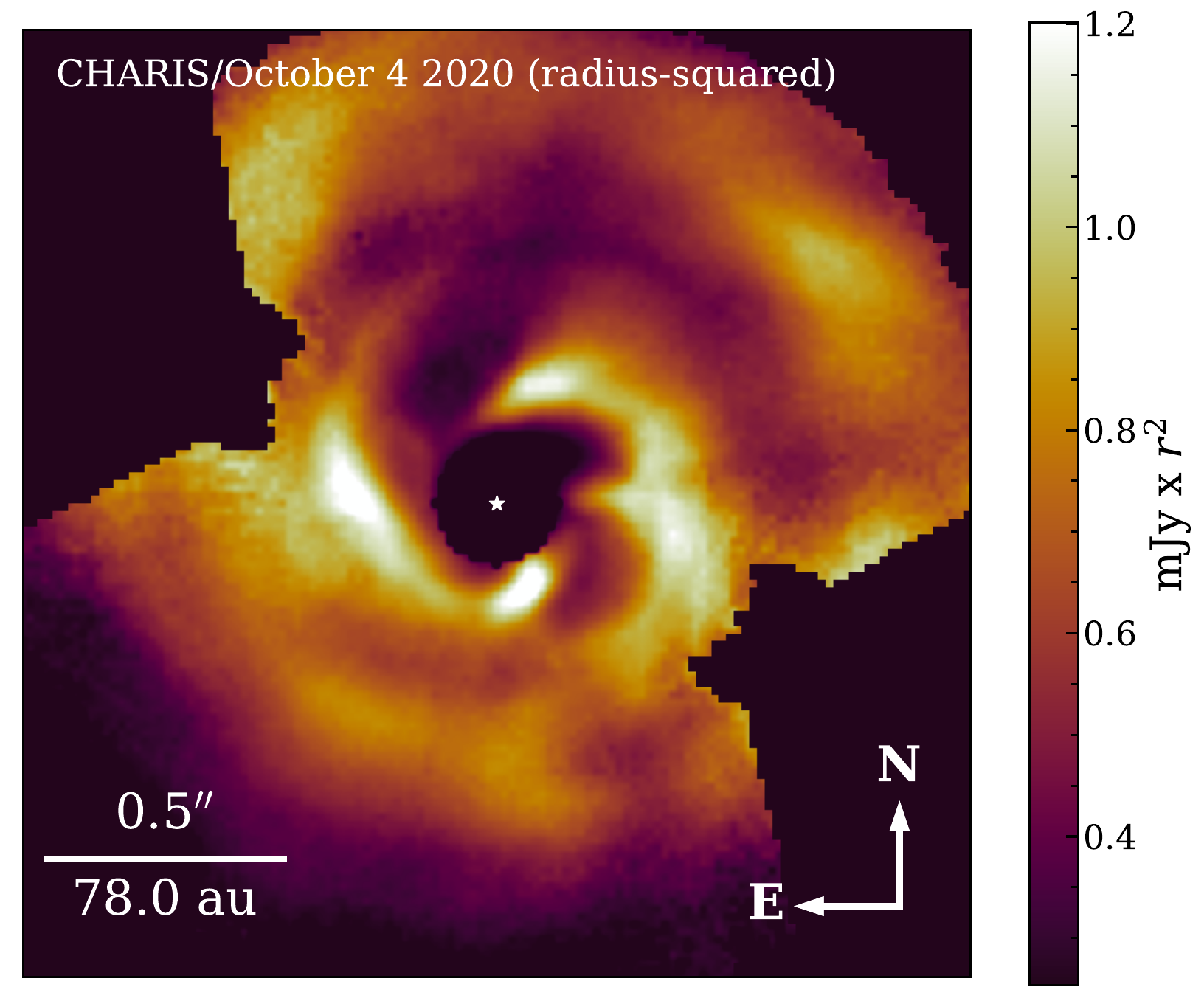}
   \vspace{-0.1in}
   \caption
   {SCExAO/CHARIS AB Aur broadband Q$_{\rm\phi}$ images (wavelength-collapsed) from 04 October 2020 obtained from PDI processing. The reduction described in Methods was used for both the linearly-scaled (left) and radius-squared-scaled(right) images shown.  The image color map intensity scaling goes from the 10th to 99.7th percentile of count values (i.e. \texttt{np.nanpercentile(image, (10, 99.7))} in the Python \texttt{NumPy} package).
   }
   \label{fig:charispdi}
   \end{figure*}

       \begin{figure*}[ht]
  \centering
   \includegraphics[width=0.325\textwidth,clip]{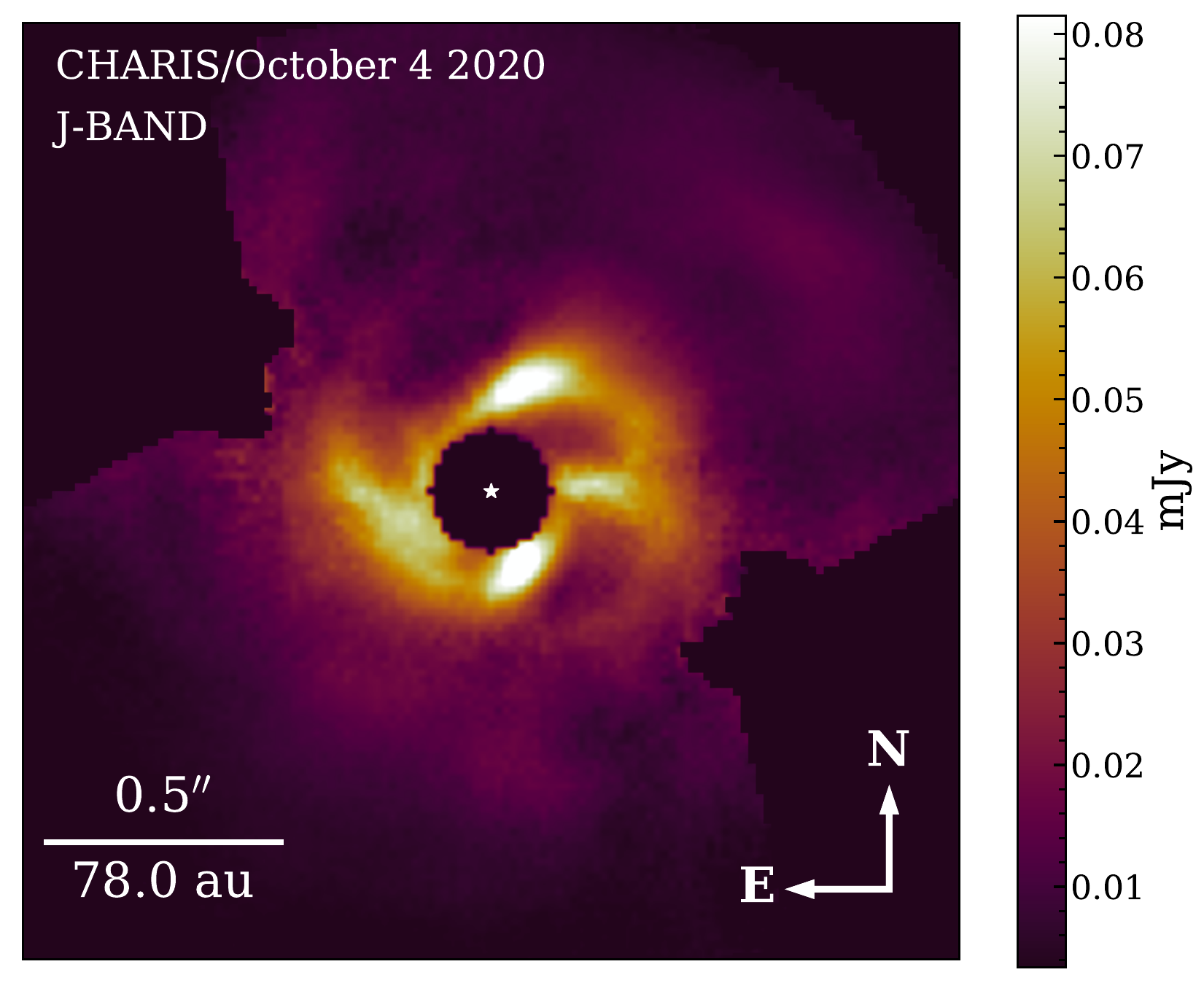}
   \includegraphics[width=0.325\textwidth,clip]{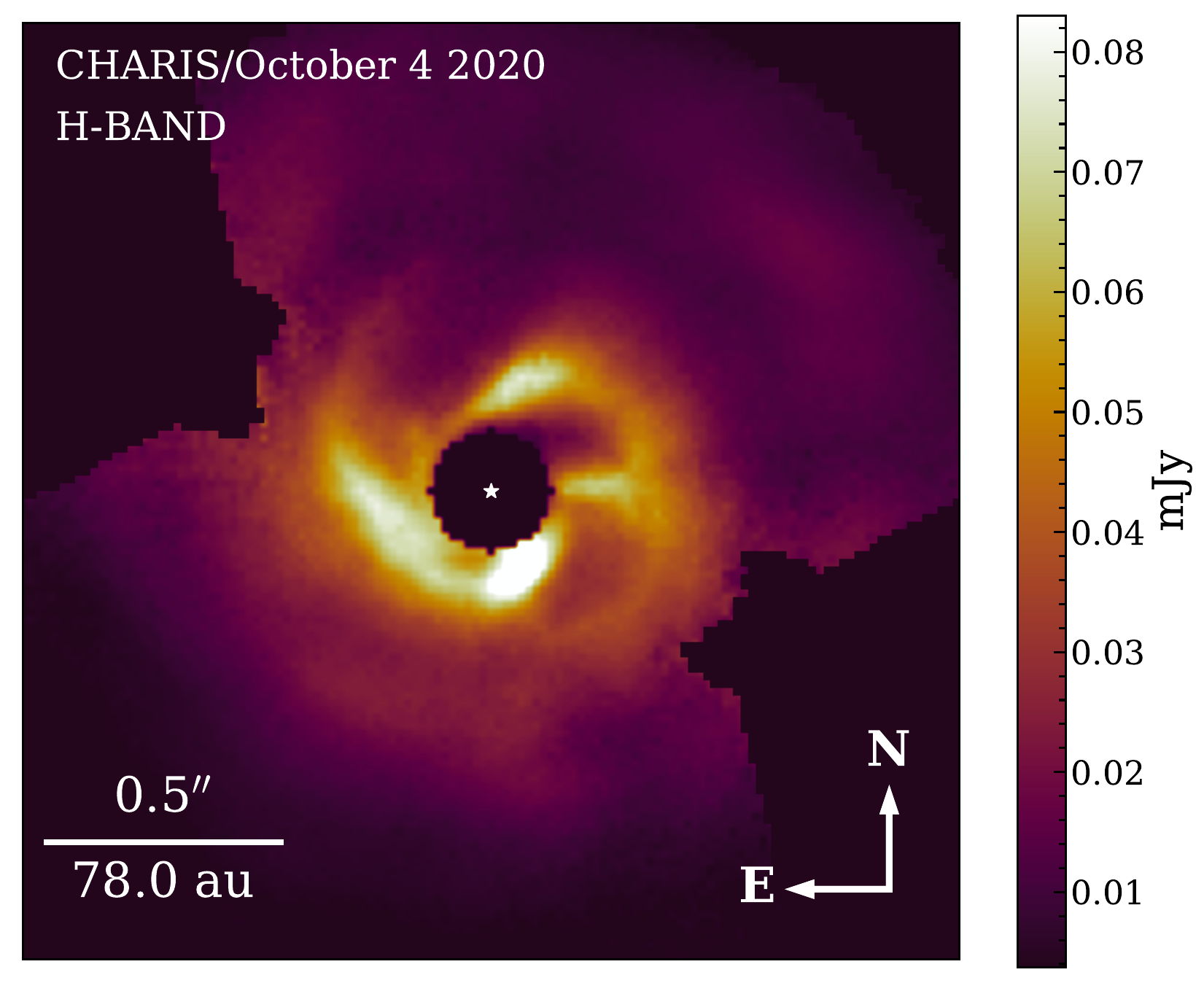}
   \includegraphics[width=0.325\textwidth,clip]{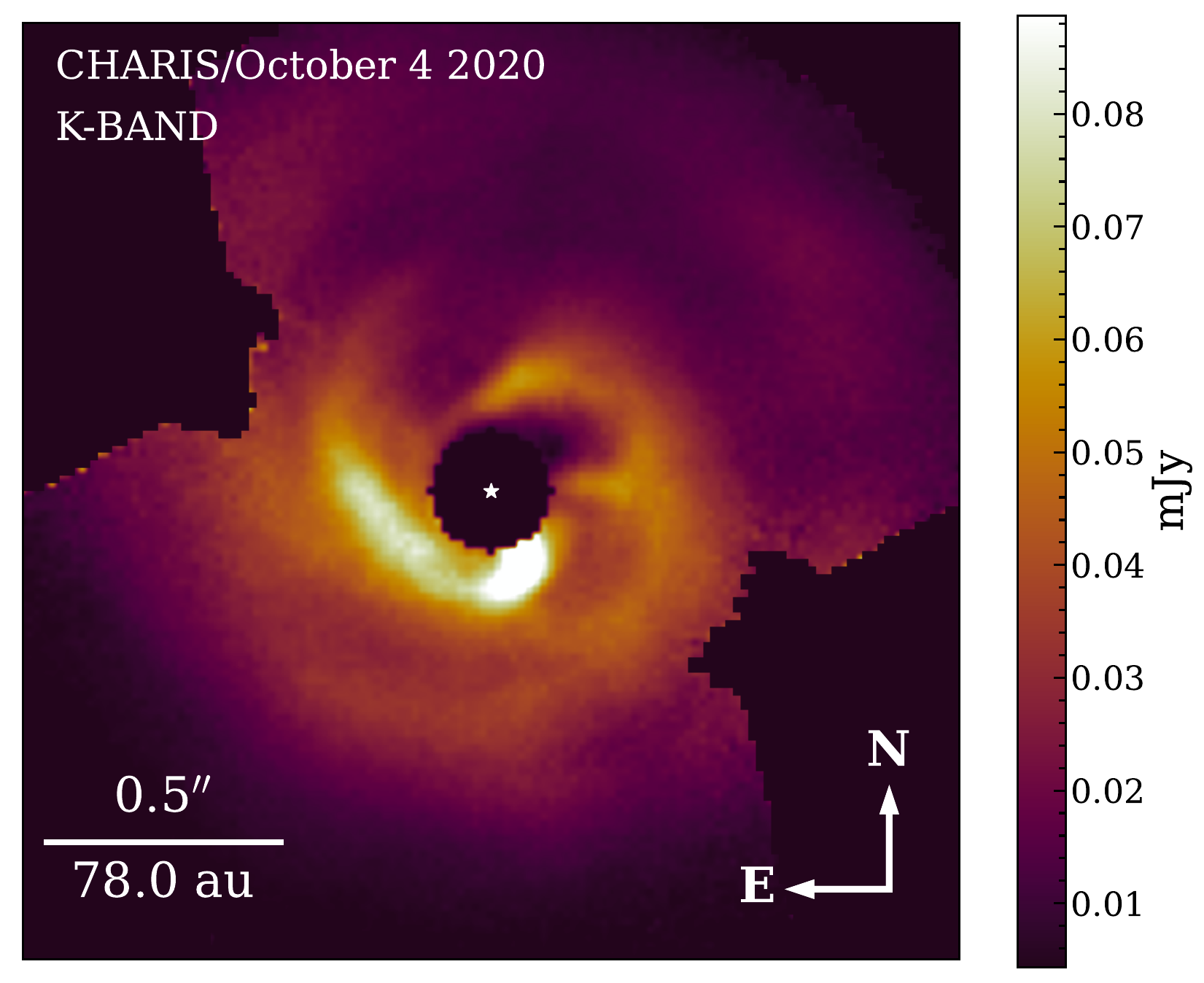}
   \includegraphics[width=0.325\textwidth,clip]{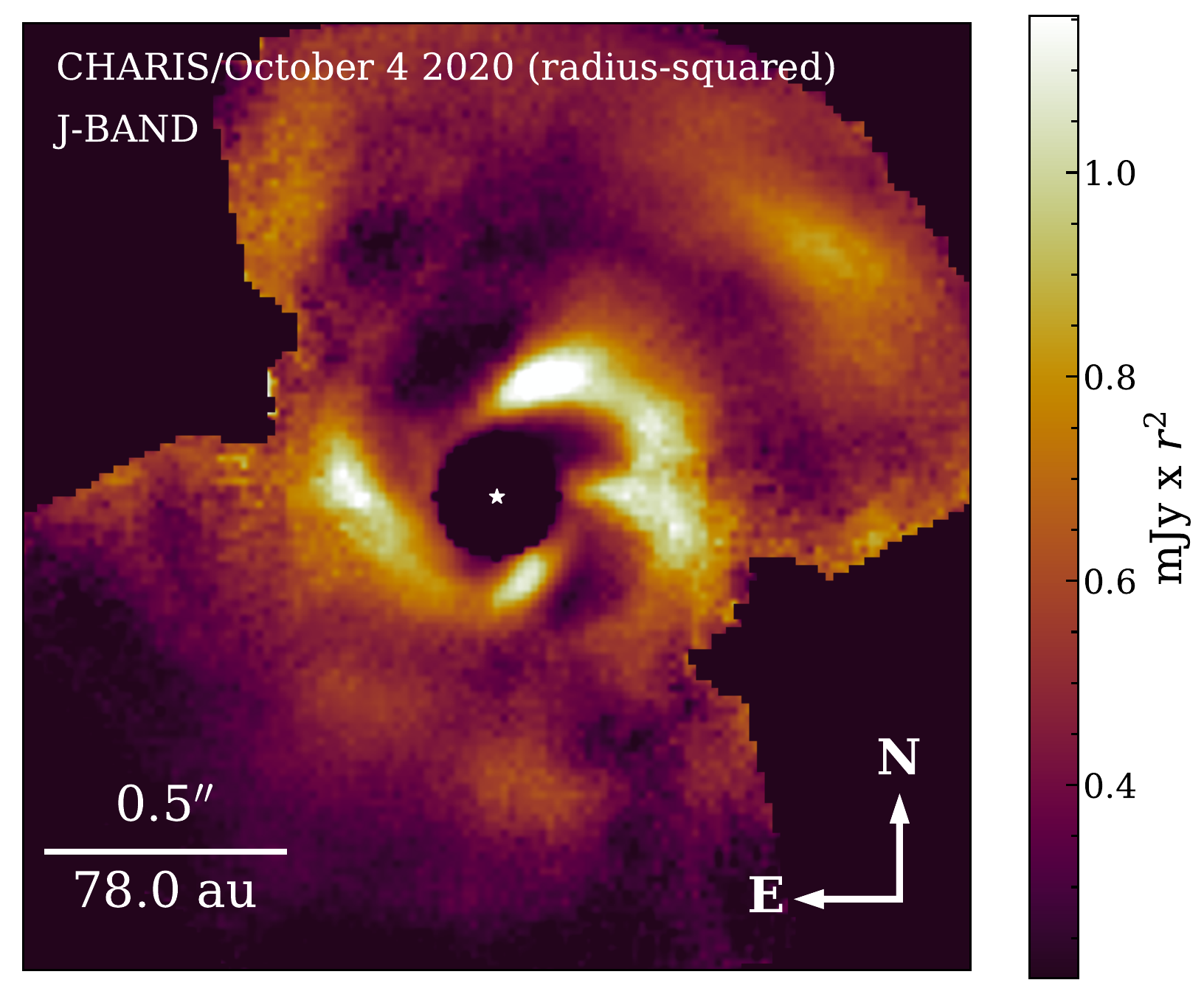}  
   \includegraphics[width=0.325\textwidth,clip]{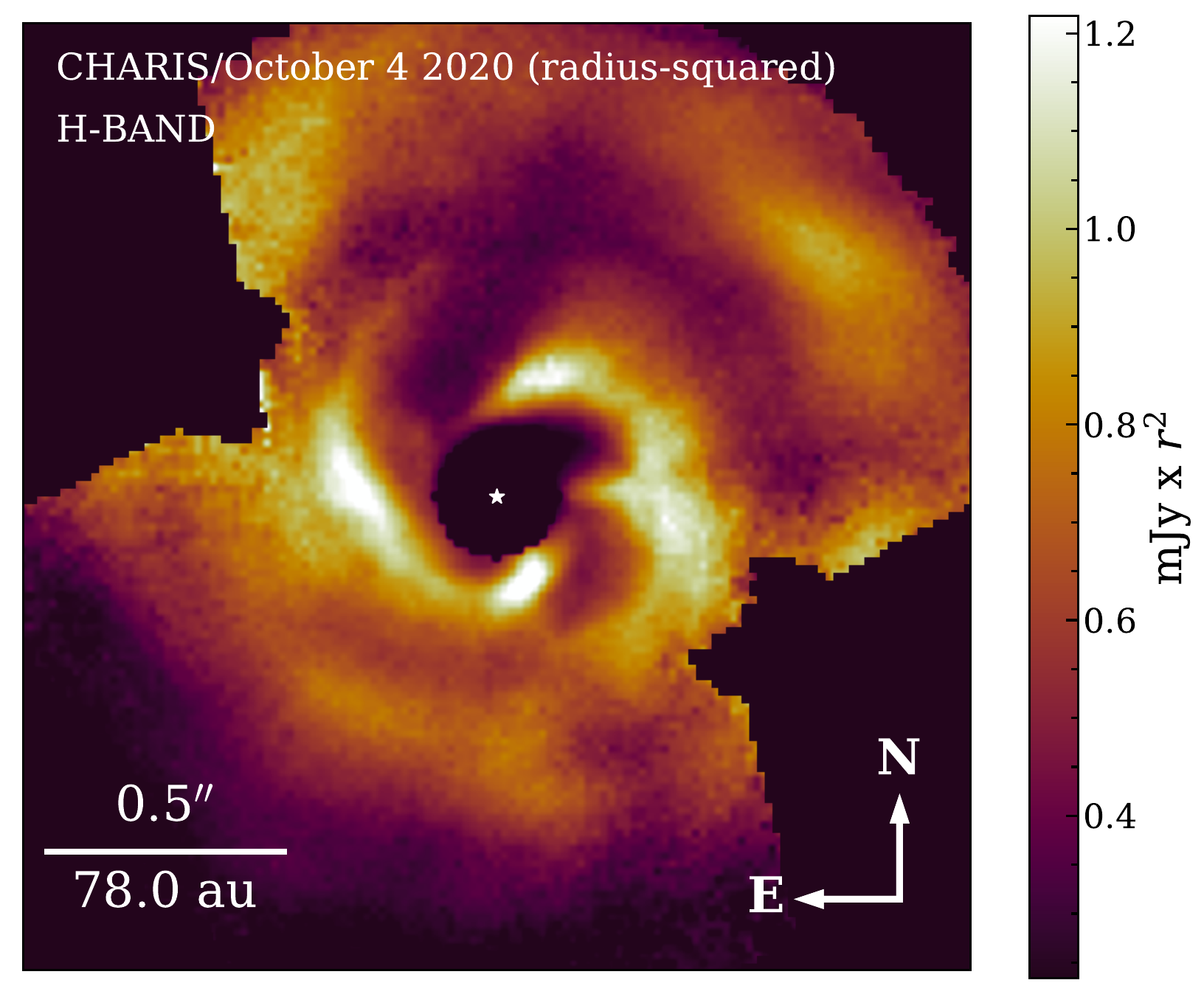}
    \includegraphics[width=0.325\textwidth,clip]{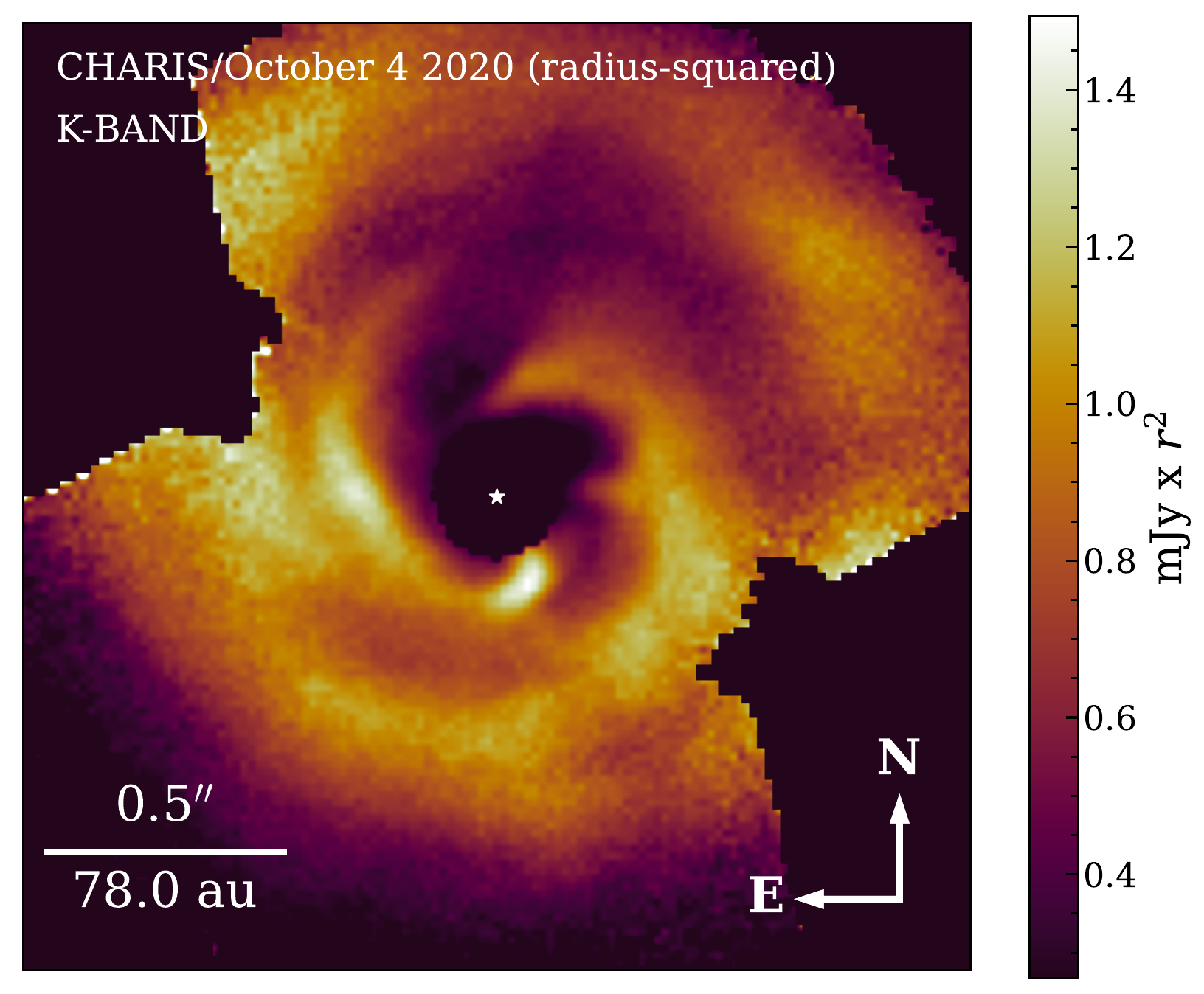}
   \vspace{-0.1in}
   \caption
   {SCExAO/CHARIS AB Aur wavelength collapsed, median combined J(right), H (center), and K(left) Q$_{\rm\phi}$ images from 04 January 2020. J, H, and K passbands were taken to be channels 1 to 6 (1.15-1.37$\mu m$), 7 to 15 (1.42-1.86$\mu m$), and 16 to 22 (1.93-2.36$\mu m$) of the PDI reduced data cubes respectively.  All images are displayed in the same percentile-based stretch from Figure \ref{fig:charispdi}.
   }
   \label{fig:charispdirsq}
   \end{figure*}

\section{Observations and Data Reduction} \label{sec:ObsDatRed}


We observed AB Aurigae with the Subaru Telescope on UT 2020 October 4 (calendar date October 3 2020), using SCExAO with CHARIS in integral field spectropolarimetry mode \citep{Jovanovic2015,Groff2016,Lawson2021} (see Table 1). The Wollaston prism at the entrance of CHARIS splits the incoming light into two 1\farcs{}0 by 2\farcs0 exposures with orthogonal polarization states that simultaneously illuminate the detector. CHARIS's broadband mode covering the $JHK$ passbands (1.1 - 2.4 $\mu m$) was used to acquire 73 exposures of 60.49 seconds each, with a total exposure time of $\approx$ 74 minutes. The observations covered a total parallactic angle rotation of 80.0$^{o}$. Seeing ranged from $\theta_{\rm V}$ $\sim$ 0\farcs{}6 to 0\farcs{}8.

Preprocessing steps largely followed those in \citet{Currie2022a} and are briefly described here. We extracted data cubes from the raw CHARIS data using the standard CHARIS cube extraction pipeline \citep{Brandt2017}, and then used the CHARIS Data Processing Pipeline to carry out sky subtraction, cube registration, and spectrophotometric calibration \citep{Currie2020}. To remove the sky background, the sky frames were scaled, weighting them by the relative median of the reddest channel. In order to spectrophotometrically calibrate the data, we used the empirical spectrum of AB Aur taken with the SpeX spectrograph on UT 2007 December 10 and presented in \citet{Currie2022a}.



After preprocessing, remaining reduction steps followed the PDI module of the CHARIS DPP \citep{Lawson2021}.  We used the double difference technique to subtract the stellar PSF from the observations \citep{Hinkley2009, Quanz2011}. We matched the corresponding half-wave plate (HWP) positions to the observations and grouped them into cycles such that for each set of four HWP positions (0$^{o}$, 45$^{o}$, 22.5$^{o}$, and 67.5$^{o}$) the parallactic angle rotation between the exposures was minimized. For each HWP cycle, double differencing allowed us to calculate the Stokes $Q$ and $U$ images, following the method and notation described by \citet{Lawson2021} and \citet{vanHolstein2020} and briefly summarized here. We first calculated the single sums and differences using the left and right sides of each exposure (corresponding to the vertical and horizontal polarization states for a given HWP angle).
\begin{equation}
X^{\pm}= I_{\rm det, L} - I_{\rm det, R}\\
\end{equation}
\begin{equation}
I_{X^{\pm}}= I_{\rm det, L} + I_{\rm det, R}
\end{equation}
$X^\pm$ here corresponds with the single differences and $I_{X\pm}$ with the single sums. We then calculated the double sums and differences using
\begin{equation}
X = \frac{1}{2}(X^+ - X^-)
\end{equation}
\begin{equation}
I_X = \frac{1}{2}(I_{X+} + I_{X-})
\end{equation}
The double difference, $X$, produces the Stokes $Q$ and $U$ parameters while the double sums give the $I_{Q}$ and $I_{U}$ components of the Stokes $I$ parameter. 
Following similar methods for polarimetry with SPHERE-IRDIS, we utilized a Mueller matrix model to account for instrumental polarization effects along the SCExAO/CHARIS optical path \citep{vanHolstein2020,Joost2021}.

The azimuthal Stokes parameters Q$_{\rm\phi}$ and U$_{\rm\phi}$ as well as the polarized intensity images ($PI$) were then determined for the final output as follows:
\begin{equation}
PI = \sqrt{Q^2 + U^2}    
\end{equation}
\begin{equation}
Q_\phi = -Q\cos(2\phi) - U\sin(2\phi)
\end{equation}
\begin{equation}
U_\phi = Q\sin(2\phi) - U\cos(2\phi)
\end{equation}
with $\phi$ being the azimuthal angle relative to the center of the image. For the reductions presented in Figures \ref{fig:charispdi} and \ref{fig:charispdirsq}, the exposures were then mean combined. Since AB Aur's disk likely has an inclination of $\lesssim$35$^{o}$ \citep{Tang2017} and no characteristic negative values are found in our Q$_{\rm\phi}$ image, we used U$_{\rm\phi}$ as a conservative noise estimate and treat Q$_{\rm\phi}$ as equivalent to PI, following previous work \citep{Canovas2015}.

To facilitate our calculation of the polarization fraction, we used SCExAO/CHARIS observations taken on 03 October 2020.This data was reduced via polarimetry constrained reference star differential (PCRDI) as described in \citet{Currie2022a}\footnote{While we did detect the disk and AB Aur b in total intensity from the Oct 4 data set used to constrain polarized emission, the resulting image quality is far poorer than with the previously-published October 3 data.} . For further specifics on the PCRDI method, see \citet{Lawson2022}.

\section{High-Contrast Polarimetric Images of AB Aurigae with SCExAO/CHARIS}




The AB Aurigae protoplanetary disk is faintly visible in raw SCExAO/CHARIS datacubes.  After PSF subtraction, the disk is resolved from the edge of the coronagraphic mask to the edge of the detector across all wavelength channels (Figure \ref{fig:charispdi}). Due to the limited parallactic angle rotation and the reduced field of view of CHARIS's PDI mode (1\arcsec{} x 2\arcsec{}), our data only extend to separations of $\approx$ 0\farcs{}5 in the 10 o'clock and 4 o'clock directions along the disk major axis.  

\begin{figure*}[]
 \centering
   \includegraphics[width=0.325\textwidth,clip]{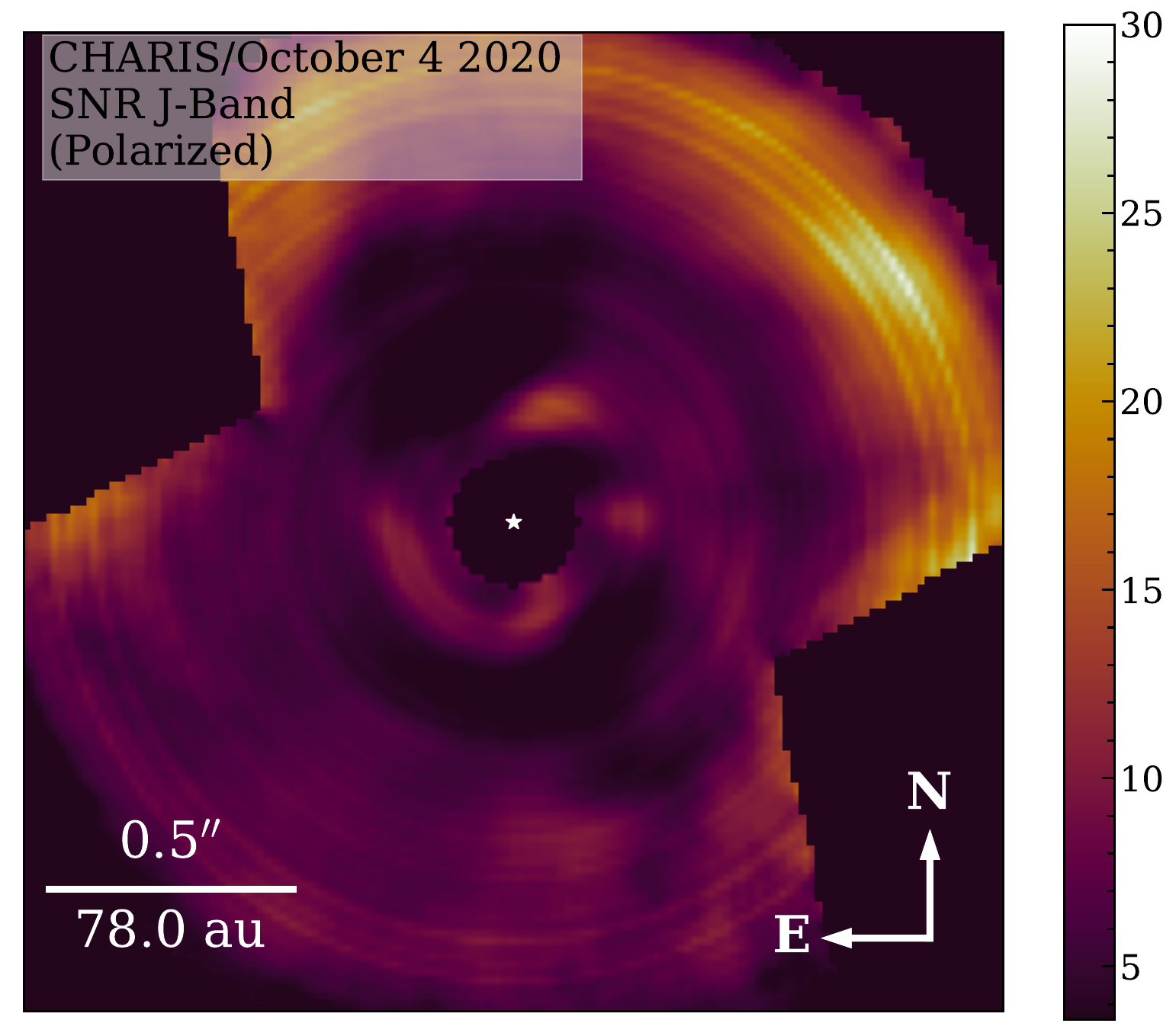}
   \includegraphics[width=0.325\textwidth,clip]{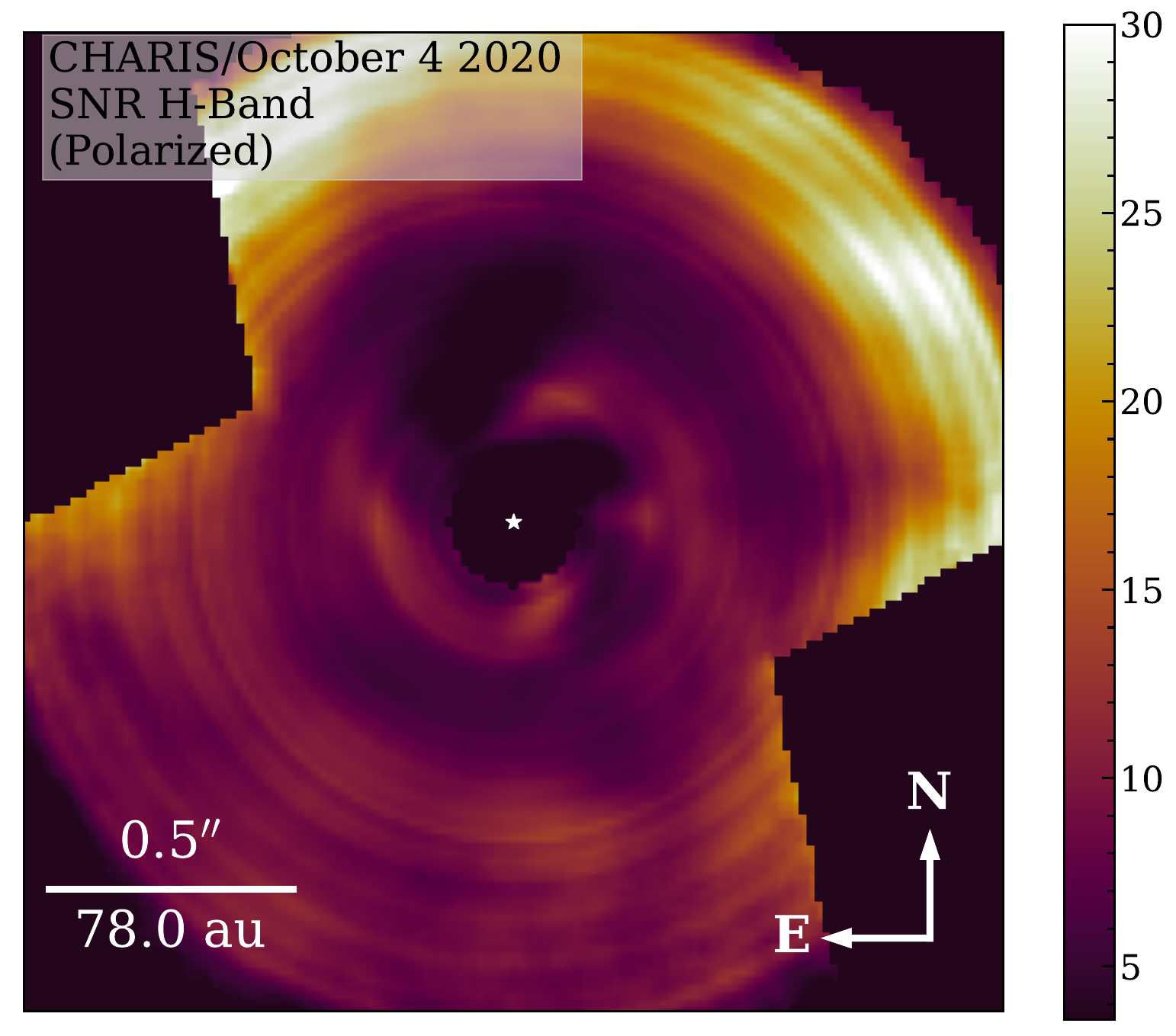}
   \includegraphics[width=0.325\textwidth,clip]{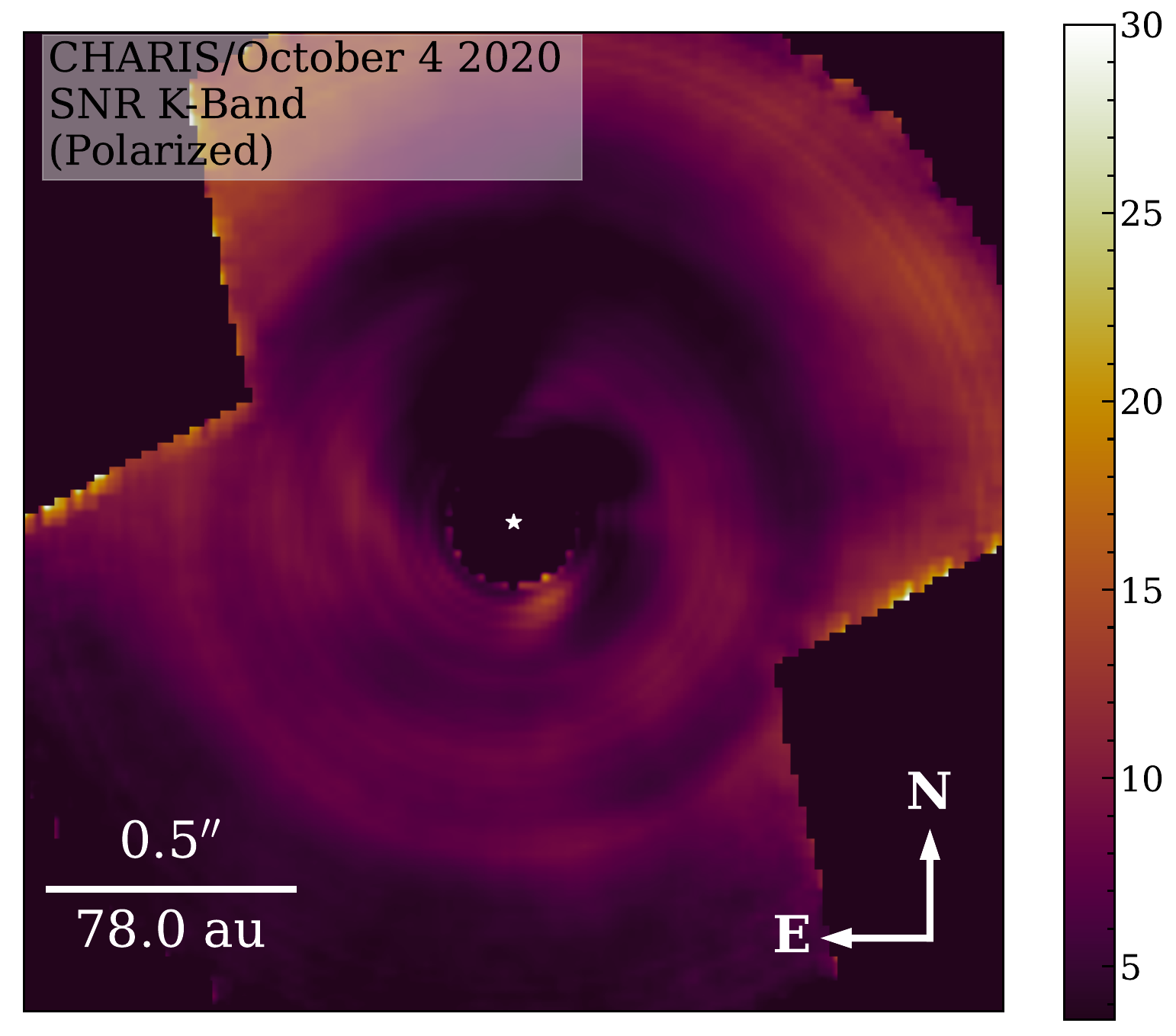}
   \includegraphics[width=0.325\textwidth,clip]{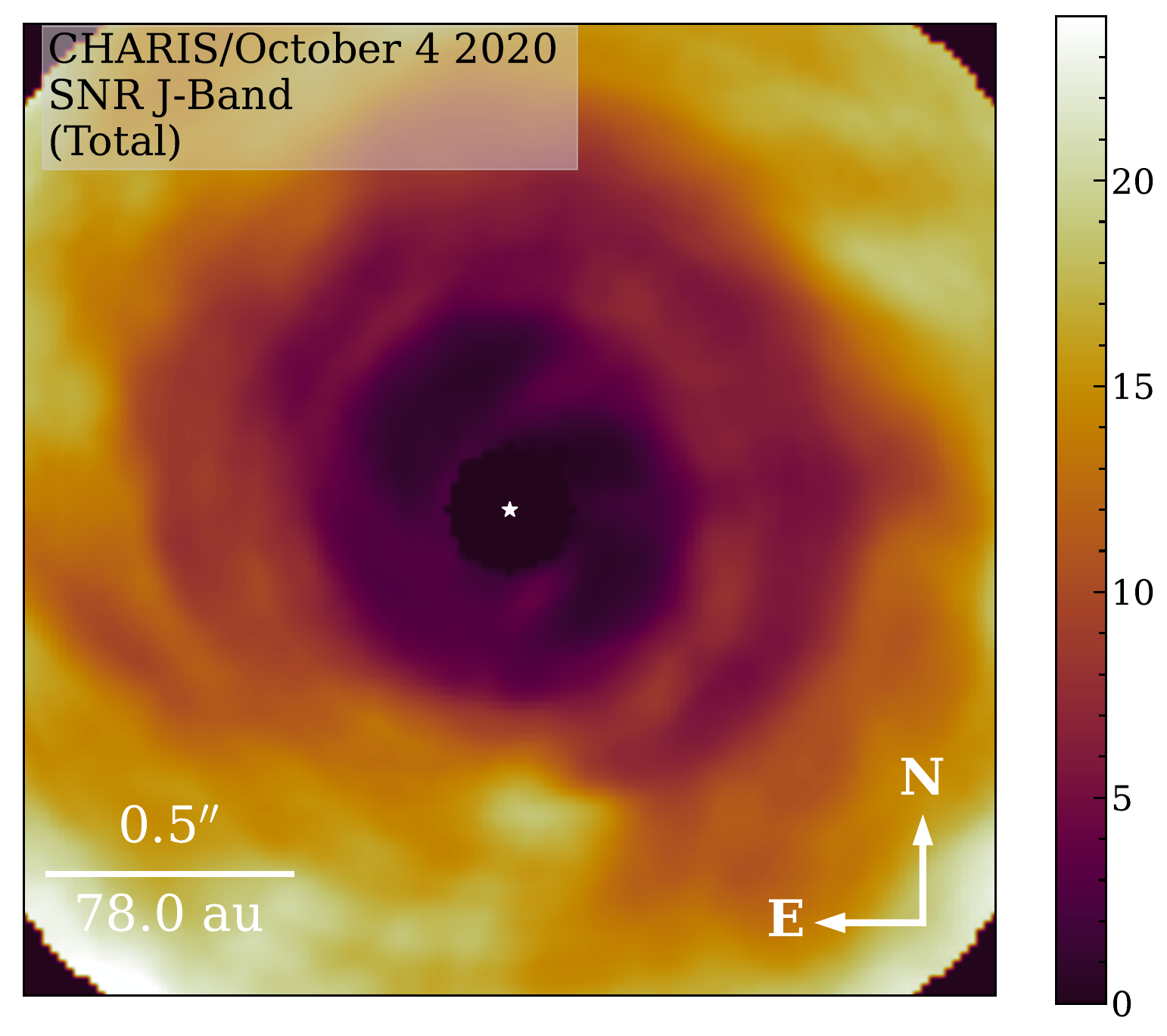}  
   \includegraphics[width=0.325\textwidth,clip]{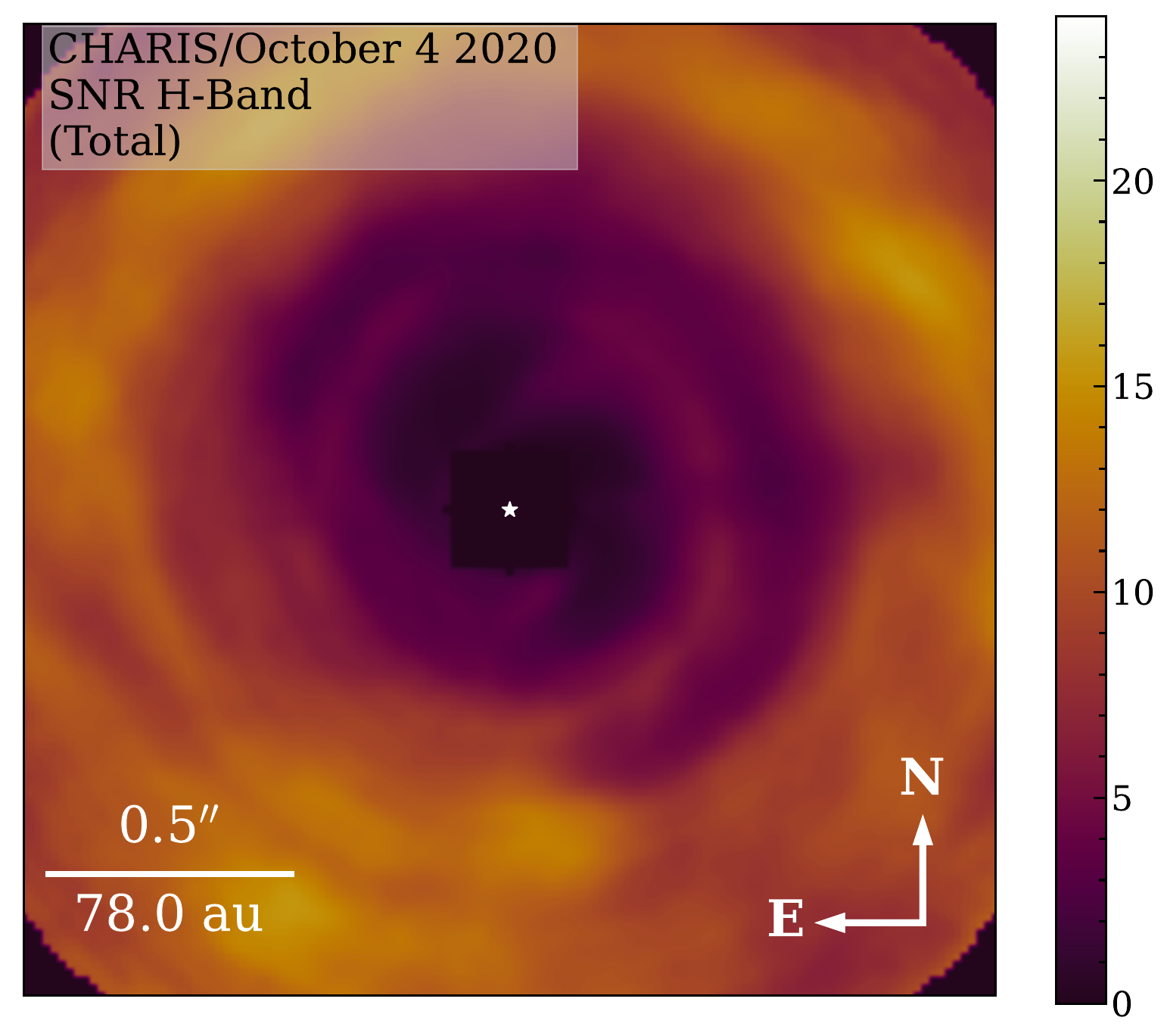}
    \includegraphics[width=0.325\textwidth,clip]{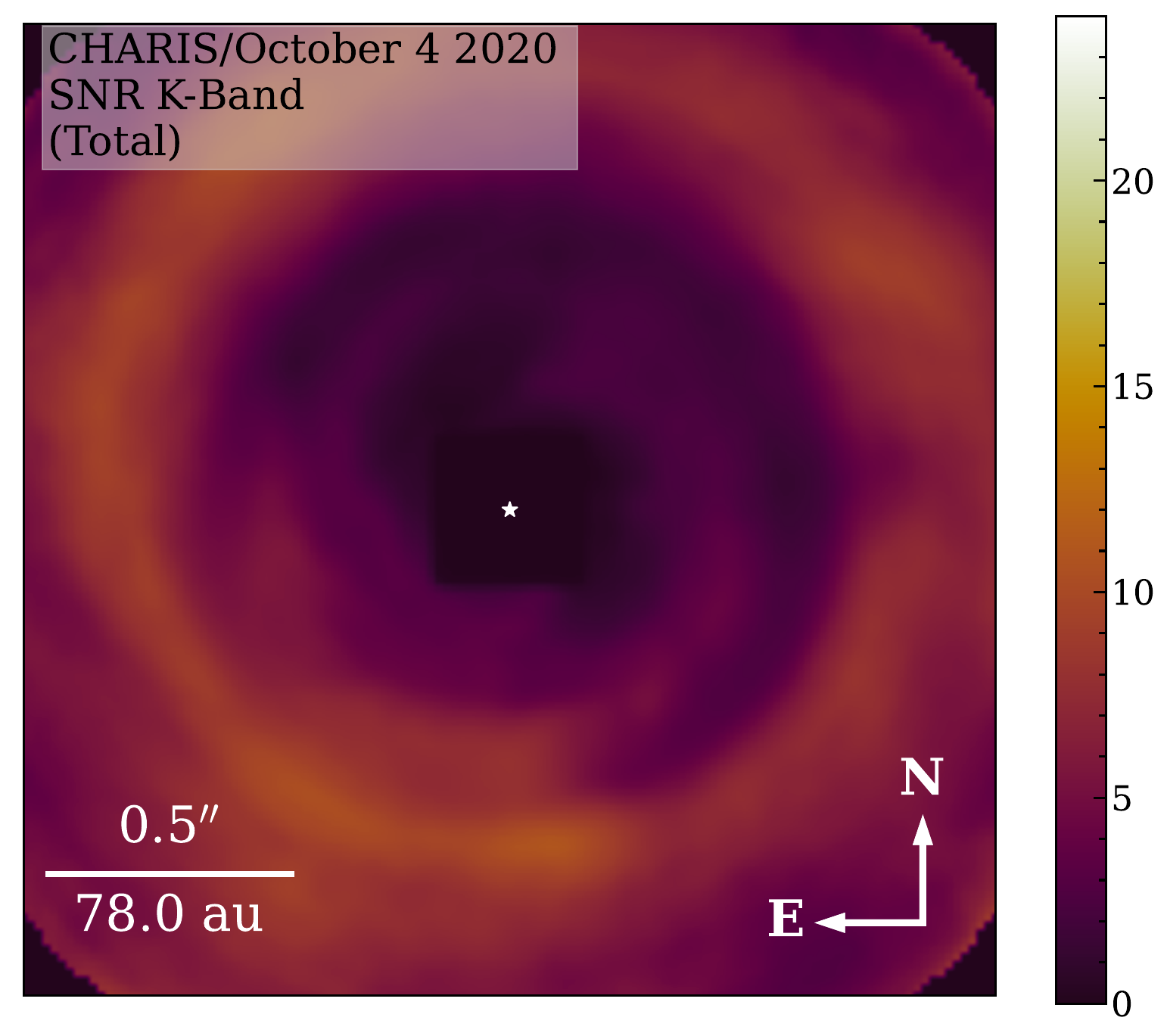}
   \vspace{-0.1in}
   \caption{Signal to noise ratio maps for polarized intensity (top row) and total intensity (bottom row) observations in $J$, $H$, and $K$ passbands.}
   \label{fig:snrmaps}
\end{figure*}

In both the native and radius-squared scaled images in Figure \ref{fig:charispdi}, we can clearly identify spiral structures at 0\farcs{}2 previously detected in polarimetric data \citep{Hashimoto2011,Boccaletti2020} and coincident with CO gas emission in the submillimeter \citep{Tang2017}.   The start of the eastern spiral coincides with the knot or ``twist`` feature seen with SPHERE polarimetry \citep{Boccaletti2020}.  The western spiral includes a bridge-like emission feature at PA = 270$^{o}$.   

\begin{figure*}[!ht]
    \includegraphics[width=\textwidth]{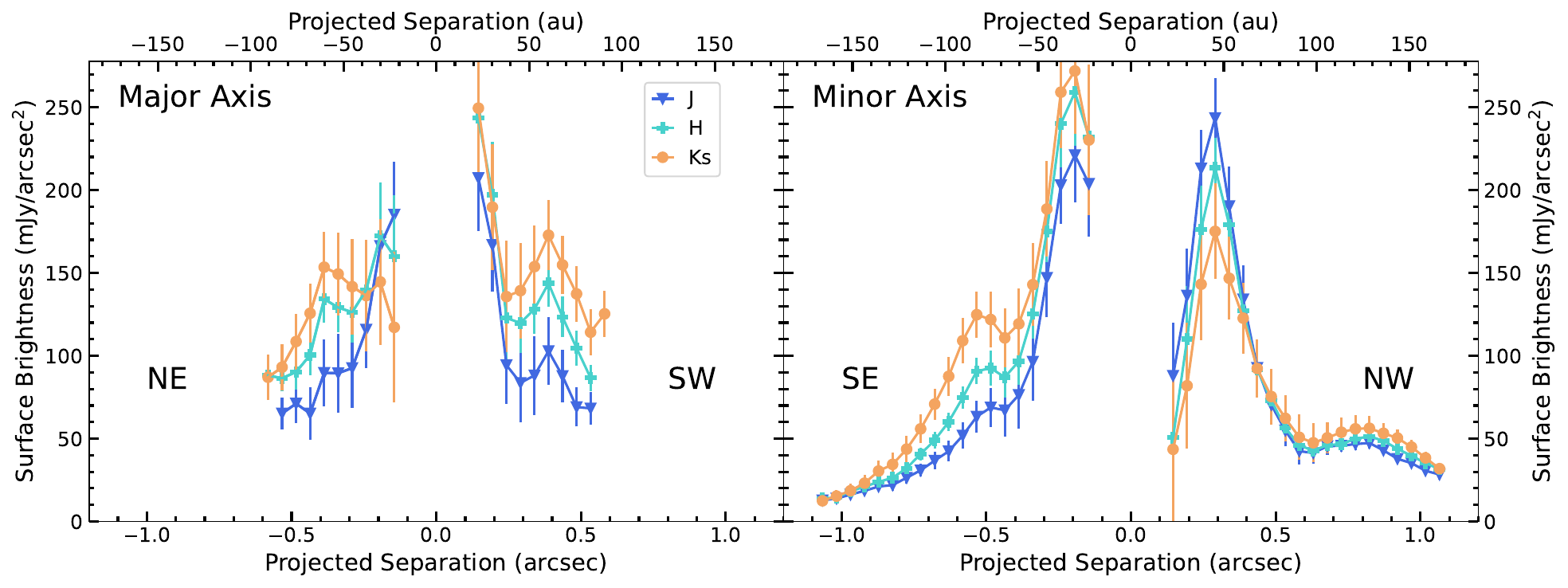}]
    \caption{Surface brightness as a function of radial separation along the major (right) and minor axis (left) of AB Aur's disk. The values for $J$, $H$, and $K$ are shown in blue, turquoise, and orange respectively.}
    \label{fig:sbJHK}
\end{figure*}

The images of AB Aur's disk in the J, H, and K passbands (Figure \ref{fig:charispdirsq}) probe the wavelength dependence of the disk's features.  In $J$ and $H$ band, signal from the western spiral terminates at $\rho$ $\sim$ 0\farcs{}2--0\farcs{}3; however, the K band image suggests that this emission extends to $\approx$ 0\farcs{}5--0\farcs{}6, just interior to and then clockwise from AB Aur b, coincident with the ALMA-identified $CO$ gas spiral.   The outer regions of the disk become progressively brighter relative to the spirals at longer wavelengths, a trend most clearly visible in the radius-squared images.   Comparing our $K$ band polarimetric data with $L_{\rm p}$ total intensity imaging \citep{Betti2022,Currie2022a} and ALMA data \citep{Tang2017} suggests that the inner boundary of the submm dust ring ($\rho$ $\approx$ 0\farcs{}7) may be visible in scattered light at $\lambda$ $\gtrsim$ 2 $\mu m$.

Some concentrated signal may be detected close to the region of AB Aur b in J band channels bluer than 1.3 $\mu m$, but the signal is featureless at H and K bands, indistinguishable from that of the surrounding disk emission, indicating that AB Aur b is not producing any detectable excess polarized signal. The f1 point-like source identified by \citet{Boccaletti2020} to the southwest of the coronagraphic mask is visible in all three passbands and consistently displays increased intensity relative to the rest of the eastern spiral (ranging from $\approx$ 1.5 times the intensity in J-band to 1.7 times the intensity in K-band). The second feature, f2, is not definitively detectable in our observations.



\section{Analysis of the AB Aurigae Protoplanetary Disk}

\subsection{Surface Brightness Profiles}

\begin{figure*}
    \includegraphics[width=\textwidth]{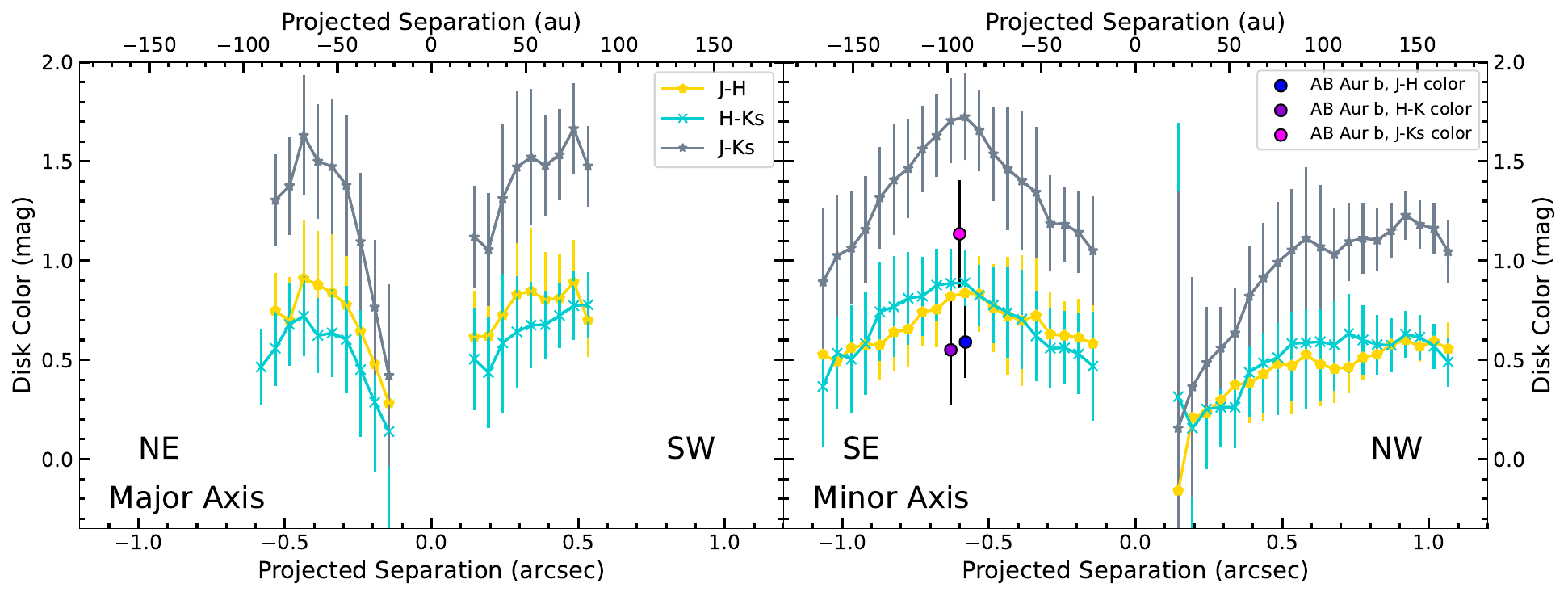}
    \includegraphics[width=\textwidth]{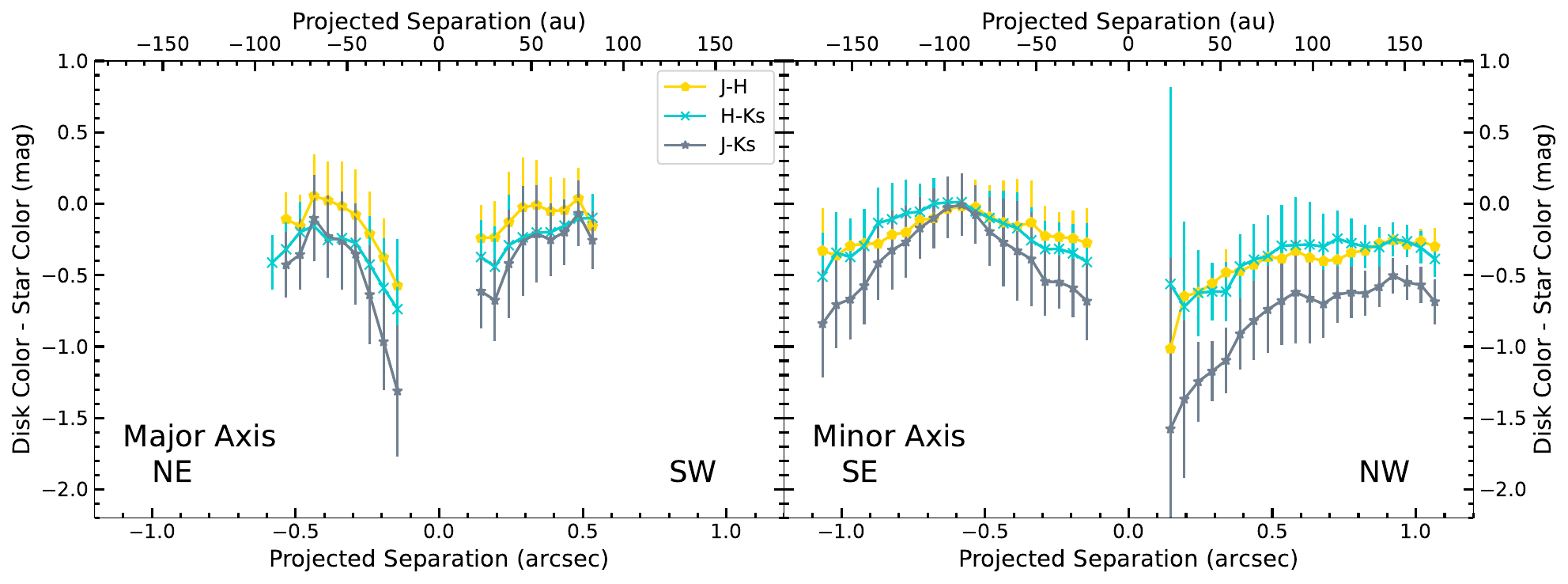}
    \caption{(Top) Color Excess as a function of radial separation for the major and minor axis. $J-H$ is shown in yellow, $H-Ks$ in blue, and $J-Ks$ in gray. AB Aur b's $J-Ks$, \textbf{$J-H$, and $H-K$} colors are also shown for reference (pink, blue, and purple dot respectively) at it's corresponding separation, though it is important to note that it does not physically lie along the minor axis. AB Aur b's $J-H$ and $H-K$ colors are offset from the actual separation value for clarity. (Bottom) Color Excess with stellar color subtracted as a function of radial separation along the major and minor axis of AB Aur's disk. $J-H$ is shown in yellow, $H-Ks$ in blue, and $J-Ks$ in gray.}
    \label{fig:colorprof}
\end{figure*}    
To compute the disk's surface brightness along its major and minor axes, we performed boxcar smoothing at the resolution scale of the median wavelength at $J$, $H$, and $K$.  We conservatively estimated spectrophotometric errors in the following manner.   First, we used the U$_{\rm\phi}$ data to calculate noise and assumed that the residual noise after PSF subtraction is a function of angular separation, similar to the noise profile expected for PSF subtracted total intensity data \citep[e.g.][]{Currie2011,Marois2008a}.   From the final U$_{\rm\phi}$ cube subdivided into the major near-IR passbands, we then computed the robust standard deviation of smoothed pixel values as a function of separation.  In total intensity, the disk signal covers the entire CHARIS field of view.  To conservatively estimate the total intensity SNR, we calculate the signal by replacing each pixel by its sum within a FWHM-sized aperture.  We then estimate the noise by derotating each PSF-subtracted exposure opposite the sign and amount needed to align the exposure to true north, median-combining the exposures, and computing the radially-dependent noise profile for each bandpass.  In both cases, we excluded data interior to the coronographic mask edge from both the brightness profile and the noise estimates.  The typical signal-to-noise ratio (SNR) of the disk in polarized light as computed from the smoothed Q$_{\rm\phi}$ and U$_{\rm\phi}$ is $\approx$10-20 per resolution element, highest in $H$ band and slightly lower in $J$ and $K$ band (Figure \ref{fig:snrmaps}).  In total intensity, the typical SNR is comparable.


AB Aur's disk is consistently brightest in K-band, especially along the forward scattering edge to the southeast (Figure \ref{fig:sbJHK}). The region interior to $\approx$ 0\farcs{}2 in the NE and $\approx$ 0\farcs{}4 in the NW where the profiles trace the spirals is a notable exception to this trend, appearing to be bluer than the rest of the disk.  These trends are consistent with a visual inspection of Figure \ref{fig:charispdirsq}, where the outer disk increases in brightness relative to the spirals at longer wavelengths.

Figure \ref{fig:colorprof} displays the disk colors (top) and the intrinsic scattering of the disk (disk colors minus the star's colors) as a function of angular separation along the major and minor axes.   The AB Aur disk has extremely red colors (e.g. $J$-$K$ $\sim$ 1--1.8) primarily due to the red intrinsic colors of the scattering source (the AB Aur primary, unresolved sub-au scale disk emission) \citep{Currie2022a}.  AB Aur b is also bluer than the colors of the average polarized intensity of the disk by $\approx$ 0.3 mag (magenta dot).  Removing the star and sub-au scale emission shows that the intrinsic scattering of the outer disk is blue.  

\subsection{Polarization Spectrum}
\begin{figure}
\centering
    \includegraphics[width=0.5\textwidth]{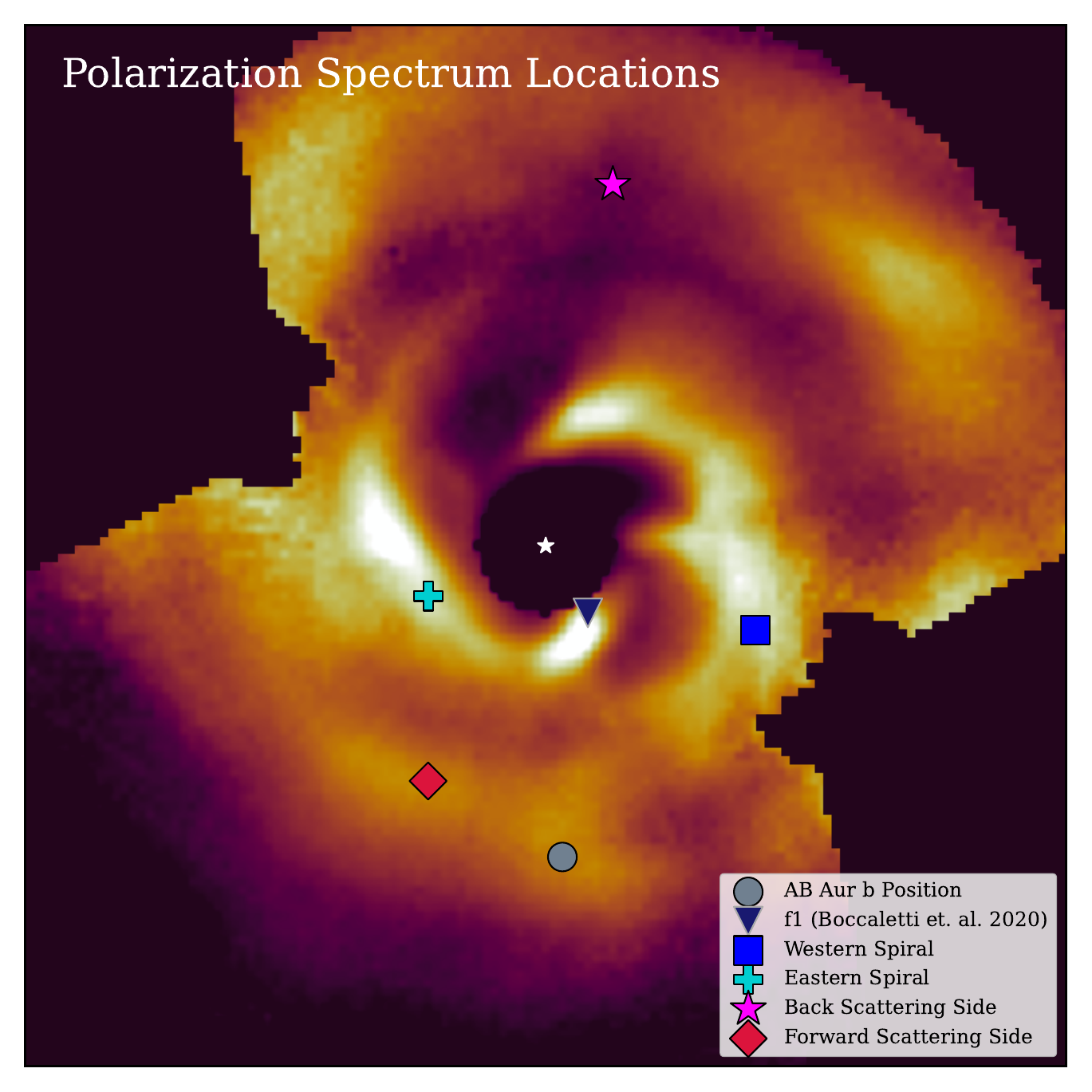}
    \caption{The radius-squared broadband image of the AB Aurigae disk with locations at which we extract polarization spectra denoted by different symbols.}
    \label{fig:specloc}
\end{figure}

We extracted polarized-light spectra of the AB Aur disk at various locations to investigate spatial variations in the disk's scattering (see Figure \ref{fig:specloc}).   For each spectral channel we apply box-car smoothing at the channel's resolution scale and extract the disk surface brightness at every location.  As before, we use the $U_{\rm \phi}$ data to estimate surface brightness uncertainties.
After applying boxcar smoothing at the resolution scale of each wavelength channel, spectrum was taken at various points throughout the disk.  

\begin{figure}
    \includegraphics[width=0.5\textwidth]{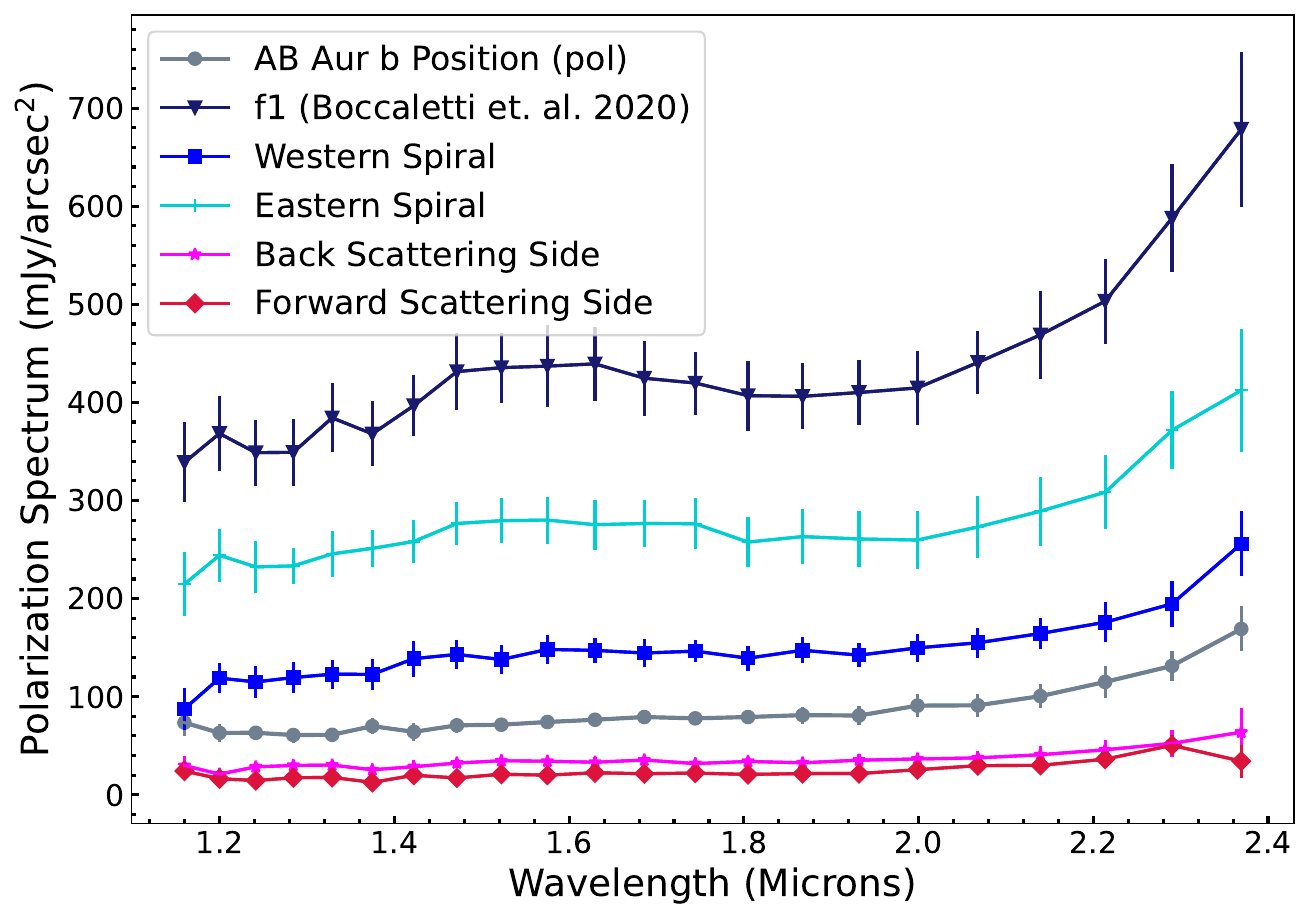}\\
    \includegraphics[width=0.5\textwidth]{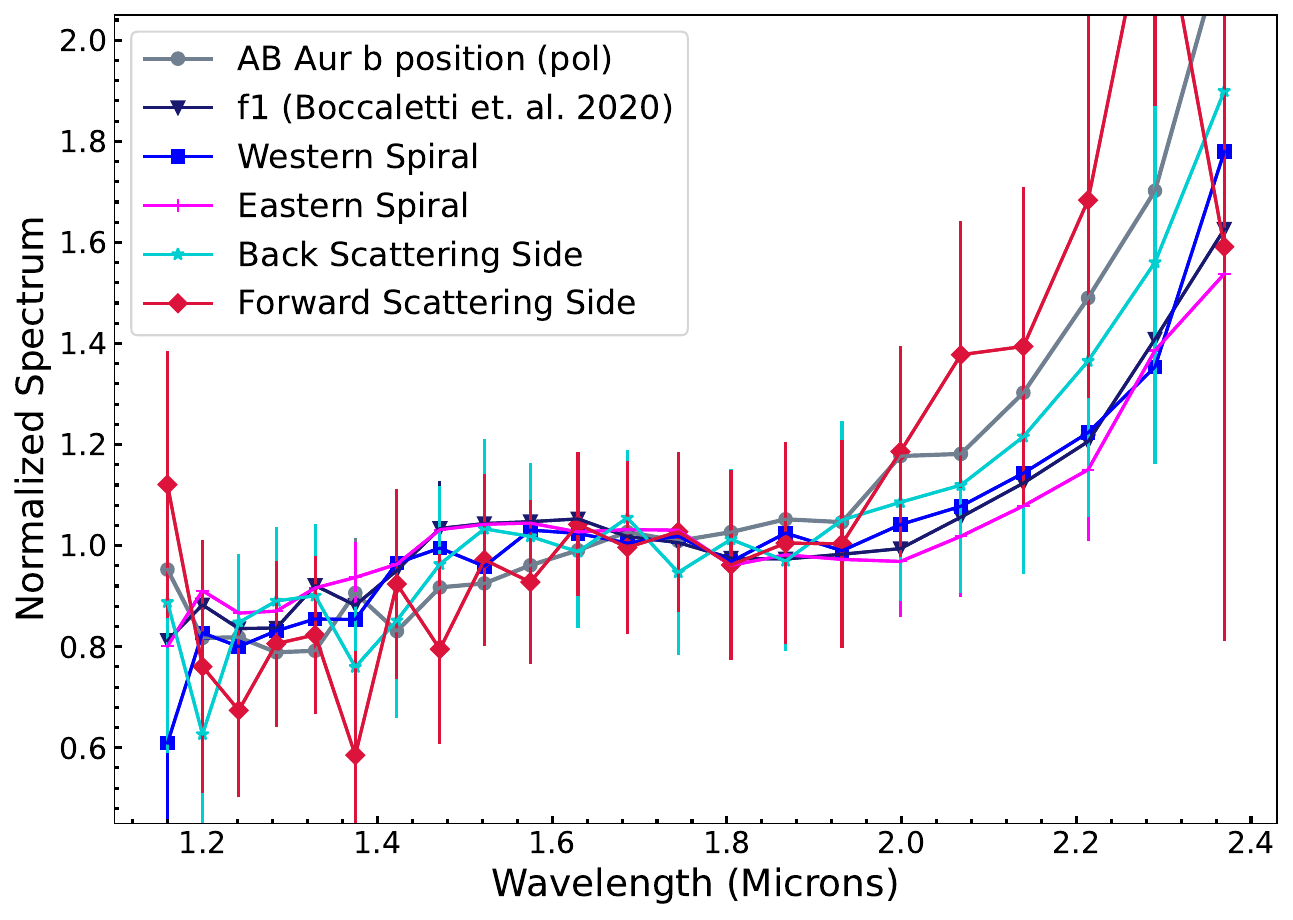}\\
    \includegraphics[width=0.5\textwidth]{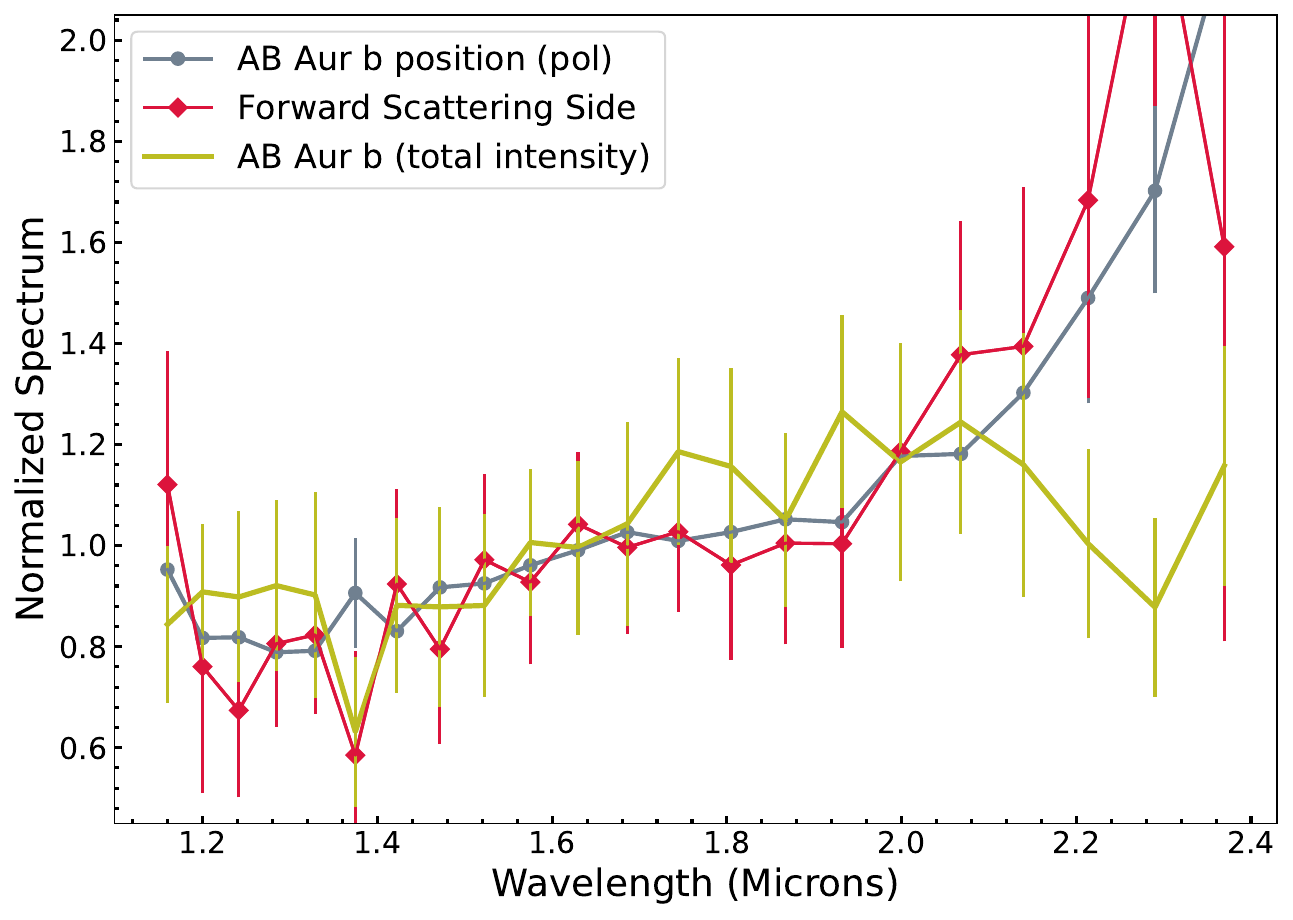}
    \caption{Measured polarization spectra (top) and normalized (by the median surface brightness for each spectrum; middle) polarization spectra for various locations in the disk listed in Table \ref{table:pdfspecpos}. Comparison of measured polarization spectra and total intensity spectra at the location of AB Aur b (bottom).}
    \label{fig:polspec}
\end{figure}

Figure \ref{fig:polspec} shows the polarized light spectrum extracted at several points in the disk.   Specifically, we consider the position of AB Aur b, the position of the f1 source from \citet{Boccaletti2020}, both spirals at $\rho$ $\le$ 0\farcs{}2, the forward-scattering edge of the outer disk, and the back-scattering side of the disk at a similar angular separation.   Table \ref{table:pdfspecpos} lists the coordinates at which we extract spectra.

\begin{deluxetable}{cccc}
     \tablewidth{0pt}
    \tablecaption{AB Aur Disk Regions Sampled for Polarization Spectra}
    \tablehead{\colhead{ID} & \colhead{E \arcsec{}} &  \colhead{N \arcsec{}} & \colhead{Description}}
    \startdata
    1 & -0.032 & -0.598 & AB Aur b position, \citet{Currie2022a}\\
    2 & -0.081 & -0.129 & \citeauthor{Boccaletti2020} ``twist" feature\\
    3 & -0.403 & -0.162 & western spiral \\
    4 &  0.226 & -0.097 & eastern spiral \\
    5 & -0.129 & 0.694 & back scattering, minor axis\\
    6 &  0.226 & -0.452 & forward-scattering peak, minor axis
    \enddata
    \tablecomments{The forward-scattering peak and back scattering disk region correspond to regions with the maximum and a minimum surface brightness along the minor axis.}
    \label{table:pdfspecpos}
    \end{deluxetable}
The polarized light spectra show several characteristics of note.   The f1 feature identified by \citet{Boccaletti2020} lies at the terminus of the eastern spiral but is brighter than the rest of the eastern spiral at all wavelengths. In polarized intensity, it is the brighest feature in the AB Aur disk.    The polarized light spectra normalized by flux are similar, especially in J and H band (Figure \ref{fig:polspec}, right panel).   However, the forward-scattering peak of the outer disk may be particularly red compared to other regions through K band, while the spirals and f1 source are the slightly bluer than other regions. The AB Aur b spectrum in total intensity is bluer than all regions of the disk, consistent with the differences seen from the disk in total intensity from \citet{Currie2022a}.  
\subsection{Polarization Fraction}
We computed the polarization fraction ($PF$) in the $J$, $H$, and $K$ bandpasses across AB Aur's disk: the ratio of the polarized intensity signal ($PI$) drawn from the $Q_{\rm \phi}$ cubes to the total intensity signal, $I$.  
The total intensity was derived from previous observations taken on UT 2020 October 3 and reduced using polarimetry-constrained reference star differential imaging \citep{Lawson2022}.  Both Q$_{\rm\phi}$ and $I$ data were smoothed at the resolution scale of the observation's median wavelength in each bandpass.   We excluded regions without signal in $PI$ data: e.g. along the major axis at $\rho$ $\ge$ 0\farcs{}6 and interior to the coronographic mask edge.  

\begin{figure*}
    \includegraphics[width=1\textwidth]{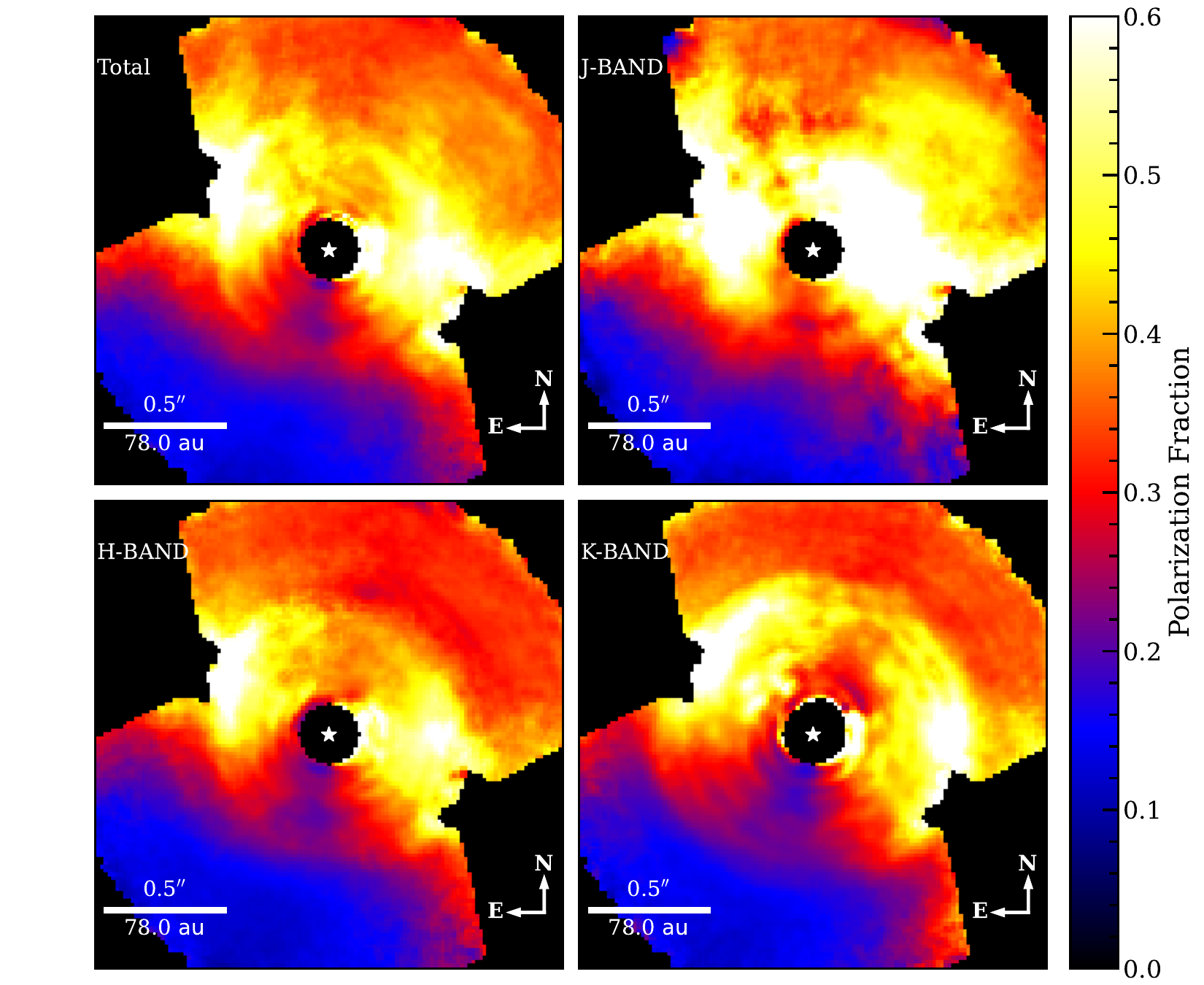}
    \caption{Wavelength collapsed, median combined polarization fraction across AB Aur's disk in broadband, $J$, $H$, and $K$ bands.}
    \label{fig:polfraction}
\end{figure*}

The forward scattering region of the disk displays the lowest polarization fraction, consistently $\approx$0.15.   The polarization fraction value in the backscattering region is about twice as high ($PF$ $\approx$ 0.3-0.4).    The disk major axis displays the greatest polarization fraction, up to $\approx$0.6 (Figure \ref{fig:polfraction}).  

\begin{figure*}
    \includegraphics[width=1\textwidth]{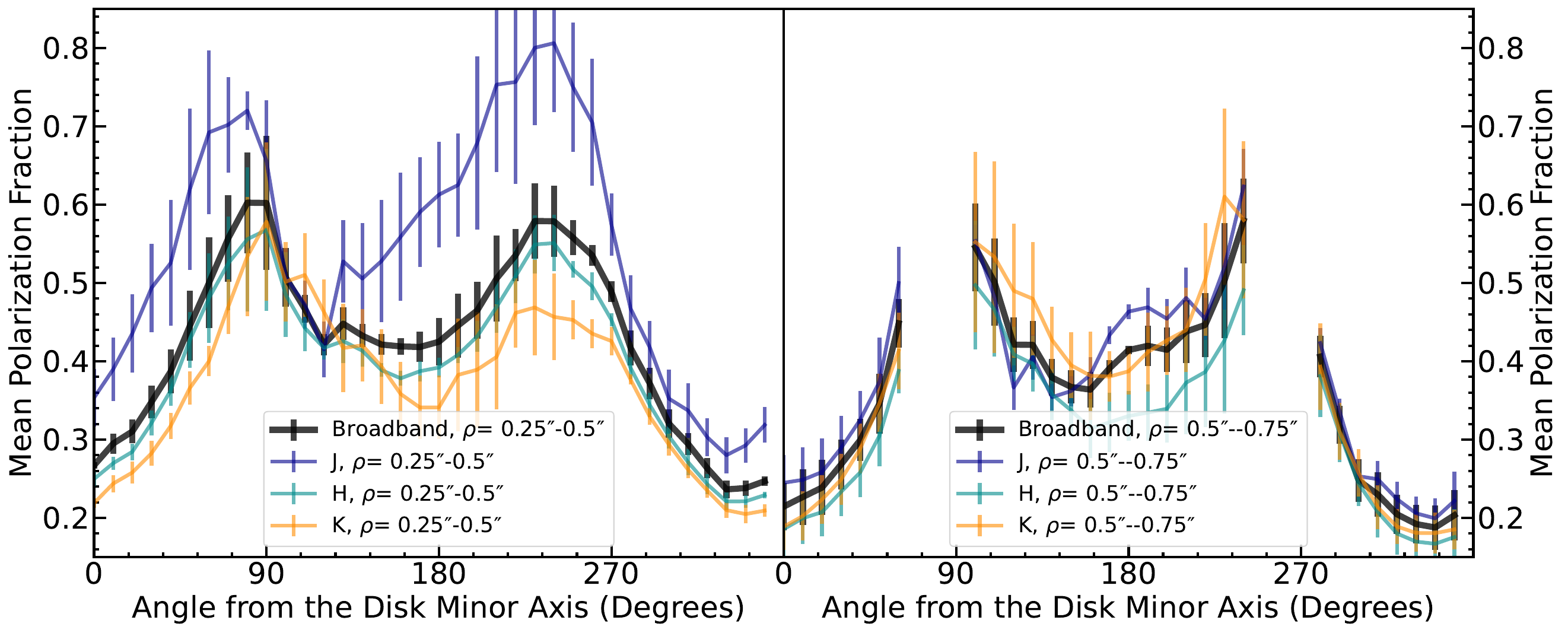}
    \caption{Mean polarization fraction vs. angle from the disk minor axis for CHARIS broadband, $J$, $H$, and $K$.   The error bars correspond to the standard deviation of the polarization at each angle.  Measurements missing from the right-hand panel (near 90 and 270 degrees) correspond to regions with limited field of view on the CHARIS detector.}
    \label{fig:polphase}
\end{figure*}

To more quantitatively assess the AB Aur disk's polarization properties, we computed the polarization fraction as a function of angle from the disk major axis (position angle of 54$^{o}$ west of north, \citealt{Tang2017}), following similar analysis in \citet{Perrin2009} for 2.0 $\mu m$. 
 Figure \ref{fig:polphase} shows the polarization fractions for the entire CHARIS bandpass and the individual $J$, $H$, an $K$ bandpasses as a function of angle from the disk minor axis at 0\farcs{}25--0\farcs{}5 (left panel) and 0\farcs{}5--0\farcs{}75 (right panel) computed in increments of 10 degrees after box-car smoothing the images by each bandpass's resolution element.   At small angular separations ($\rho$ $\approx$ 0\farcs{}25--0\farcs{}5), the $PF$ value is slightly higher along the major axis in the $J$ band image than at $H$ and $K$ band or in broadband.   At face value, this trend suggests that the polarization fraction decreases with wavelength.   However, at wider separations, evidence for a wavelength dependent polarization fraction is weaker as $J$ band has only marginally higher polarizability than redder passbands.  The trend may reflect a real difference in polarization with wavelength or reflect challenges in measuring polarization fraction at $J$ band (e.g. due to an underestimate of the total intensity signal).
Otherwise, we find no clear trends in polarization fraction vs wavelength.    

Our $PF$ levels at $K$ band generally agree with those derived from 2 $\mu m$ polarimetry at 0\farcs{}7--1\farcs{}5 obtained from HST data and presented in \citet{Perrin2009}.   CHARIS's limited field-of-view prevents us from directly comparing our peak PF at 0\farcs{}5--0\farcs{}75 near 90 and 270$^{o}$ (i.e. along the major axis) to theirs.  However, measurements at other angles -- e.g. along the disk minor axis and angles in between the major and minor axes -- match values from \citeauthor{Perrin2009} to within errors.

\section{Modeling the Scattering Properties of the AB Aurigae Disk}
The polarization of AB Aur's disk seen in scattered light likely depends on both its structure and its dust grain properties (composition, porosity) \citep[e.g.][]{Tazaki2019,Min2009}.   To investigate the AB Aur disk structure and dust grain composition we therefore modeled its scattered-light images and spectral energy distribution (SED) using the MCFOST 3D Monte Carlo Radiative Transfer code with ray tracing \citep{Pinte2006,Pinte2009}, the same code used in prior investigations \citep{Perrin2009,Betti2022}.

MCFOST computes the disk thermal profile from user-specified power-law fits to the disk density profiles and scale height, radial extents, overall dust/gas mass, and dust properties.  For the latter, MCFOST nominally uses Mie theory to calculate the dust's optical properties, treating the grains as spheres with different porosities, species, and size distributions.  The assumption of simple spheres with Mie-derived optical properties likely limits the accuracy of synthetic scattered-light images compared to approaches using the discrete dipole approximation (DDA; \citealt{Draine1994}) or more recent work that models the dust as (highly-porous) aggregates consisting of nanometer-scale monomers with fractal dimensions \citep{Tazaki2019}.   The aggregate dust grain models show that both the polarization fraction and azimuthal profile can change substantially depending on the assumed dust structure.  Even within the porous aggregate formalism, the disk polarization fraction varies with the monomer size \citep{Tazaki2023}.   

Because the disk scattered light appearance is dependent on the dust structure (which can only be assumed, not directly measured) and the method for computing optical constants\footnote{The AB Aurigae disk is morphologically complex: another reason why a detailed and quantitative comparison with radiative transfer model predictions for morphologically simple models is not justified.  MCFOST can impute disk models with arbitrary density profiles (e.g. those driving large-scale spirals as seen in AB Aur's disk), but we consider this modeling to be beyond the scope of our analysis, whose sole focus is to qualitatively investigate the dust properties of the disk.} (which vary in sophistication and ease of implementation), we focused on qualitative comparisons between MCFOST models and the CHARIS data beyond simply computing the $\chi^{2}$ statistic comparing the models to data.  Disks comprised of highly-porous dust grains have higher polarization fractions than more compact dust, less porous dust.  With high-fidelity images in both total and polarized intensity, we can therefore coarsely investigate the properties of dust grains in AB Aur's disk responsible for reflecting starlight.  
\begin{table*}
\caption{MCFOST Model Parameter Grid}
\label{table:modelgrid}
\begin{tabular}{cccccccccccc}
\hline
\multicolumn{5}{c}{Fixed Parameters} & \multicolumn{3}{c}{Value} & {Reference}\\
\hline
\multicolumn{5}{c}{$\Rstar$ } & \multicolumn{3}{c}{2.7 $\Rsun$} & \citet{Currie2022a}\\

\multicolumn{5}{c}{$T_{eff}$$^{1}$} & \multicolumn{3}{c}{9800 K} & \citet{Currie2022a,Betti2022}\\

\multicolumn{5}{c}{$\Mstar$} & \multicolumn{3}{c}{2.4 $\Msun$} & \citet{Currie2022a}\\

\multicolumn{5}{c}{Distance} & 
\multicolumn{3}{c}{155.9 pc} & \citet{gaiadr3}\\



\multicolumn{5}{c}{Surface density exponent, $\epsilon$} & 
\multicolumn{3}{c}{1.0} &\citet{Pinte2009,Perrin2009}\\

\multicolumn{5}{c}{Reference radius, $R_0$} & 
\multicolumn{3}{c}{1 au} & \citet{Pinte2009}, this work\\

\multicolumn{5}{c}{Disk flaring index, $\beta$} & 
\multicolumn{3}{c}{1.3} & \citet{Perrin2009}\\

\multicolumn{5}{c}{Disk inner radius, $R_{in}$$^{2}$} & 
\multicolumn{3}{c}{0.15 au} & this work\\

\multicolumn{5}{c}{Disk outer radius, $R_{out}$} & 
\multicolumn{3}{c}{300 au} & this work\\

\multicolumn{5}{c}{Disk Position Angle, $PA$} & 
\multicolumn{3}{c}{234\degree} & \citet{Tang2017,Betti2022}\\


\multicolumn{5}{c}{Maximum grain size, $a_{\rm max}$} & 
\multicolumn{3}{c}{1 $\mu m$} & this work, \citet{Currie2022a}\\

\multicolumn{5}{c}{Grain size power-law, $p$} & 
\multicolumn{3}{c}{-3.5} & \citet{Pinte2009}\\

\hline
\multicolumn{4}{c}{Varied Parameters} & \multicolumn{2}{c}{Range} \\
\hline



\multicolumn{4}{c}{Disk Mass$^{3}$} & 
\multicolumn{2}{c}{[1,2.5, 5, 7.5] $\times$ 10$^{-4}$ M$\sun$ } \\

\multicolumn{4}{c}{Disk Inclination} & 
\multicolumn{2}{c}{[15, 25, 35, 45]$\degree$ } \\

\multicolumn{4}{c}{Reference scale height, $h_0$} & 
\multicolumn{2}{c}{[0.0125,0.025,0.035,0.05] au} \\

\multicolumn{4}{c}{Minimum grain size} & 
\multicolumn{2}{c}{[0.1,0.25,0.5] $\mu m$} \\


\multicolumn{4}{c}{Dust grain porosity} & \multicolumn{2}{c}{[0.3,0.5,0.6,0.7,0.8]} \\




\hline
\end{tabular}\\
{1) } {As described in the main text, the spectral energy distribution of AB Aurigae is modeled as as a 9800 K Kurucz atmosphere plus a sub-au component responsible for unresolved emission}
{2) }{
See discussion in main text.}
{3) }{Assumes a gas-to-dust ratio of 100.}
\end{table*}

\subsection{Modeling Strategy and Parameters}

Our radiative transfer modeling approach leverages heavily on approaches taken in \citet{Perrin2009}, \citet{Betti2022}, and \citet{Currie2022a}.  Table \ref{table:modelgrid} lists the fixed and varied model parameters, which are detailed below.

We adopted a composite emitting source to model the thermal emission from the star and unresolved, sub-au scale gas.  For the star, we adopt a 9800 $K$ Phoenix NEXT-GEN atmosphere model appropriate for an A0V spectral type.  To simulate the gas emission near the magnetospheric truncation radius, as found from interferometric data \citep{Tannirkulam2008}, we added a second 1950 $K$ blackbody source, following \citet{Betti2022}.   We scale this emission so that the composite model matches AB Aur's 1--5 $\mu m$ SED after adding in the protoplanetary dust component\footnote{As shown in \citet{Tannirkulam2008}, gas emission dominates the near-IR SED of AB Aurigae.  Over our entire grid of MCFOST disk model parameter space considered, we find that dust disk model parameters (e.g. inclination, scale height, dust mass) have a negligible effect on the near-to-mid IR SED.   So long as the inner regions of the disk are optically thick, the predicted SED for a given inclination angle and scale height is negligibly sensitive to the dust mass over values we consider.   }\footnote{This second emission component was not considered in \citet{Perrin2009} but was modeled in \citet{Betti2022}.}.   The practical consequence of including this gas emission is an increase in both the broadband emission and scattered-light brightness (total and polarized intensity) of the disk even at wide separations for a given disk mass since the disk surface is now reflecting more light.

Literature results guide our settings for a number of disk parameters.   The surface density and disk scale height power laws as a function of radius, $r$ ($\Sigma$ $\propto$ $r^{-\alpha}$ and $H$ $\propto$ $r^{\beta}$) follow assumptions in standard steady-state, flared disk models \citep[e.g.][]{KenyonHartmann1987,ChiangGoldreich1997}.  We set the inner and outer disk radii to be 0.15 au and 300 au, respectively.   While the inner radius is slightly smaller than found from interferometry \citep{Tannirkulam2008} we found little dependence on the output spectral energy distribution and scattered-light intensity of the AB Aur disk from varying the radius over a range of 0.1--0.3 au.  For a given surface density profile, different values for the outer radius are degenerate with the disk's dust mass.  We set the disk position angle to be 234\degree \citep{Betti2022,Tang2017}.   We adopt a standard \citet{Mathis1989} power law for the dust size distriubtion and set the maximum grain size to be 1 $\mu m$.

We varied disk parameters that are not directly constrained or poorly constrained by previous literature and have a key effect on the scattered-light intensity of AB Aur's disk.   The disk mass affects the optical depth of the disk as a function of radius.   Our range in disk masses brackets the value found from a different code (MCMax3D) in \citet{Currie2022a}.   The disk inclination affects the disk polarization curve and brightnesses at forward and backscattering angles.   As reported disk inclinations vary substantially in the literature \citep[see discussion in][]{Perrin2009}, we consider a range of values between 15\degree and 45\degree.  The reference scale height of the disk, $H_{\rm o}$ (i.e. the scale height at a reference radius of 1 au), affects how much light is intercepted by the disk at a given radius: our range brackets the value adopted in \citet{Currie2022a}.   We varied the minimum dust grain size -- 0.1, 0.25, and 0.5 $\mu m$ -- as it affects the level of dust forward-scattering. Finally, we convolve the model image with the CHARIS PSF. 

The key variable we consider is the dust grain porosity, since it has a strong effect on the polarized intensity emission of the disk and polarization fraction \citep[e.g.][]{Tazaki2019}.   For the dust grain composition, we consider the \citeauthor{DraineLee1984} astronomical silicates.   The range in porosity we consider (0.3--0.8) covers ranges considered by \citet{Perrin2009} in modeling HST/NICMOS polarimetry of the AB Aur disk.   

We chose H band for our full grid of model comparisons, as $J$ band data of the disk (in particular at small angular separations) is more limited by residual speckles and $K$ band sensitivity is limited by thermal background emission.  However, for the core results presented below we find consistent trends at $J$ and $K$.




\begin{figure*}
    \includegraphics[width=1\textwidth]{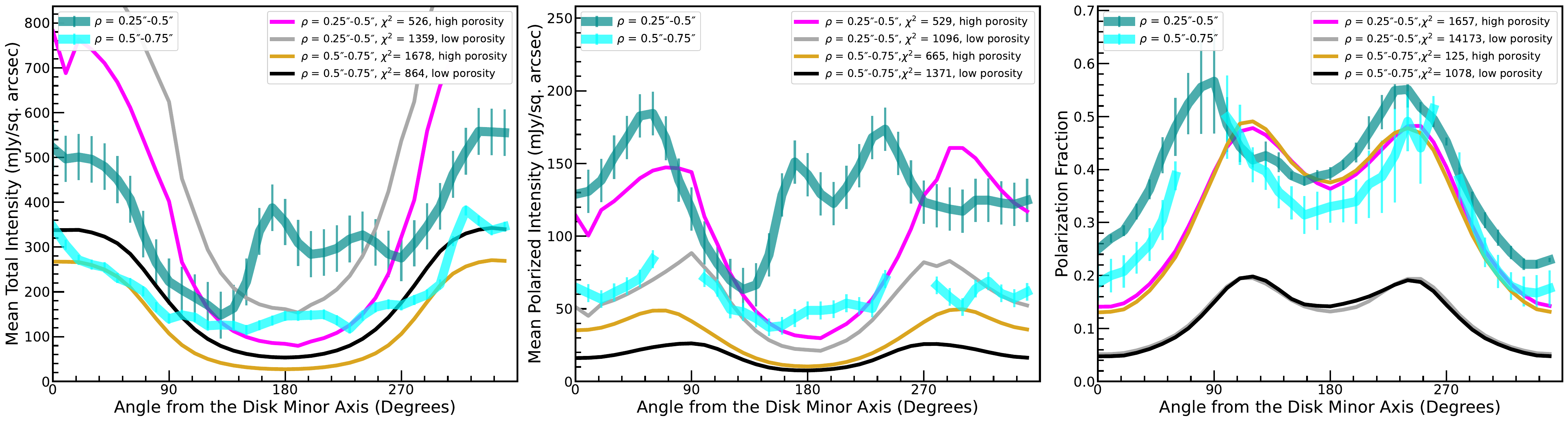}
    \caption{Model azimuthal dependent values for total intensity, polarized intensity, and polarization fraction compared to measured values at 0\farcs{}25--0\farcs{}5 and 0\farcs{}5--0\farcs{}75 for a model of high porosity ($p$=0.7) and low porosity ($p$=0.3).  All other parameters for these models are the same (see main text).   }
    \label{fig:phasefunccomp}
\end{figure*}

\begin{figure*}
    \includegraphics[width=0.33\textwidth]{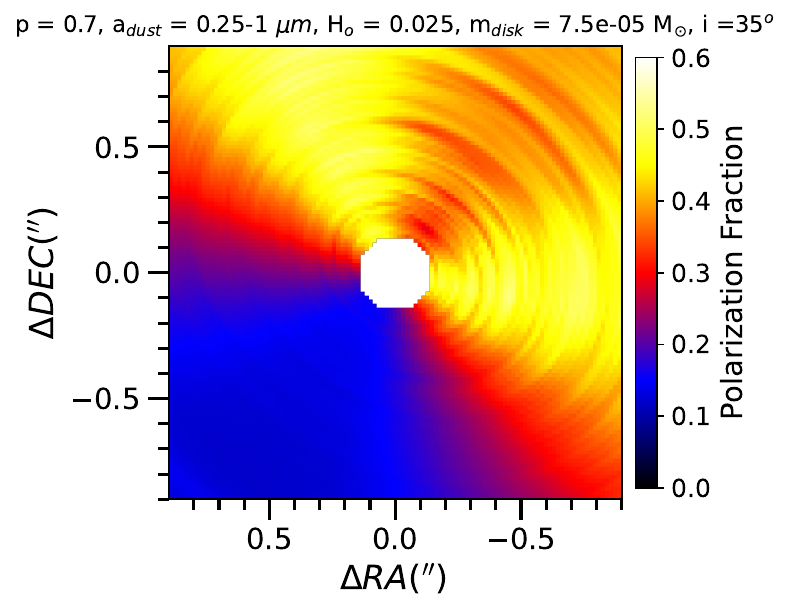}
        \includegraphics[width=0.33\textwidth]{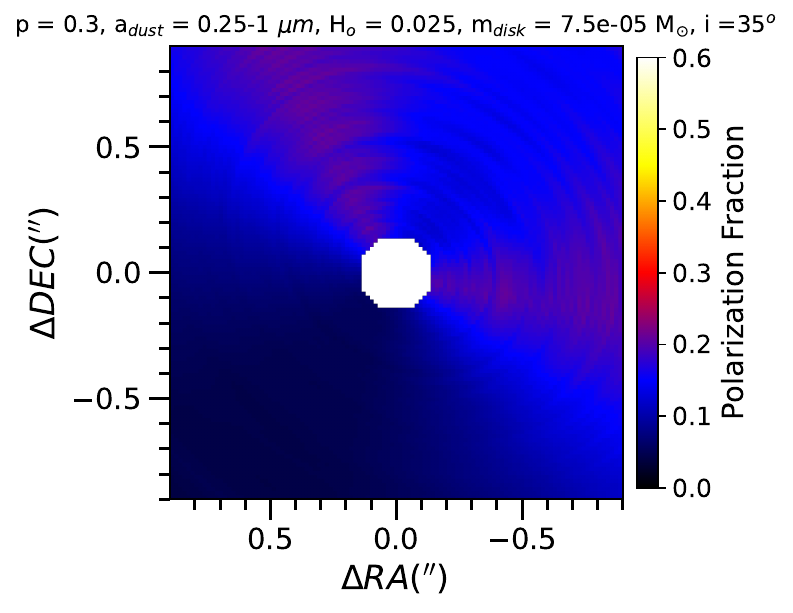}
            \includegraphics[width=0.31\textwidth]{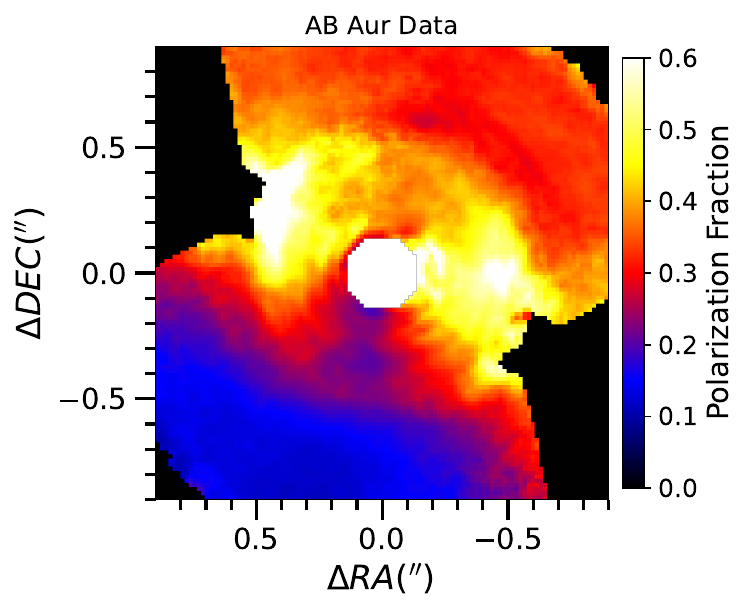}    
            \vspace{-0.1in}
    \caption{Synthetic polarization fraction maps at $H$ band for the high and low porosity models (left, middle) compared to the measured polarization map (right).}
    \label{fig:polfracmap}
\end{figure*}

\begin{figure*}
    \centering
    \includegraphics[width=0.6\textwidth]{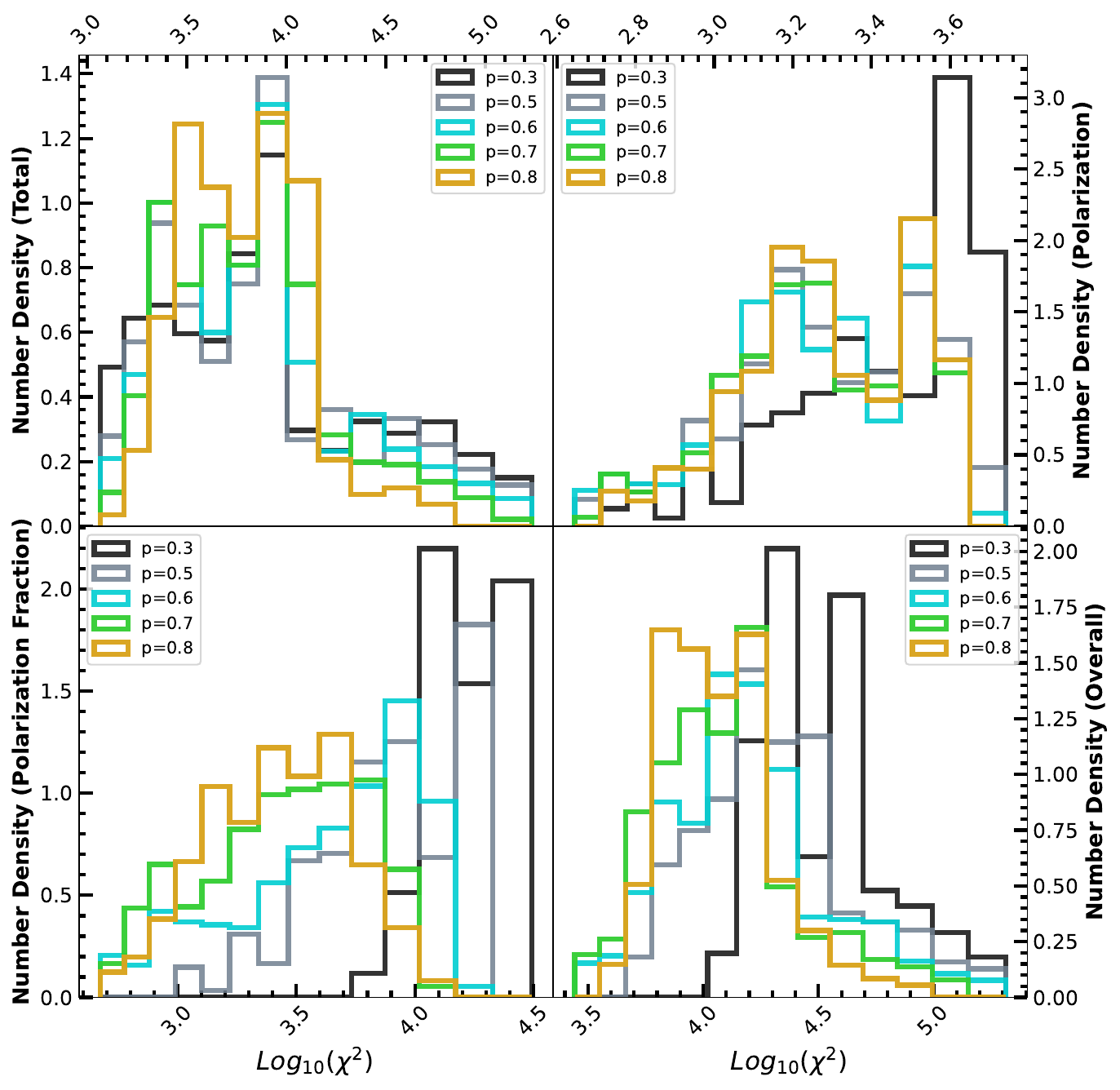}
    \caption{The $\chi^{2}$ distributions for models of varying dust grain porosities for total intensity, polarized intensity, and polarization fraction (top left, top right, and lower left respectively). The cumulative $\chi^{2}$ across all three data types is shown in the lower right panel. All $\chi^{2}$ values are plotted in logarimethic scale due to the large range in values.}
    \label{fig:chisquarehist}
\end{figure*}

\subsection{Model Results}

Figure \ref{fig:phasefunccomp} compares the measured mean total intensity, polarized intensity, and polarization fraction at $H$ band as a function of angle from the disk minor axis with a representative high porosity disk model ($p$ = 0.7) and low porosity model ($p$ = 0.3) and depicts our primary modeling result. Figure \ref{fig:polfracmap} shows synthetic polarization fraction images for these two models compared to the empirical H-band polarization fraction map. For both models, the disk mass is 7.5$\times$10$^{-5}$ $M_{\rm \odot}$, reference scale height is $H_{\rm o}$ = 0.025, the minimum dust size is 0.25 $\mu m$, and the inclination is 35$^{o}$\footnote{There is a small amount of field rotation between the +/- frames used in double differencing in the real data that is not simulated in our models. However, as described in \citet{Lawson2021} this effect is negligible.}.  

 At small and large angles (i.e. near the disk minor axis), both representative models shown in Figure \ref{fig:phasefunccomp} coarsely reproduce the total intensity azimuthal profile at small separations (0\farcs{}25--0\farcs{}5) and large separations (0\farcs{}5--0\farcs{}75) to within 50-75\%\footnote{An exception is the region near 180$^{o}$ in both total and polarized intensity. This region, however, overlaps with spiral structure in AB Aur's disk, which is not modeled in our MCFOST grid.}.  However, the low porosity model predicts substantially lower polarized intensity emission.  Consequently, its polarization fraction peaks at $\approx$0.2, a factor of 3 lower than that measured for AB Aur's disk.   Over the entire range of model parameter space explored, all low porosity models have polarization fractions that peak at less than 0.25.

In contrast, the higher-porosity model ($p$=0.7) reproduces the azimuthal profiles of both total \textit{and} polarized intensity to within 50-75\%. The polarization fraction curve peak value of $\sim$0.6 matches that measured for AB Aur's disk. The model polarization fraction peaks appear at angles of $\sim$110$^{o}$ and $\sim$250$^{o}$, similar to the measured peaks at $\sim$90$^{o}$ and $\sim$250$^{o}$.  A wide range of parameter space for models with $p$ = 0.6--0.8 match the observed peak polarization fraction and locate these peaks at angles near the measured peaks.

As with the representative models, we find that the $\chi^{2}$ values for our entire model grid show similiar distributions regardless of porosity when compared to the observed total and polarized intensities, as seen in Figure \ref{fig:chisquarehist}. However, when compared to the polarization fraction, which is more directly related to the porosity, only models with $p$=0.6--0.8 produce the lowest $\chi^{2}$. Similarly, the lowest overall $\chi^{2}$ (i.e. the sum of the $\chi^{2}$ of total intensities, polarized intensity, and polarization fraction comparisions) are produced by models with $p$=0.6 -- 0.7. 

Figures in the Appendix show how different parameters in the models affect the resulting azimuthal profiles and, subseqeuntly, the resulting $\chi^{2}$ distributions.  Porosities of $p$=0.6--0.8 reproduce the AB Aur disk's polarization fraction curve better than lower porosity models.  Higher-inclination models and those with a lower scale height ($H_{\rm o}$ = 0.0125--0.025) yield a better match for the angles of these peaks.  However, highly settled disks -- i.e. those with scale heights of $H_{\rm o}$ = 0.0125 -- predict substantially lower total and polarized intensity surface brightnesses than observed. Models with lower inclinations ($i <$ 45$\degree$) better fit the polarized intensity and polarization fraction, though the difference in $\chi^{2}$ is less than the trends for scale height. Minimum grain sizes less than 0.5 $\mu m$ yield better fits to the polarized intensity azimuthal profile and the polarization fraction.

The supplemental online material compares all model azimuthal profiles to data.  No model reproduces all AB Aur disk azimuthal profiles in total and polarized intensity to within 20\% at all angles.  Future work considering models with more complex density and temperature profiles, incorporating spiral structure, adding multiple dust components that may be misaligned (e.g. \citealt{Currie2022a, Hashimoto2011}) and utilizing more realistic dust models \citep[e.g.][]{Tazaki2019,Tazaki2023} will likely yield better agreement with the data.  

\section{Discussion}

Using SCExAO/CHARIS, we resolve AB Aur's protoplanetary disk in polarized light across 22 wavelength channels spanning the major near-IR passbands ($\lambda_{\rm o}$ = 1.1--2.4 $\mu m$).   The morphology and colors of the disk's polarized light signal show some wavelength dependence.   At $K$ band, the western spiral seen in prior polarimetry data and in ALMA data at $\rho$ $\le$ 0\farcs{}3 extends to larger angular separations coincident with the $CO$ gas spiral.  The inner boundary of the dust continuum ring seen with ALMA data may be visible with our data \citep[see ][]{Boccaletti2020, Tang2017}.  All regions in the disk increase in intensity at longer wavelengths (i.e. the disk has a red apparent color), consistent with unresolved sub au-scale gas emission dominating the IR light incident upon the disk, although the intrinsic scattering of the outer disk is blue.


At $\lambda$ $>$ 1.3 $\mu m$, the polarized light signal near AB Aur b's position is featureless, and its spectrum is indistinguishable from that of other regions.  In contrast, AB Aur b's emission seen at 1.1--2.4 $\mu m$ in total intensity is concentrated, not featureless, with a spectrum bluer than those found for from pure disk regions.  These two qualities -- low polarized intensity, bluer total intensity spectrum -- favor interpreting AB Aur b as the location of an embedded protoplanet, as found in previous work \citep[see][]{Currie2022a,Lawson2022} and supported by detailed follow-up analysis (Hashimoto et al. in prep.; Currie et al. in prep.).  At all wavelengths, our data recover the \citet{Boccaletti2020} spiral/``twist" ``f1" feature near the coronagraphic mask, while a second feature at wider separations (``f2") is not definitively detected.


Combining our polarized intensity data with prior total intensity data yields the polarization fraction map of AB Aur's disk.
The polarization fraction peaks at $\approx$ 0.6 along the major axis. It appears wavelength independent at angular separations of $\rho$ = 0\farcs{}5--0\farcs{}75 and possibly also at smaller separations of 0\farcs{}25--0\
\farcs{}5.  Radiative transfer modeling of AB Aur's disk with MCFOST shows dust grains with high porosities ($p$ = 0.6--0.8) better reproduce the observed polarization fraction as a function of position angle than do low porosity models ($p$ = 0.3).

We can directly compare our results to the recently-published CHARIS spectropolarimetry analysis of the HD 34700 A disk \citep{Chen2024}.  AB Aur's disk fractional polarization peaks at slightly slower values than HD 34700 A's.  Like the more complex AB Aur disk, HD 34700 A's ring-like disk cannot be simultaneously fit in total and polarized intensity by any MCFOST model.  

While CHARIS's IFSP mode's novelty makes direct comparison to other high-contrast polarimetric observations difficult, we can examine general trends. The AB Aur disk's K-band fractional polarization is comparable to or higher than those inferred for nearly all disks studied in \citet{Ren2023}.   CHARIS IFSP observations of these and similar disks may allow us to better identify trends in the disks' dust scattering properties.  Protoplanetary disks around K/M stars studied by \citet{Avenhaus2018} have much redder $J$-$H$ color in polarized light than AB Aur's disk, albeit with substantially larger errors than AB Aur's.  
The AB Aur system bears some resemblance to that of HD 100546, which also has spiral structure in its outer disk, a hot inner disk component, a gap between the inner and outer disk, and embedded protoplanet candidates \citep{Boccaletti2013,Quanz2015,Currie2015}. \citet{Mulders2013} found that HD 100546's outer disk has red scattering colors: very small grains or very large ($>$ 2.5 $\mu m$) but extremely porous grains (a filling factor of $ff$ $\lesssim$ 0.01) may explain the disk's colors and brightness asymmetry.  Our modeling suggests that the dust in AB Aur's disk is porous but the $\chi^{2}$ of our models becomes slightly worse at the highest porosities considered ($p$ = 0.8).  Similarly, our modeling disfavors grains with minimum sizes greater than 0.5 $\mu m$.

This study demonstrates the power of combining polarimetry with an integral field spectrograph to resolve structures in protoplanetary disks and characterize the disks' dust properties.  Most studies of disks in polarized light focus on a single photometric bandpass and few have quantified the disk polarization fraction from both polarized intensity and total intensity data \citep[e.g.][]{Avenhaus2018,Rich2019,Ren2023}.  However, SCExAO's integral field spectropolarimetry mode yields polarized light disk images and spectra in 22 channels across $J$, $H$, and $K$.   Adding CHARIS total intensity data, we can then determine the polarization fraction of the disk as a function of wavelength and position.   Modeling analysis of total intensity imaging, polarized intensity imaging, and the polarization fraction map qualitatively constrains the dust porosities in AB Aur's disk despite its complex structures (e.g. spirals).   Analyses of disks that are morphologically simpler than AB Aurigae's and thus more amenable to detailed model comparisons may yield strong constraints on the structure of disks (e.g. scale height, flaring, inclination, etc.), in addition to shedding light on dust porosities.

Upgrades to the Subaru Telescope's AO capabilities will extend our analyses to a much more diverse sample of planet-forming environments.   SCExAO provides a high-order correction of wavefront errors that are initially partially mitigated by Subaru's facility AO system, AO-188 \citep{Hayano2008}, which does wavefront sensing in blue optical wavelengths.  Subaru has now upgraded AO-188 to ``AO-3000", replacing its current 188-element deformable mirror with a 3200-actuator mirror and utilizing a near-IR Pyramid wavefront sensor (NIR PyWFS) \citep{Lozi2022}.  The NIR PyWFS will allow AO-3000 to achieve sharp images of optically faint stars currently inaccessible with AO-188 and thus SCExAO/CHARIS\footnote{The SCExAO deformable mirror has limited stroke (i.e. a limited actuator displacement range from a flat shape) to correct for wavefront aberrations.   Thus, if AO-188's partial correction of atmospheric turbulence is poor or absent, SCExAO's wavefront control will be ineffective.}.   For the nearest star-forming regions (e.g. Taurus, Ophiucus), these upgrades will allow SCExAO/CHARIS to probe the structure and dust properties of numerous protoplanetary disks around young, solar mass and subsolar-mass stars, complementing studies of disks around more massive stars like AB Aurigae.

\begin{acknowledgments}
\indent We thank Christophe Pinte for assistance with setting up and running MCFOST and Ryo Tazaki for helpful discussions about dust scattering effects.

 \indent The authors wish to acknowledge the very significant cultural role and reverence that the summit of Mauna Kea holds within the Hawaiian community.  We are most fortunate to have the opportunity to conduct observations from this mountain.

\indent T.C. was supported in part by NASA/Keck grant LK-2663-948181.  \\
\indent The development of SCExAO was supported by JSPS (Grant-in-Aid for Research \#23340051, \#26220704 \& \#23103002), Astrobiology Center of NINS, Japan, the Mt Cuba Foundation, and the director's contingency fund at Subaru Telescope.  CHARIS was developed under the support by the Grant-in-Aid for Scientific Research on Innovative Areas \#2302.   Some of the data presented herein were obtained at the W. M. Keck Observatory, which is operated as a scientific partnership among the California Institute of Technology, the University of California and the National Aeronautics and Space Administration. The Observatory was made possible by the generous financial support of the W. M. Keck Foundation.

Software: \texttt{MCFOST} \citep{Pinte2006,Pinte2009}, \texttt{NumPy} 
\citep{numpy}, \texttt{Jupyter Notebooks} \citep{jupyter}, \texttt{Matplotlib} \citep{matplotlib},  \texttt{Astropy} \citep{astropy:2013,astropy:2018,astropy:2022}, \texttt{SciPy} \citep{scipy}, \texttt{CHARIS DRP} \citep{Brandt2017}, \texttt{CHARIS DPP} \citep{Currie2020,Lawson2021}.
\end{acknowledgments}

\bibliography{abaur}   

\begin{thebibliography}{}
\expandafter\ifx\csname natexlab\endcsname\relax\def\natexlab#1{#1}\fi
\providecommand{\url}[1]{\href{#1}{#1}}
\providecommand{\dodoi}[1]{doi:~\href{http://doi.org/#1}{\nolinkurl{#1}}}
\providecommand{\doeprint}[1]{\href{http://ascl.net/#1}{\nolinkurl{http://ascl.net/#1}}}
\providecommand{\doarXiv}[1]{\href{https://arxiv.org/abs/#1}{\nolinkurl{https://arxiv.org/abs/#1}}}

\bibitem[{{Astropy Collaboration} {et~al.}(2013){Astropy Collaboration}, {Robitaille}, {Tollerud}, {Greenfield}, {Droettboom}, {Bray}, {Aldcroft}, {Davis}, {Ginsburg}, {Price-Whelan}, {Kerzendorf}, {Conley}, {Crighton}, {Barbary}, {Muna}, {Ferguson}, {Grollier}, {Parikh}, {Nair}, {Unther}, {Deil}, {Woillez}, {Conseil}, {Kramer}, {Turner}, {Singer}, {Fox}, {Weaver}, {Zabalza}, {Edwards}, {Azalee Bostroem}, {Burke}, {Casey}, {Crawford}, {Dencheva}, {Ely}, {Jenness}, {Labrie}, {Lim}, {Pierfederici}, {Pontzen}, {Ptak}, {Refsdal}, {Servillat}, \& {Streicher}}]{astropy:2013}
{Astropy Collaboration}, {Robitaille}, T.~P., {Tollerud}, E.~J., {et~al.} 2013, \aap, 558, A33, \dodoi{10.1051/0004-6361/201322068}

\bibitem[{{Astropy Collaboration} {et~al.}(2022){Astropy Collaboration}, {Price-Whelan}, {Lim}, {Earl}, {Starkman}, {Bradley}, {Shupe}, {Patil}, {Corrales}, {Brasseur}, {N{\"o}the}, {Donath}, {Tollerud}, {Morris}, {Ginsburg}, {Vaher}, {Weaver}, {Tocknell}, {Jamieson}, {van Kerkwijk}, {Robitaille}, {Merry}, {Bachetti}, {G{\"u}nther}, {Aldcroft}, {Alvarado-Montes}, {Archibald}, {B{\'o}di}, {Bapat}, {Barentsen}, {Baz{\'a}n}, {Biswas}, {Boquien}, {Burke}, {Cara}, {Cara}, {Conroy}, {Conseil}, {Craig}, {Cross}, {Cruz}, {D'Eugenio}, {Dencheva}, {Devillepoix}, {Dietrich}, {Eigenbrot}, {Erben}, {Ferreira}, {Foreman-Mackey}, {Fox}, {Freij}, {Garg}, {Geda}, {Glattly}, {Gondhalekar}, {Gordon}, {Grant}, {Greenfield}, {Groener}, {Guest}, {Gurovich}, {Handberg}, {Hart}, {Hatfield-Dodds}, {Homeier}, {Hosseinzadeh}, {Jenness}, {Jones}, {Joseph}, {Kalmbach}, {Karamehmetoglu}, {Ka{\l}uszy{\'n}ski}, {Kelley}, {Kern}, {Kerzendorf}, {Koch}, {Kulumani}, {Lee}, {Ly}, {Ma}, {MacBride}, {Maljaars}, {Muna}, {Murphy}, {Norman},
  {O'Steen}, {Oman}, {Pacifici}, {Pascual}, {Pascual-Granado}, {Patil}, {Perren}, {Pickering}, {Rastogi}, {Roulston}, {Ryan}, {Rykoff}, {Sabater}, {Sakurikar}, {Salgado}, {Sanghi}, {Saunders}, {Savchenko}, {Schwardt}, {Seifert-Eckert}, {Shih}, {Jain}, {Shukla}, {Sick}, {Simpson}, {Singanamalla}, {Singer}, {Singhal}, {Sinha}, {Sip{\H{o}}cz}, {Spitler}, {Stansby}, {Streicher}, {{\v{S}}umak}, {Swinbank}, {Taranu}, {Tewary}, {Tremblay}, {de Val-Borro}, {Van Kooten}, {Vasovi{\'c}}, {Verma}, {de Miranda Cardoso}, {Williams}, {Wilson}, {Winkel}, {Wood-Vasey}, {Xue}, {Yoachim}, {Zhang}, {Zonca}, \& {Astropy Project Contributors}}]{astropy:2022}
{Astropy Collaboration}, {Price-Whelan}, A.~M., {Lim}, P.~L., {et~al.} 2022, \apj, 935, 167, \dodoi{10.3847/1538-4357/ac7c74}

\bibitem[{{Avenhaus} {et~al.}(2018){Avenhaus}, {Quanz}, {Garufi}, {Perez}, {Casassus}, {Pinte}, {Bertrang}, {Caceres}, {Benisty}, \& {Dominik}}]{Avenhaus2018}
{Avenhaus}, H., {Quanz}, S.~P., {Garufi}, A., {et~al.} 2018, \apj, 863, 44, \dodoi{10.3847/1538-4357/aab846}

\bibitem[{{Benisty} {et~al.}(2017){Benisty}, {Stolker}, {Pohl}, {de Boer}, {Lesur}, {Dominik}, {Dullemond}, {Langlois}, {Min}, {Wagner}, {Henning}, {Juhasz}, {Pinilla}, {Facchini}, {Apai}, {van Boekel}, {Garufi}, {Ginski}, {M{\'e}nard}, {Pinte}, {Quanz}, {Zurlo}, {Boccaletti}, {Bonnefoy}, {Beuzit}, {Chauvin}, {Cudel}, {Desidera}, {Feldt}, {Fontanive}, {Gratton}, {Kasper}, {Lagrange}, {LeCoroller}, {Mouillet}, {Mesa}, {Sissa}, {Vigan}, {Antichi}, {Buey}, {Fusco}, {Gisler}, {Llored}, {Magnard}, {Moeller-Nilsson}, {Pragt}, {Roelfsema}, {Sauvage}, \& {Wildi}}]{Benisty2017}
{Benisty}, M., {Stolker}, T., {Pohl}, A., {et~al.} 2017, \aap, 597, A42, \dodoi{10.1051/0004-6361/201629798}

\bibitem[{{Benisty} {et~al.}(2022){Benisty}, {Dominik}, {Follette}, {Garufi}, {Ginski}, {Hashimoto}, {Keppler}, {Kley}, \& {Monnier}}]{Benisty2022}
{Benisty}, M., {Dominik}, C., {Follette}, K., {et~al.} 2022, arXiv e-prints, arXiv:2203.09991, \dodoi{10.48550/arXiv.2203.09991}

\bibitem[{{Betti} {et~al.}(2022){Betti}, {Follette}, {Jorquera}, {Duch{\^e}ne}, {Mazoyer}, {Bonnefoy}, {Chauvin}, {P{\'e}rez}, {Boccaletti}, {Pinte}, {Weinberger}, {Grady}, {Close}, {Defr{\`e}re}, {Downey}, {Hinz}, {M{\'e}nard}, {Schneider}, {Skemer}, \& {Vaz}}]{Betti2022}
{Betti}, S.~K., {Follette}, K., {Jorquera}, S., {et~al.} 2022, \aj, 163, 145, \dodoi{10.3847/1538-3881/ac4d9b}

\bibitem[{{Boccaletti} {et~al.}(2013){Boccaletti}, {Pantin}, {Lagrange}, {Augereau}, {Meheut}, \& {Quanz}}]{Boccaletti2013}
{Boccaletti}, A., {Pantin}, E., {Lagrange}, A.~M., {et~al.} 2013, \aap, 560, A20, \dodoi{10.1051/0004-6361/201322365}

\bibitem[{{Boccaletti} {et~al.}(2020){Boccaletti}, {Di Folco}, {Pantin}, {Dutrey}, {Guilloteau}, {Tang}, {Pi{\'e}tu}, {Habart}, {Milli}, {Beck}, \& {Maire}}]{Boccaletti2020}
{Boccaletti}, A., {Di Folco}, E., {Pantin}, E., {et~al.} 2020, \aap, 637, L5, \dodoi{10.1051/0004-6361/202038008}

\bibitem[{{Boss}(1997)}]{Boss1997}
{Boss}, A.~P. 1997, Science, 276, 1836, \dodoi{10.1126/science.276.5320.1836}

\bibitem[{{Brandt} {et~al.}(2017){Brandt}, {Rizzo}, {Groff}, {Chilcote}, {Greco}, {Kasdin}, {Limbach}, {Galvin}, {Loomis}, {Knapp}, {McElwain}, {Jovanovic}, {Currie}, {Mede}, {Tamura}, {Takato}, \& {Hayashi}}]{Brandt2017}
{Brandt}, T.~D., {Rizzo}, M., {Groff}, T., {et~al.} 2017, Journal of Astronomical Telescopes, Instruments, and Systems, 3, 048002, \dodoi{10.1117/1.JATIS.3.4.048002}

\bibitem[{{Canovas} {et~al.}(2015){Canovas}, {M{\'e}nard}, {de Boer}, {Pinte}, {Avenhaus}, \& {Schreiber}}]{Canovas2015}
{Canovas}, H., {M{\'e}nard}, F., {de Boer}, J., {et~al.} 2015, \aap, 582, L7, \dodoi{10.1051/0004-6361/201527267}

\bibitem[{{Carson} {et~al.}(2013){Carson}, {Thalmann}, {Janson}, {Kozakis}, {Bonnefoy}, {Biller}, {Schlieder}, {Currie}, {McElwain}, {Goto}, {Henning}, {Brandner}, {Feldt}, {Kandori}, {Kuzuhara}, {Stevens}, {Wong}, {Gainey}, {Fukagawa}, {Kuwada}, {Brand t}, {Kwon}, {Abe}, {Egner}, {Grady}, {Guyon}, {Hashimoto}, {Hayano}, {Hayashi}, {Hayashi}, {Hodapp}, {Ishii}, {Iye}, {Knapp}, {Kudo}, {Kusakabe}, {Matsuo}, {Miyama}, {Morino}, {Moro-Martin}, {Nishimura}, {Pyo}, {Serabyn}, {Suto}, {Suzuki}, {Takami}, {Takato}, {Terada}, {Tomono}, {Turner}, {Watanabe}, {Wisniewski}, {Yamada}, {Takami}, {Usuda}, \& {Tamura}}]{Carson2013}
{Carson}, J., {Thalmann}, C., {Janson}, M., {et~al.} 2013, \apjl, 763, L32, \dodoi{10.1088/2041-8205/763/2/L32}

\bibitem[{{Chauvin} {et~al.}(2017){Chauvin}, {Desidera}, {Lagrange}, {Vigan}, {Gratton}, {Langlois}, {Bonnefoy}, {Beuzit}, {Feldt}, {Mouillet}, {Meyer}, {Cheetham}, {Biller}, {Boccaletti}, {D'Orazi}, {Galicher}, {Hagelberg}, {Maire}, {Mesa}, {Olofsson}, {Samland}, {Schmidt}, {Sissa}, {Bonavita}, {Charnay}, {Cudel}, {Daemgen}, {Delorme}, {Janin-Potiron}, {Janson}, {Keppler}, {Le Coroller}, {Ligi}, {Marleau}, {Messina}, {Molli{\`e}re}, {Mordasini}, {M{\"u}ller}, {Peretti}, {Perrot}, {Rodet}, {Rouan}, {Zurlo}, {Dominik}, {Henning}, {Menard}, {Schmid}, {Turatto}, {Udry}, {Vakili}, {Abe}, {Antichi}, {Baruffolo}, {Baudoz}, {Baudrand}, {Blanchard}, {Bazzon}, {Buey}, {Carbillet}, {Carle}, {Charton}, {Cascone}, {Claudi}, {Costille}, {Deboulbe}, {De Caprio}, {Dohlen}, {Fantinel}, {Feautrier}, {Fusco}, {Gigan}, {Giro}, {Gisler}, {Gluck}, {Hubin}, {Hugot}, {Jaquet}, {Kasper}, {Madec}, {Magnard}, {Martinez}, {Maurel}, {Le Mignant}, {M{\"o}ller-Nilsson}, {Llored}, {Moulin}, {Orign{\'e}}, {Pavlov}, {Perret}, {Petit},
  {Pragt}, {Puget}, {Rabou}, {Ramos}, {Rigal}, {Rochat}, {Roelfsema}, {Rousset}, {Roux}, {Salasnich}, {Sauvage}, {Sevin}, {Soenke}, {Stadler}, {Suarez}, {Weber}, {Wildi}, {Antoniucci}, {Augereau}, {Baudino}, {Brandner}, {Engler}, {Girard}, {Gry}, {Kral}, {Kopytova}, {Lagadec}, {Milli}, {Moutou}, {Schlieder}, {Szul{\'a}gyi}, {Thalmann}, \& {Wahhaj}}]{Chauvin2017}
{Chauvin}, G., {Desidera}, S., {Lagrange}, A.~M., {et~al.} 2017, \aap, 605, L9, \dodoi{10.1051/0004-6361/201731152}

\bibitem[{{Chen} {et~al.}(2024){Chen}, {Lawson}, {Brandt}, {Lewis}, {Uyama}, {Millar-Blanchaer}, {Tazaki}, \& {Currie}}]{Chen2024}
{Chen}, M., {Lawson}, K., {Brandt}, T.~D., {et~al.} 2024, arXiv e-prints, arXiv:2408.09038.
\newblock \doarXiv{2408.09038}

\bibitem[{{Chiang} \& {Goldreich}(1997)}]{ChiangGoldreich1997}
{Chiang}, E.~I., \& {Goldreich}, P. 1997, \apj, 490, 368, \dodoi{10.1086/304869}

\bibitem[{{Currie}(2024)}]{Currie2024}
{Currie}, T. 2024, Research Notes of the American Astronomical Society, 8, 146, \dodoi{10.3847/2515-5172/ad50ce}

\bibitem[{{Currie} {et~al.}(2023{\natexlab{a}}){Currie}, {Biller}, {Lagrange}, {Marois}, {Guyon}, {Nielsen}, {Bonnefoy}, \& {De Rosa}}]{Currie2023b}
{Currie}, T., {Biller}, B., {Lagrange}, A., {et~al.} 2023{\natexlab{a}}, in Astronomical Society of the Pacific Conference Series, Vol. 534, Protostars and Planets VII, ed. S.~{Inutsuka}, Y.~{Aikawa}, T.~{Muto}, K.~{Tomida}, \& M.~{Tamura}, 799, \dodoi{10.48550/arXiv.2205.05696}

\bibitem[{{Currie} {et~al.}(2015){Currie}, {Cloutier}, {Brittain}, {Grady}, {Burrows}, {Muto}, {Kenyon}, \& {Kuchner}}]{Currie2015}
{Currie}, T., {Cloutier}, R., {Brittain}, S., {et~al.} 2015, \apjl, 814, L27, \dodoi{10.1088/2041-8205/814/2/L27}

\bibitem[{{Currie} {et~al.}(2011){Currie}, {Burrows}, {Itoh}, {Matsumura}, {Fukagawa}, {Apai}, {Madhusudhan}, {Hinz}, {Rodigas}, {Kasper}, {Pyo}, \& {Ogino}}]{Currie2011}
{Currie}, T., {Burrows}, A., {Itoh}, Y., {et~al.} 2011, \apj, 729, 128, \dodoi{10.1088/0004-637X/729/2/128}

\bibitem[{{Currie} {et~al.}(2019){Currie}, {Marois}, {Cieza}, {Mulders}, {Lawson}, {Caceres}, {Rodriguez-Ruiz}, {Wisniewski}, {Guyon}, {Brandt}, {Kasdin}, {Groff}, {Lozi}, {Chilcote}, {Hodapp}, {Jovanovic}, {Martinache}, {Skaf}, {Lyra}, {Tamura}, {Asensio-Torres}, {Dong}, {Grady}, {Gerard}, {Fukagawa}, {Hand}, {Hayashi}, {Henning}, {Kudo}, {Kuzuhara}, {Kwon}, {McElwain}, \& {Uyama}}]{Currie2019a}
{Currie}, T., {Marois}, C., {Cieza}, L., {et~al.} 2019, \apjl, 877, L3, \dodoi{10.3847/2041-8213/ab1b42}

\bibitem[{{Currie} {et~al.}(2020){Currie}, {Guyon}, {Lozi}, {Sahoo}, {Vievard}, {Deo}, {Chilcote}, {Groff}, {Brandt}, {Lawson}, {Skaf}, {Martinache}, \& {Kasdin}}]{Currie2020}
{Currie}, T., {Guyon}, O., {Lozi}, J., {et~al.} 2020, in Society of Photo-Optical Instrumentation Engineers (SPIE) Conference Series, Vol. 11448, Adaptive Optics Systems VII, ed. L.~{Schreiber}, D.~{Schmidt}, \& E.~{Vernet}, 114487H, \dodoi{10.1117/12.2576349}

\bibitem[{{Currie} {et~al.}(2022){Currie}, {Lawson}, {Schneider}, {Lyra}, {Wisniewski}, {Grady}, {Guyon}, {Tamura}, {Kotani}, {Kawahara}, {Brandt}, {Uyama}, {Muto}, {Dong}, {Kudo}, {Hashimoto}, {Fukagawa}, {Wagner}, {Lozi}, {Chilcote}, {Tobin}, {Groff}, {Ward-Duong}, {Januszewski}, {Norris}, {Tuthill}, {van der Marel}, {Sitko}, {Deo}, {Vievard}, {Jovanovic}, {Martinache}, \& {Skaf}}]{Currie2022a}
{Currie}, T., {Lawson}, K., {Schneider}, G., {et~al.} 2022, Nature Astronomy, 6, 751, \dodoi{10.1038/s41550-022-01634-x}

\bibitem[{{Currie} {et~al.}(2023{\natexlab{b}}){Currie}, {Brandt}, {Brandt}, {Lacy}, {Burrows}, {Guyon}, {Tamura}, {Liu}, {Sagynbayeva}, {Tobin}, {Chilcote}, {Groff}, {Marois}, {Thompson}, {Murphy}, {Kuzuhara}, {Lawson}, {Lozi}, {Deo}, {Vievard}, {Skaf}, {Uyama}, {Jovanovic}, {Martinache}, {Kasdin}, {Kudo}, {McElwain}, {Janson}, {Wisniewski}, {Hodapp}, {Nishikawa}, {He{\l}miniak}, {Kwon}, \& {Hayashi}}]{Currie2023}
{Currie}, T., {Brandt}, G.~M., {Brandt}, T.~D., {et~al.} 2023{\natexlab{b}}, Science, 380, 198, \dodoi{10.1126/science.abo6192}

\bibitem[{{Draine} \& {Flatau}(1994)}]{Draine1994}
{Draine}, B.~T., \& {Flatau}, P.~J. 1994, Journal of the Optical Society of America A, 11, 1491, \dodoi{10.1364/JOSAA.11.001491}

\bibitem[{{Draine} \& {Lee}(1984)}]{DraineLee1984}
{Draine}, B.~T., \& {Lee}, H.~M. 1984, \apj, 285, 89, \dodoi{10.1086/162480}

\bibitem[{{Francis} \& {van der Marel}(2020)}]{Francisvandermarel2020}
{Francis}, L., \& {van der Marel}, N. 2020, \apj, 892, 111, \dodoi{10.3847/1538-4357/ab7b63}

\bibitem[{{Fukagawa} {et~al.}(2004){Fukagawa}, {Hayashi}, {Tamura}, {Itoh}, {Hayashi}, {Oasa}, {Takeuchi}, {Morino}, {Murakawa}, {Oya}, {Yamashita}, {Suto}, {Mayama}, {Naoi}, {Ishii}, {Pyo}, {Nishikawa}, {Takato}, {Usuda}, {Ando}, {Iye}, {Miyama}, \& {Kaifu}}]{Fukagawa2004}
{Fukagawa}, M., {Hayashi}, M., {Tamura}, M., {et~al.} 2004, \apjl, 605, L53, \dodoi{10.1086/420699}

\bibitem[{{Gaia Collaboration}(2022)}]{gaiadr3}
{Gaia Collaboration}. 2022, VizieR Online Data Catalog, I/355

\bibitem[{{Garufi} {et~al.}(2018){Garufi}, {Benisty}, {Pinilla}, {Tazzari}, {Dominik}, {Ginski}, {Henning}, {Kral}, {Langlois}, {M{\'e}nard}, {Stolker}, {Szulagyi}, {Villenave}, \& {van der Plas}}]{Garufi2018}
{Garufi}, A., {Benisty}, M., {Pinilla}, P., {et~al.} 2018, \aap, 620, A94, \dodoi{10.1051/0004-6361/201833872}

\bibitem[{{Grady} {et~al.}(1999){Grady}, {Woodgate}, {Bruhweiler}, {Boggess}, {Plait}, {Lindler}, {Clampin}, \& {Kalas}}]{Grady1999}
{Grady}, C.~A., {Woodgate}, B., {Bruhweiler}, F.~C., {et~al.} 1999, \apjl, 523, L151, \dodoi{10.1086/312270}

\bibitem[{{Groff} {et~al.}(2016){Groff}, {Chilcote}, {Kasdin}, {Galvin}, {Loomis}, {Carr}, {Brand t}, {Knapp}, {Limbach}, {Guyon}, {Jovanovic}, {McElwain}, {Takato}, \& {Hayashi}}]{Groff2016}
{Groff}, T.~D., {Chilcote}, J., {Kasdin}, N.~J., {et~al.} 2016, in Society of Photo-Optical Instrumentation Engineers (SPIE) Conference Series, Vol. 9908, Ground-based and Airborne Instrumentation for Astronomy VI, 99080O, \dodoi{10.1117/12.2233447}

\bibitem[{{Hashimoto} {et~al.}(2011){Hashimoto}, {Tamura}, {Muto}, {Kudo}, {Fukagawa}, {Fukue}, {Goto}, {Grady}, {Henning}, {Hodapp}, {Honda}, {Inutsuka}, {Kokubo}, {Knapp}, {McElwain}, {Momose}, {Ohashi}, {Okamoto}, {Takami}, {Turner}, {Wisniewski}, {Janson}, {Abe}, {Brandner}, {Carson}, {Egner}, {Feldt}, {Golota}, {Guyon}, {Hayano}, {Hayashi}, {Hayashi}, {Ishii}, {Kandori}, {Kusakabe}, {Matsuo}, {Mayama}, {Miyama}, {Morino}, {Moro-Martin}, {Nishimura}, {Pyo}, {Suto}, {Suzuki}, {Takato}, {Terada}, {Thalmann}, {Tomono}, {Watanabe}, {Yamada}, {Takami}, \& {Usuda}}]{Hashimoto2011}
{Hashimoto}, J., {Tamura}, M., {Muto}, T., {et~al.} 2011, \apjl, 729, L17, \dodoi{10.1088/2041-8205/729/2/L17}

\bibitem[{{Hayano} {et~al.}(2008){Hayano}, {Takami}, {Guyon}, {Oya}, {Hattori}, {Saito}, {Watanabe}, {Murakami}, {Minowa}, {Ito}, {Colley}, {Eldred}, {Golota}, {Dinkins}, {Kashikawa}, \& {Iye}}]{Hayano2008}
{Hayano}, Y., {Takami}, H., {Guyon}, O., {et~al.} 2008, in Society of Photo-Optical Instrumentation Engineers (SPIE) Conference Series, Vol. 7015, Adaptive Optics Systems, ed. N.~{Hubin}, C.~E. {Max}, \& P.~L. {Wizinowich}, 701510, \dodoi{10.1117/12.789992}

\bibitem[{{Hinkley} {et~al.}(2009){Hinkley}, {Oppenheimer}, {Soummer}, {Brenner}, {Graham}, {Perrin}, {Sivaramakrishnan}, {Lloyd}, {Roberts}, \& {Kuhn}}]{Hinkley2009}
{Hinkley}, S., {Oppenheimer}, B.~R., {Soummer}, R., {et~al.} 2009, \apj, 701, 804, \dodoi{10.1088/0004-637X/701/1/804}

\bibitem[{{Hunter}(2007)}]{matplotlib}
{Hunter}, J.~D. 2007, Computing in Science and Engineering, 9, 90, \dodoi{10.1109/MCSE.2007.55}

\bibitem[{Jones {et~al.}(2001)Jones, Oliphant, Peterson, {et~al.}}]{scipy}
Jones, E., Oliphant, T., Peterson, P., {et~al.} 2001, {SciPy}: Open source scientific tools for {Python}.
\newblock \url{http://www.scipy.org/}

\bibitem[{{Joost 't Hart} {et~al.}(2021){Joost 't Hart}, {van Holstein}, {Bos}, {Ruigrok}, {Snik}, {Lozi}, {Guyon}, {Kudo}, {Zhang}, {Jovanovic}, {Norris}, {Martinod}, {Groff}, {Chilcote}, {Currie}, {Tamura}, {Vievard}, {Sahoo}, {Deo}, {Ahn}, {Martinache}, \& {Kasdin}}]{Joost2021}
{Joost 't Hart}, G.~J., {van Holstein}, R.~G., {Bos}, S.~P., {et~al.} 2021, arXiv e-prints, arXiv:2108.04833, \dodoi{10.48550/arXiv.2108.04833}

\bibitem[{{Jovanovic} {et~al.}(2015){Jovanovic}, {Martinache}, {Guyon}, {Clergeon}, {Singh}, {Kudo}, {Garrel}, {Newman}, {Doughty}, {Lozi}, {Males}, {Minowa}, {Hayano}, {Takato}, {Morino}, {Kuhn}, {Serabyn}, {Norris}, {Tuthill}, {Schworer}, {Stewart}, {Close}, {Huby}, {Perrin}, {Lacour}, {Gauchet}, {Vievard}, {Murakami}, {Oshiyama}, {Baba}, {Matsuo}, {Nishikawa}, {Tamura}, {Lai}, {Marchis}, {Duchene}, {Kotani}, \& {Woillez}}]{Jovanovic2015}
{Jovanovic}, N., {Martinache}, F., {Guyon}, O., {et~al.} 2015, \pasp, 127, 890, \dodoi{10.1086/682989}

\bibitem[{{Kenyon} \& {Hartmann}(1987)}]{KenyonHartmann1987}
{Kenyon}, S.~J., \& {Hartmann}, L. 1987, \apj, 323, 714, \dodoi{10.1086/165866}

\bibitem[{{Keppler} {et~al.}(2018){Keppler}, {Benisty}, {M{\"u}ller}, {Henning}, {van Boekel}, {Cantalloube}, {Ginski}, {van Holstein}, {Maire}, {Pohl}, {Samland }, {Avenhaus}, {Baudino}, {Boccaletti}, {de Boer}, {Bonnefoy}, {Chauvin}, {Desidera}, {Langlois}, {Lazzoni}, {Marleau}, {Mordasini}, {Pawellek}, {Stolker}, {Vigan}, {Zurlo}, {Birnstiel}, {Brandner}, {Feldt}, {Flock}, {Girard}, {Gratton}, {Hagelberg}, {Isella}, {Janson}, {Juhasz}, {Kemmer}, {Kral}, {Lagrange}, {Launhardt}, {Matter}, {M{\'e}nard}, {Milli}, {Molli{\`e}re}, {Olofsson}, {P{\'e}rez}, {Pinilla}, {Pinte}, {Quanz}, {Schmidt}, {Udry}, {Wahhaj}, {Williams}, {Buenzli}, {Cudel}, {Dominik}, {Galicher}, {Kasper}, {Lannier}, {Mesa}, {Mouillet}, {Peretti}, {Perrot}, {Salter}, {Sissa}, {Wildi}, {Abe}, {Antichi}, {Augereau}, {Baruffolo}, {Baudoz}, {Bazzon}, {Beuzit}, {Blanchard}, {Brems}, {Buey}, {De Caprio}, {Carbillet}, {Carle}, {Cascone}, {Cheetham}, {Claudi}, {Costille}, {Delboulb{\'e}}, {Dohlen}, {Fantinel}, {Feautrier}, {Fusco}, {Giro}, {Gluck},
  {Gry}, {Hubin}, {Hugot}, {Jaquet}, {Le Mignant}, {Llored}, {Madec}, {Magnard}, {Martinez}, {Maurel}, {Meyer}, {M{\"o}ller-Nilsson}, {Moulin}, {Mugnier}, {Orign{\'e}}, {Pavlov}, {Perret}, {Petit}, {Pragt}, {Puget}, {Rabou}, {Ramos}, {Rigal}, {Rochat}, {Roelfsema}, {Rousset}, {Roux}, {Salasnich}, {Sauvage}, {Sevin}, {Soenke}, {Stadler}, {Suarez}, {Turatto}, \& {Weber}}]{Keppler2018}
{Keppler}, M., {Benisty}, M., {M{\"u}ller}, A., {et~al.} 2018, \aap, 617, A44, \dodoi{10.1051/0004-6361/201832957}

\bibitem[{Kluyver {et~al.}(2016)Kluyver, Ragan-Kelley, P{\'e}rez, Granger, Bussonnier, Frederic, Kelley, Hamrick, Grout, Corlay, {et~al.}}]{jupyter}
Kluyver, T., Ragan-Kelley, B., P{\'e}rez, F., {et~al.} 2016, in ELPUB, 87--90

\bibitem[{{Lada} {et~al.}(2006){Lada}, {Muench}, {Luhman}, {Allen}, {Hartmann}, {Megeath}, {Myers}, {Fazio}, {Wood}, {Muzerolle}, {Rieke}, {Siegler}, \& {Young}}]{Lada2006}
{Lada}, C.~J., {Muench}, A.~A., {Luhman}, K.~L., {et~al.} 2006, \aj, 131, 1574, \dodoi{10.1086/499808}

\bibitem[{{Lagrange} {et~al.}(2010){Lagrange}, {Bonnefoy}, {Chauvin}, {Apai}, {Ehrenreich}, {Boccaletti}, {Gratadour}, {Rouan}, {Mouillet}, {Lacour}, \& {Kasper}}]{Lagrange2010}
{Lagrange}, A.~M., {Bonnefoy}, M., {Chauvin}, G., {et~al.} 2010, Science, 329, 57, \dodoi{10.1126/science.1187187}

\bibitem[{{Lawson} {et~al.}(2022){Lawson}, {Currie}, {Wisniewski}, {Groff}, {McElwain}, \& {Schlieder}}]{Lawson2022}
{Lawson}, K., {Currie}, T., {Wisniewski}, J.~P., {et~al.} 2022, \apjl, 935, L25, \dodoi{10.3847/2041-8213/ac853b}

\bibitem[{{Lawson} {et~al.}(2021){Lawson}, {Currie}, {Wisniewski}, {Hashimoto}, {Guyon}, {Kasdin}, {Groff}, {Lozi}, {Brandt}, {Chilcote}, {Deo}, {Uyama}, \& {Vievard}}]{Lawson2021}
{Lawson}, K., {Currie}, T., {Wisniewski}, J.~P., {et~al.} 2021, in Society of Photo-Optical Instrumentation Engineers (SPIE) Conference Series, Vol. 11823, Techniques and Instrumentation for Detection of Exoplanets X, ed. S.~B. {Shaklan} \& G.~J. {Ruane}, 118230D, \dodoi{10.1117/12.2594819}

\bibitem[{{Lozi} {et~al.}(2022){Lozi}, {Ahn}, {Clergeon}, {Deo}, {Guyon}, {Hattori}, {Minowa}, {Nishiyama}, {Ono}, \& {Vievard}}]{Lozi2022}
{Lozi}, J., {Ahn}, K., {Clergeon}, C., {et~al.} 2022, in Society of Photo-Optical Instrumentation Engineers (SPIE) Conference Series, Vol. 12185, Adaptive Optics Systems VIII, ed. L.~{Schreiber}, D.~{Schmidt}, \& E.~{Vernet}, 1218533, \dodoi{10.1117/12.2630634}

\bibitem[{{Luhman} {et~al.}(2010){Luhman}, {Allen}, {Espaillat}, {Hartmann}, \& {Calvet}}]{Luhman2010}
{Luhman}, K.~L., {Allen}, P.~R., {Espaillat}, C., {Hartmann}, L., \& {Calvet}, N. 2010, The Astrophysical Journal Supplement Series, 186, 111, \dodoi{10.1088/0067-0049/186/1/111}

\bibitem[{{Marois} {et~al.}(2008){Marois}, {Macintosh}, {Barman}, {Zuckerman}, {Song}, {Patience}, {Lafreni{\`e}re}, \& {Doyon}}]{Marois2008a}
{Marois}, C., {Macintosh}, B., {Barman}, T., {et~al.} 2008, Science, 322, 1348, \dodoi{10.1126/science.1166585}

\bibitem[{{Mathis} \& {Whiffen}(1989)}]{Mathis1989}
{Mathis}, J.~S., \& {Whiffen}, G. 1989, \apj, 341, 808, \dodoi{10.1086/167538}

\bibitem[{{Min} {et~al.}(2009){Min}, {Dullemond}, {Dominik}, {de Koter}, \& {Hovenier}}]{Min2009}
{Min}, M., {Dullemond}, C.~P., {Dominik}, C., {de Koter}, A., \& {Hovenier}, J.~W. 2009, \aap, 497, 155, \dodoi{10.1051/0004-6361/200811470}

\bibitem[{{Mulders} {et~al.}(2013){Mulders}, {Min}, {Dominik}, {Debes}, \& {Schneider}}]{Mulders2013}
{Mulders}, G.~D., {Min}, M., {Dominik}, C., {Debes}, J.~H., \& {Schneider}, G. 2013, \aap, 549, A112, \dodoi{10.1051/0004-6361/201219522}

\bibitem[{{Muto} {et~al.}(2012){Muto}, {Grady}, {Hashimoto}, {Fukagawa}, {Hornbeck}, {Sitko}, {Russell}, {Werren}, {Cur{\'e}}, {Currie}, {Ohashi}, {Okamoto}, {Momose}, {Honda}, {Inutsuka}, {Takeuchi}, {Dong}, {Abe}, {Brandner}, {Brandt}, {Carson}, {Egner}, {Feldt}, {Fukue}, {Goto}, {Guyon}, {Hayano}, {Hayashi}, {Hayashi}, {Henning}, {Hodapp}, {Ishii}, {Iye}, {Janson}, {Kandori}, {Knapp}, {Kudo}, {Kusakabe}, {Kuzuhara}, {Matsuo}, {Mayama}, {McElwain}, {Miyama}, {Morino}, {Moro-Martin}, {Nishimura}, {Pyo}, {Serabyn}, {Suto}, {Suzuki}, {Takami}, {Takato}, {Terada}, {Thalmann}, {Tomono}, {Turner}, {Watanabe}, {Wisniewski}, {Yamada}, {Takami}, {Usuda}, \& {Tamura}}]{Muto2012}
{Muto}, T., {Grady}, C.~A., {Hashimoto}, J., {et~al.} 2012, \apjl, 748, L22, \dodoi{10.1088/2041-8205/748/2/L22}

\bibitem[{{Oppenheimer} {et~al.}(2008){Oppenheimer}, {Brenner}, {Hinkley}, {Zimmerman}, {Sivaramakrishnan}, {Soummer}, {Kuhn}, {Graham}, {Perrin}, {Lloyd}, {Roberts}, \& {Harrington}}]{Oppenheimer2008}
{Oppenheimer}, B.~R., {Brenner}, D., {Hinkley}, S., {et~al.} 2008, \apj, 679, 1574, \dodoi{10.1086/587778}

\bibitem[{{Perrin} {et~al.}(2009){Perrin}, {Schneider}, {Duchene}, {Pinte}, {Grady}, {Wisniewski}, \& {Hines}}]{Perrin2009}
{Perrin}, M.~D., {Schneider}, G., {Duchene}, G., {et~al.} 2009, \apjl, 707, L132, \dodoi{10.1088/0004-637X/707/2/L132}

\bibitem[{{Pinte} {et~al.}(2009){Pinte}, {Harries}, {Min}, {Watson}, {Dullemond}, {Woitke}, {M{\'e}nard}, \& {Dur{\'a}n-Rojas}}]{Pinte2009}
{Pinte}, C., {Harries}, T.~J., {Min}, M., {et~al.} 2009, \aap, 498, 967, \dodoi{10.1051/0004-6361/200811555}

\bibitem[{{Pinte} {et~al.}(2006){Pinte}, {M{\'e}nard}, {Duch{\^e}ne}, \& {Bastien}}]{Pinte2006}
{Pinte}, C., {M{\'e}nard}, F., {Duch{\^e}ne}, G., \& {Bastien}, P. 2006, \aap, 459, 797, \dodoi{10.1051/0004-6361:20053275}

\bibitem[{{Price-Whelan} {et~al.}(2018){Price-Whelan}, {Sip{\H{o}}cz}, {G{\"u}nther}, {Lim}, {Crawford}, {Conseil}, {Shupe}, {Craig}, {Dencheva}, {Ginsburg}, {VanderPlas}, {Bradley}, {P{\'e}rez-Su{\'a}rez}, {de Val-Borro}, {Paper Contributors}, {Aldcroft}, {Cruz}, {Robitaille}, {Tollerud}, {Coordination Committee}, {Ardelean}, {Babej}, {Bach}, {Bachetti}, {Bakanov}, {Bamford}, {Barentsen}, {Barmby}, {Baumbach}, {Berry}, {Biscani}, {Boquien}, {Bostroem}, {Bouma}, {Brammer}, {Bray}, {Breytenbach}, {Buddelmeijer}, {Burke}, {Calderone}, {Cano Rodr{\'\i}guez}, {Cara}, {Cardoso}, {Cheedella}, {Copin}, {Corrales}, {Crichton}, {D{\textquoteright}Avella}, {Deil}, {Depagne}, {Dietrich}, {Donath}, {Droettboom}, {Earl}, {Erben}, {Fabbro}, {Ferreira}, {Finethy}, {Fox}, {Garrison}, {Gibbons}, {Goldstein}, {Gommers}, {Greco}, {Greenfield}, {Groener}, {Grollier}, {Hagen}, {Hirst}, {Homeier}, {Horton}, {Hosseinzadeh}, {Hu}, {Hunkeler}, {Ivezi{\'c}}, {Jain}, {Jenness}, {Kanarek}, {Kendrew}, {Kern}, {Kerzendorf}, {Khvalko},
  {King}, {Kirkby}, {Kulkarni}, {Kumar}, {Lee}, {Lenz}, {Littlefair}, {Ma}, {Macleod}, {Mastropietro}, {McCully}, {Montagnac}, {Morris}, {Mueller}, {Mumford}, {Muna}, {Murphy}, {Nelson}, {Nguyen}, {Ninan}, {N{\"o}the}, {Ogaz}, {Oh}, {Parejko}, {Parley}, {Pascual}, {Patil}, {Patil}, {Plunkett}, {Prochaska}, {Rastogi}, {Reddy Janga}, {Sabater}, {Sakurikar}, {Seifert}, {Sherbert}, {Sherwood-Taylor}, {Shih}, {Sick}, {Silbiger}, {Singanamalla}, {Singer}, {Sladen}, {Sooley}, {Sornarajah}, {Streicher}, {Teuben}, {Thomas}, {Tremblay}, {Turner}, {Terr{\'o}n}, {van Kerkwijk}, {de la Vega}, {Watkins}, {Weaver}, {Whitmore}, {Woillez}, {Zabalza}, \& {Contributors}}]{astropy:2018}
{Price-Whelan}, A.~M., {Sip{\H{o}}cz}, B.~M., {G{\"u}nther}, H.~M., {et~al.} 2018, \aj, 156, 123, \dodoi{10.3847/1538-3881/aabc4f}

\bibitem[{{Quanz} {et~al.}(2015){Quanz}, {Amara}, {Meyer}, {Girard}, {Kenworthy}, \& {Kasper}}]{Quanz2015}
{Quanz}, S.~P., {Amara}, A., {Meyer}, M.~R., {et~al.} 2015, \apj, 807, 64, \dodoi{10.1088/0004-637X/807/1/64}

\bibitem[{{Quanz} {et~al.}(2011){Quanz}, {Schmid}, {Geissler}, {Meyer}, {Henning}, {Brandner}, \& {Wolf}}]{Quanz2011}
{Quanz}, S.~P., {Schmid}, H.~M., {Geissler}, K., {et~al.} 2011, \apj, 738, 23, \dodoi{10.1088/0004-637X/738/1/23}

\bibitem[{{Ren} {et~al.}(2023){Ren}, {Benisty}, {Ginski}, {Tazaki}, {Wallack}, {Milli}, {Garufi}, {Bae}, {Facchini}, {M{\'e}nard}, {Pinilla}, {Swastik}, {Teague}, \& {Wahhaj}}]{Ren2023}
{Ren}, B.~B., {Benisty}, M., {Ginski}, C., {et~al.} 2023, \aap, 680, A114, \dodoi{10.1051/0004-6361/202347353}

\bibitem[{{Rich} {et~al.}(2019){Rich}, {Wisniewski}, {Currie}, {Fukagawa}, {Grady}, {Sitko}, {Pikhartova}, {Hashimoto}, {Abe}, {Brand ner}, {Brandt}, {Carson}, {Chilcote}, {Dong}, {Feldt}, {Goto}, {Groff}, {Guyon}, {Hayano}, {Hayashi}, {Hayashi}, {Henning}, {Hodapp}, {Ishii}, {Iye}, {Janson}, {Jovanovic}, {Kand ori}, {Kasdin}, {Knapp}, {Kudo}, {Kusakabe}, {Kuzuhara}, {Kwon}, {Lozi}, {Martinache}, {Matsuo}, {Mayama}, {McElwain}, {Miyama}, {Morino}, {Moro-Martin}, {Nakagawa}, {Nishimura}, {Pyo}, {Serabyn}, {Suto}, {Russel}, {Suzuki}, {Takami}, {Takato}, {Terada}, {Thalmann}, {Turner}, {Uyama}, {Wagner}, {Watanabe}, {Yamada}, {Takami}, {Usuda}, \& {Tamura}}]{Rich2019}
{Rich}, E.~A., {Wisniewski}, J.~P., {Currie}, T., {et~al.} 2019, \apj, 875, 38, \dodoi{10.3847/1538-4357/ab0f3b}

\bibitem[{{Rich} {et~al.}(2022){Rich}, {Monnier}, {Aarnio}, {Laws}, {Setterholm}, {Wilner}, {Calvet}, {Harries}, {Miller}, {Davies}, {Adams}, {Andrews}, {Bae}, {Espaillat}, {Greenbaum}, {Hinkley}, {Kraus}, {Hartmann}, {Isella}, {McClure}, {Oppenheimer}, {P{\'e}rez}, \& {Zhu}}]{Rich2022}
{Rich}, E.~A., {Monnier}, J.~D., {Aarnio}, A., {et~al.} 2022, \aj, 164, 109, \dodoi{10.3847/1538-3881/ac7be4}

\bibitem[{{Speedie} {et~al.}(2024){Speedie}, {Dong}, {Hall}, {Longarini}, {Veronesi}, {Paneque-Carre{\~n}o}, {Lodato}, {Tang}, {Teague}, \& {Hashimoto}}]{Speedie2024}
{Speedie}, J., {Dong}, R., {Hall}, C., {et~al.} 2024, \nat, 633, 58, \dodoi{10.1038/s41586-024-07877-0}

\bibitem[{{Tang} {et~al.}(2012){Tang}, {Guilloteau}, {Pi{\'e}tu}, {Dutrey}, {Ohashi}, \& {Ho}}]{Tang2012}
{Tang}, Y.~W., {Guilloteau}, S., {Pi{\'e}tu}, V., {et~al.} 2012, \aap, 547, A84, \dodoi{10.1051/0004-6361/201219414}

\bibitem[{{Tang} {et~al.}(2017){Tang}, {Guilloteau}, {Dutrey}, {Muto}, {Shen}, {Gu}, {Inutsuka}, {Momose}, {Pietu}, {Fukagawa}, {Chapillon}, {Ho}, {di Folco}, {Corder}, {Ohashi}, \& {Hashimoto}}]{Tang2017}
{Tang}, Y.-W., {Guilloteau}, S., {Dutrey}, A., {et~al.} 2017, \apj, 840, 32, \dodoi{10.3847/1538-4357/aa6af7}

\bibitem[{{Tannirkulam} {et~al.}(2008){Tannirkulam}, {Monnier}, {Harries}, {Millan-Gabet}, {Zhu}, {Pedretti}, {Ireland}, {Tuthill}, {ten Brummelaar}, {McAlister}, {Farrington}, {Goldfinger}, {Sturmann}, {Sturmann}, \& {Turner}}]{Tannirkulam2008}
{Tannirkulam}, A., {Monnier}, J.~D., {Harries}, T.~J., {et~al.} 2008, \apj, 689, 513, \dodoi{10.1086/592346}

\bibitem[{{Tazaki} {et~al.}(2023){Tazaki}, {Ginski}, \& {Dominik}}]{Tazaki2023}
{Tazaki}, R., {Ginski}, C., \& {Dominik}, C. 2023, \apjl, 944, L43, \dodoi{10.3847/2041-8213/acb824}

\bibitem[{{Tazaki} {et~al.}(2019){Tazaki}, {Tanaka}, {Muto}, {Kataoka}, \& {Okuzumi}}]{Tazaki2019}
{Tazaki}, R., {Tanaka}, H., {Muto}, T., {Kataoka}, A., \& {Okuzumi}, S. 2019, \mnras, 485, 4951, \dodoi{10.1093/mnras/stz662}

\bibitem[{{Thalmann} {et~al.}(2016){Thalmann}, {Janson}, {Garufi}, {Boccaletti}, {Quanz}, {Sissa}, {Gratton}, {Salter}, {Benisty}, {Bonnefoy}, {Chauvin}, {Daemgen}, {Desidera}, {Dominik}, {Engler}, {Feldt}, {Henning}, {Lagrange}, {Langlois}, {Lannier}, {Le Coroller}, {Ligi}, {M{\'e}nard}, {Mesa}, {Meyer}, {Mulders}, {Olofsson}, {Pinte}, {Schmid}, {Vigan}, \& {Zurlo}}]{Thalmann2016}
{Thalmann}, C., {Janson}, M., {Garufi}, A., {et~al.} 2016, \apjl, 828, L17, \dodoi{10.3847/2041-8205/828/2/L17}

\bibitem[{{Tobin} {et~al.}(2024){Tobin}, {Currie}, {Li}, {Chilcote}, {Brandt}, {Lacy}, {Kuzuhara}, {Vincent}, {El Morsy}, {Deo}, {Williams}, {Guyon}, {Lozi}, {Vievard}, {Skaf}, {Ahn}, {Groff}, {Kasdin}, {Uyama}, {Tamura}, {Gibbs}, {Lewis}, {Bowens-Rubin}, {Salama}, {An}, \& {Chen}}]{Tobin2024}
{Tobin}, T.~L., {Currie}, T., {Li}, Y., {et~al.} 2024, \aj, 167, 205, \dodoi{10.3847/1538-3881/ad3077}

\bibitem[{{van der Walt} {et~al.}(2011){van der Walt}, {Colbert}, \& {Varoquaux}}]{numpy}
{van der Walt}, S., {Colbert}, S.~C., \& {Varoquaux}, G. 2011, Computing in Science and Engineering, 13, 22, \dodoi{10.1109/MCSE.2011.37}

\bibitem[{{van Holstein} {et~al.}(2020){van Holstein}, {Bos}, {Ruigrok}, {Lozi}, {Guyon}, {Norris}, {Snik}, {Chilcote}, {Currie}, {Groff}, {'t Hart}, {Jovanovic}, {Kasdin}, {Kudo}, {Martinache}, {Mazin}, {Sahoo}, {Tamura}, {Vievard}, {Walter}, \& {Zhang}}]{vanHolstein2020}
{van Holstein}, R.~G., {Bos}, S.~P., {Ruigrok}, J., {et~al.} 2020, in Society of Photo-Optical Instrumentation Engineers (SPIE) Conference Series, Vol. 11447, Ground-based and Airborne Instrumentation for Astronomy VIII, ed. C.~J. {Evans}, J.~J. {Bryant}, \& K.~{Motohara}, 114475B, \dodoi{10.1117/12.2576188}

\end{thebibliography}
\bibliographystyle{aasjournal}{}                                                              

\appendix \label{app:charis}
\section{MCFOST Model Azimuthal Profile Dependencies on Different Input Parameters} 

\begin{figure*}[!h]
    \includegraphics[width=.95\textwidth]{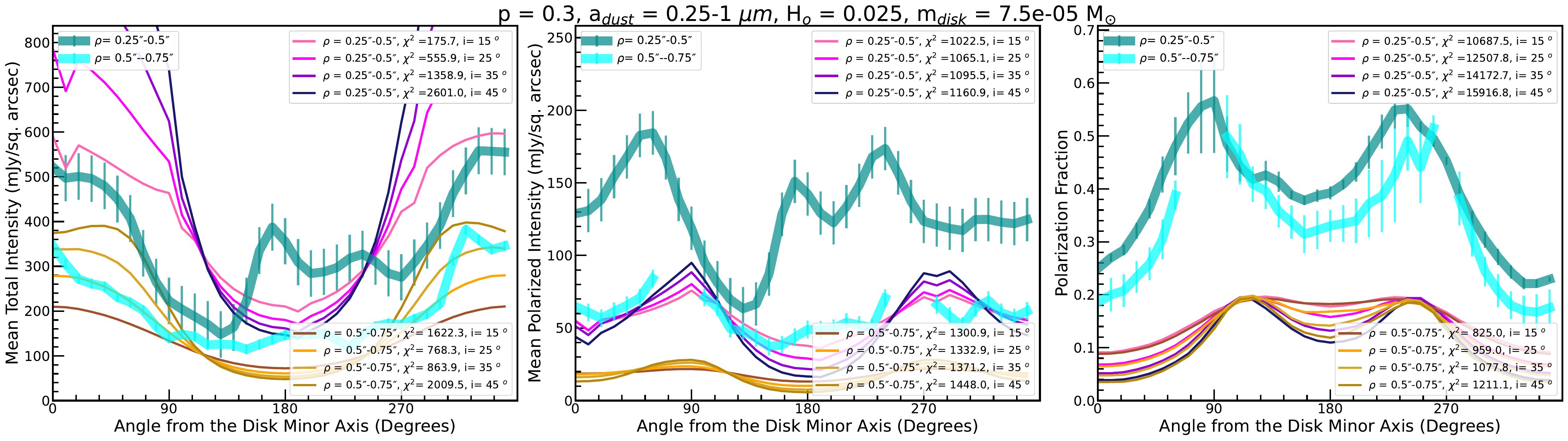}
      \vspace{-0.0in}
    \includegraphics[width=.95\textwidth]{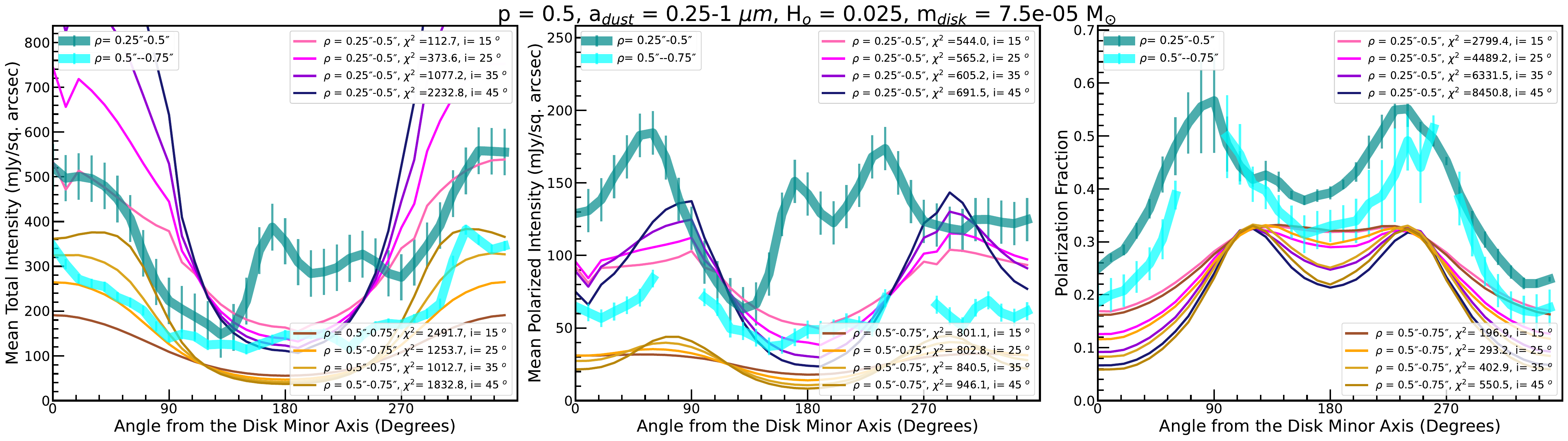}
      \vspace{-0.0in}
    \includegraphics[width=0.95\textwidth]{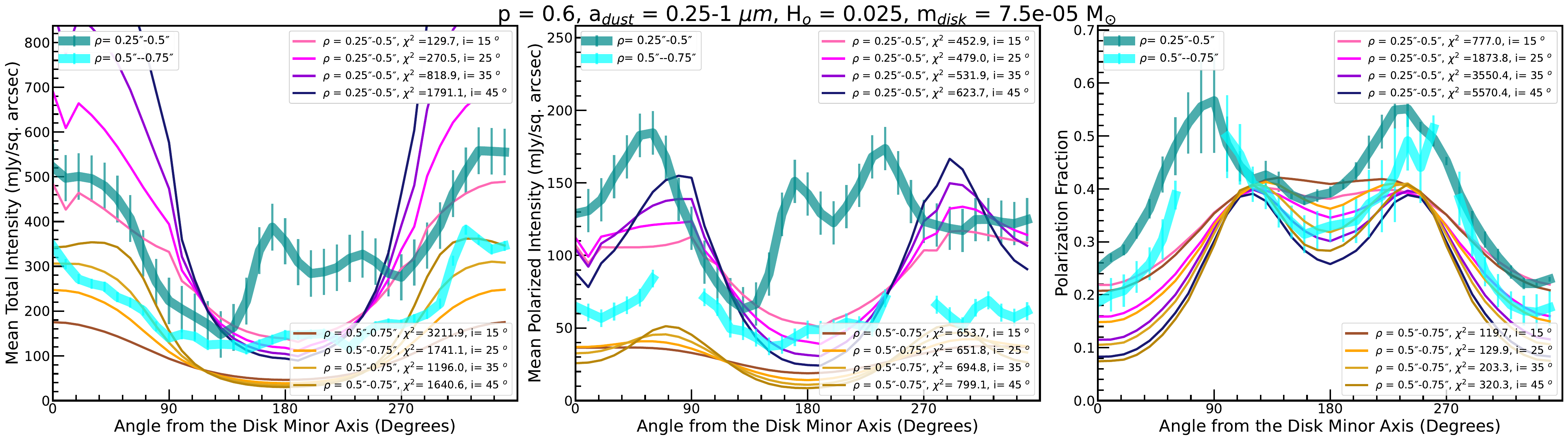}
      \vspace{-0.0in}
    \includegraphics[width=0.95\textwidth]{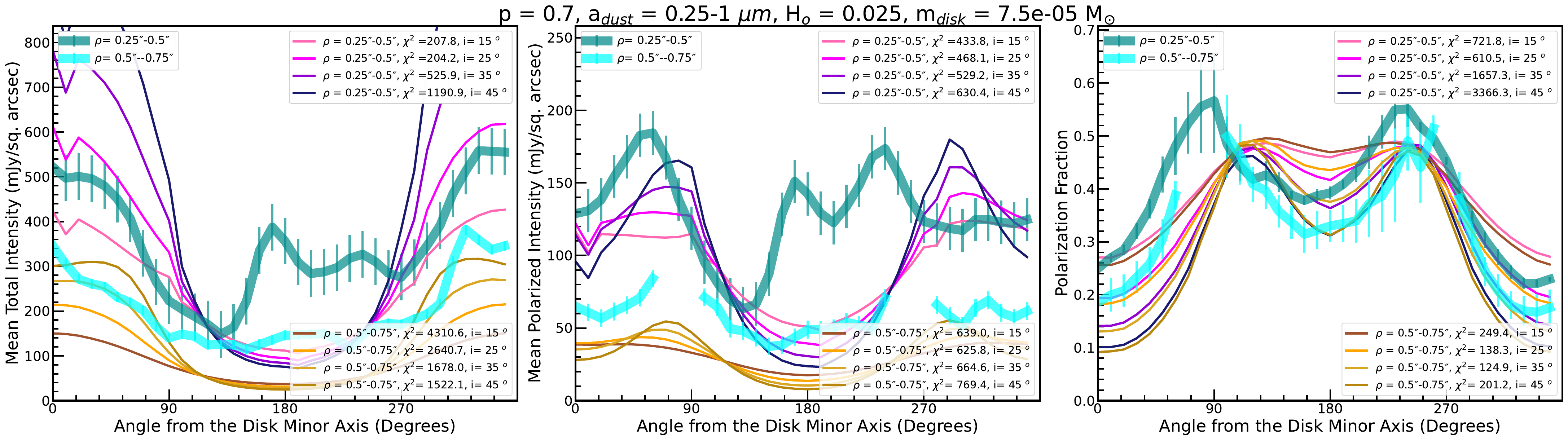}
      \vspace{-0.0in}
    \includegraphics[width=0.95\textwidth]{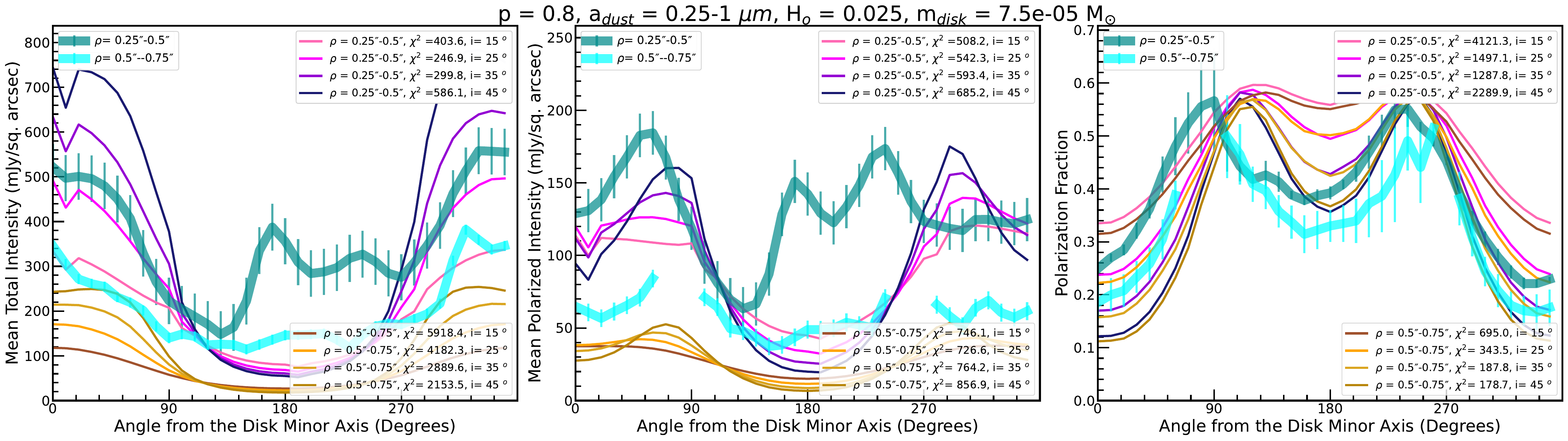}
    \vspace{-0.1in}
    \caption{Gallery of synthetic polarization fraction maps for different models at inclinations of 15--45$^{o}$, showing the effect of varying the dust porosity from $p$ = 0.3 to $p$ = 0.8.}
    \label{fig:phasefuncgallery}
\end{figure*}

\begin{figure*}
    \includegraphics[width=.95\textwidth]{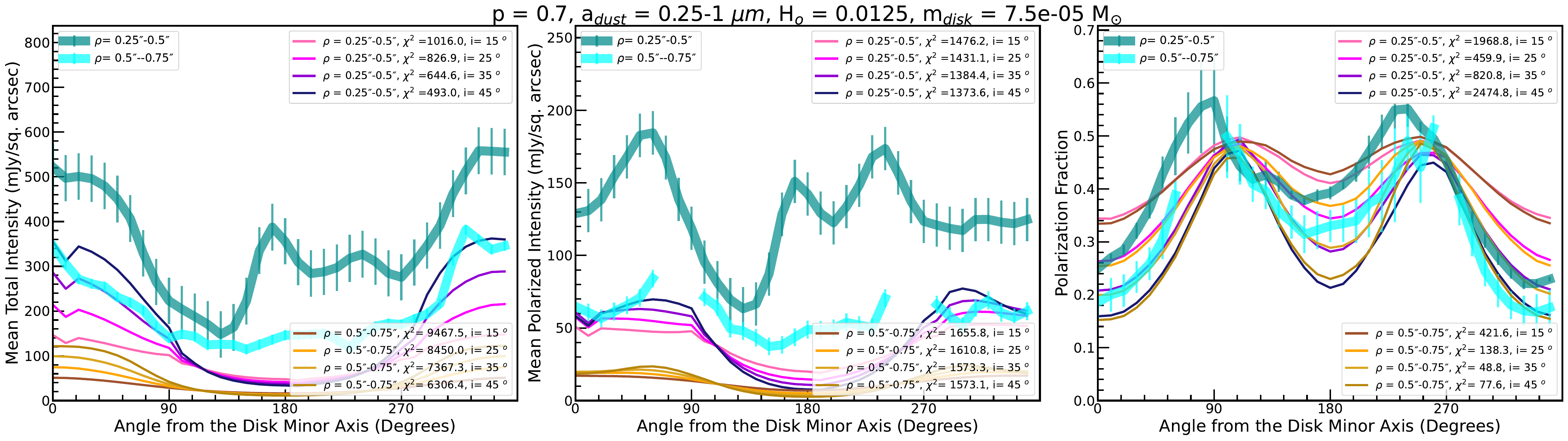}
      \vspace{-0.0in}
    \includegraphics[width=.95\textwidth]{phasefunc_comp_07_75e-05_0025_025_chisquare.pdf}
      \vspace{-0.0in}
    \includegraphics[width=0.95\textwidth]{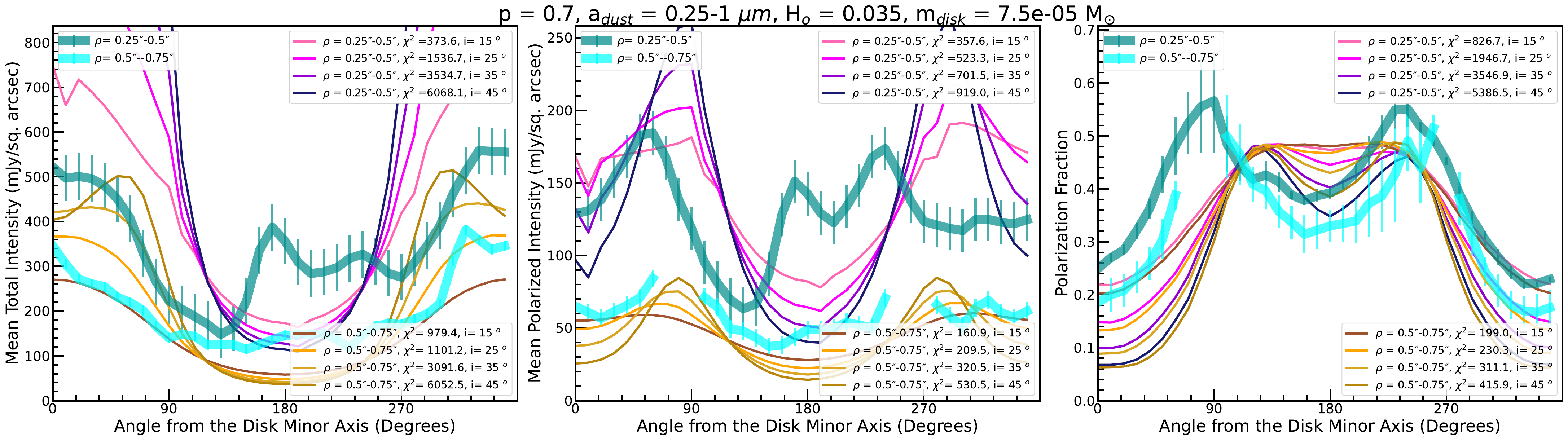}
      \vspace{-0.0in}
    \includegraphics[width=0.95\textwidth]{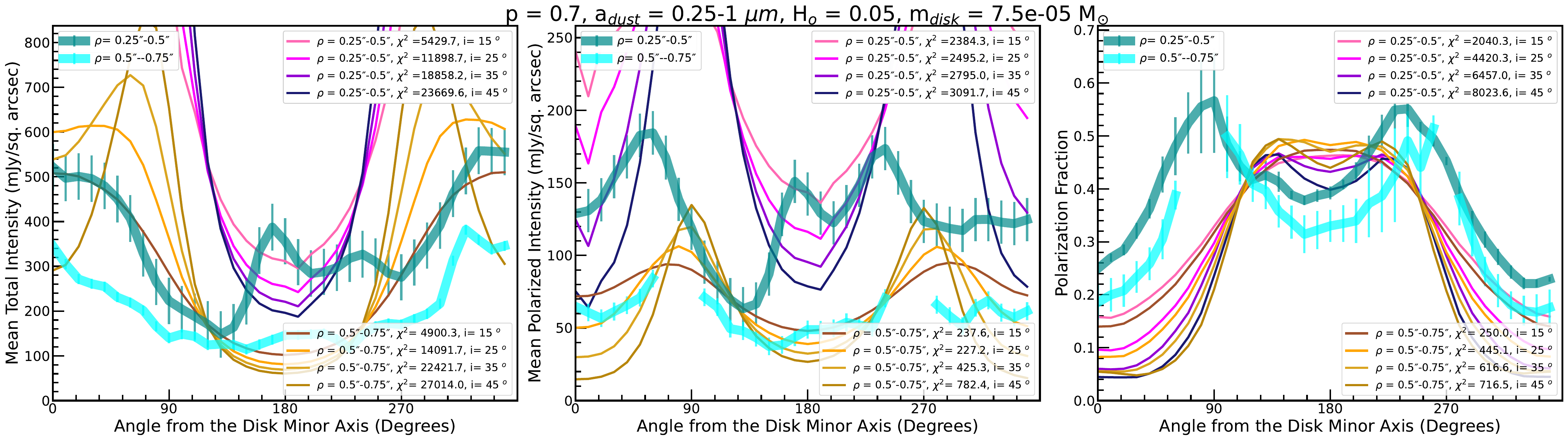}
      \vspace{-0.0in}
    \vspace{-0.025in}
    \caption{Gallery of synthetic polarization fraction maps for different models at inclinations of 15--45$^{o}$, showing the effect of varying the reference scale height from 0.0125 to 0.05.}
    \label{fig:phasefuncgallery2}
\end{figure*}

\begin{figure*}
    \includegraphics[width=.95\textwidth]{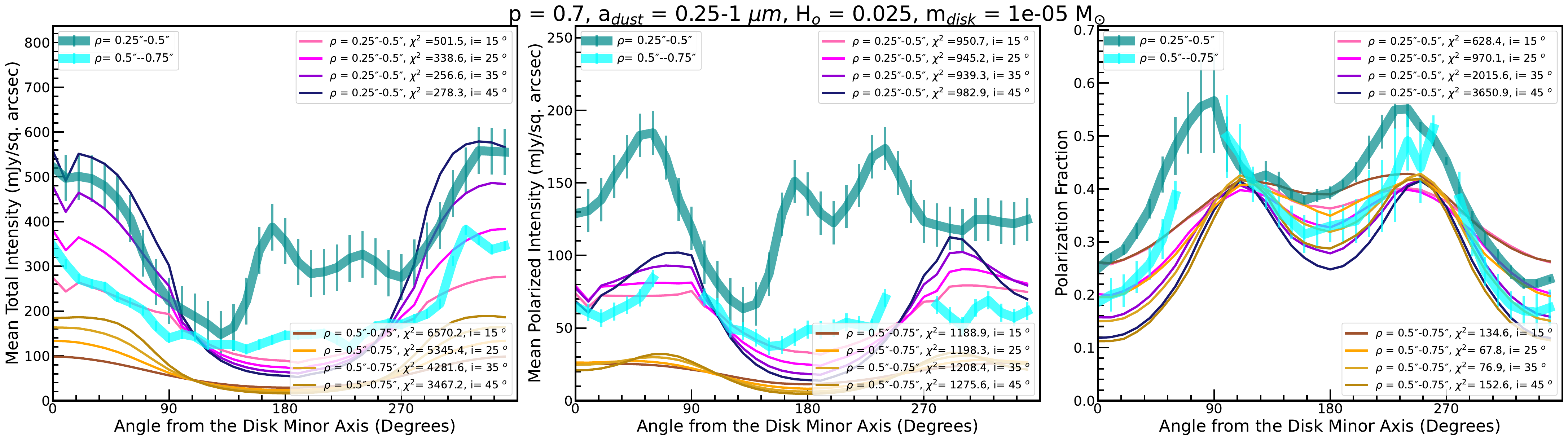}
      \vspace{-0.0in}
    \includegraphics[width=.95\textwidth]{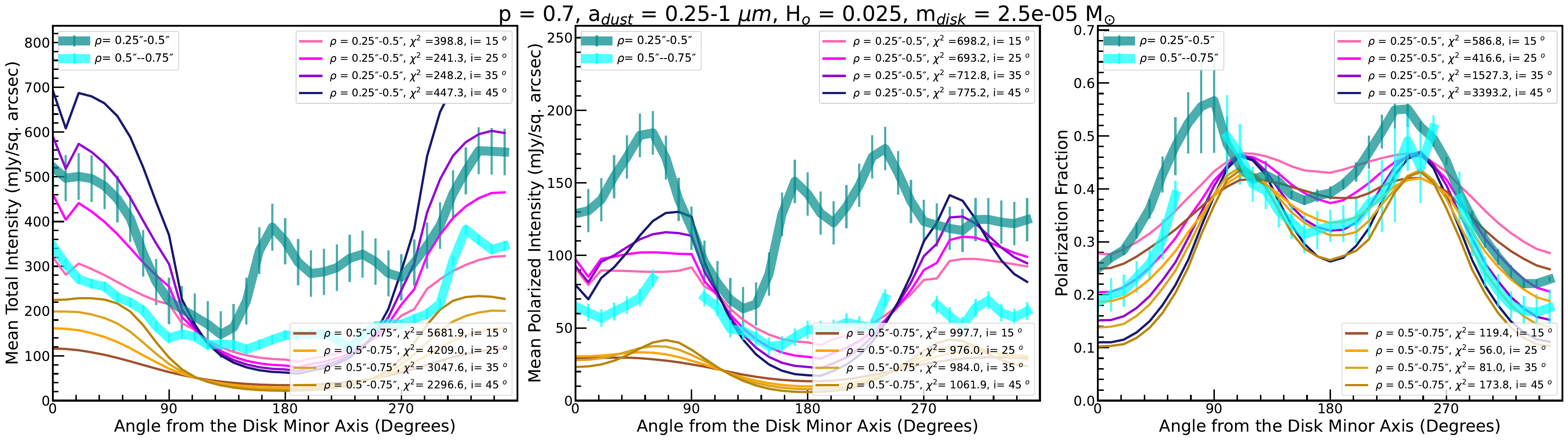}
      \vspace{-0.0in}
    \includegraphics[width=0.95\textwidth]{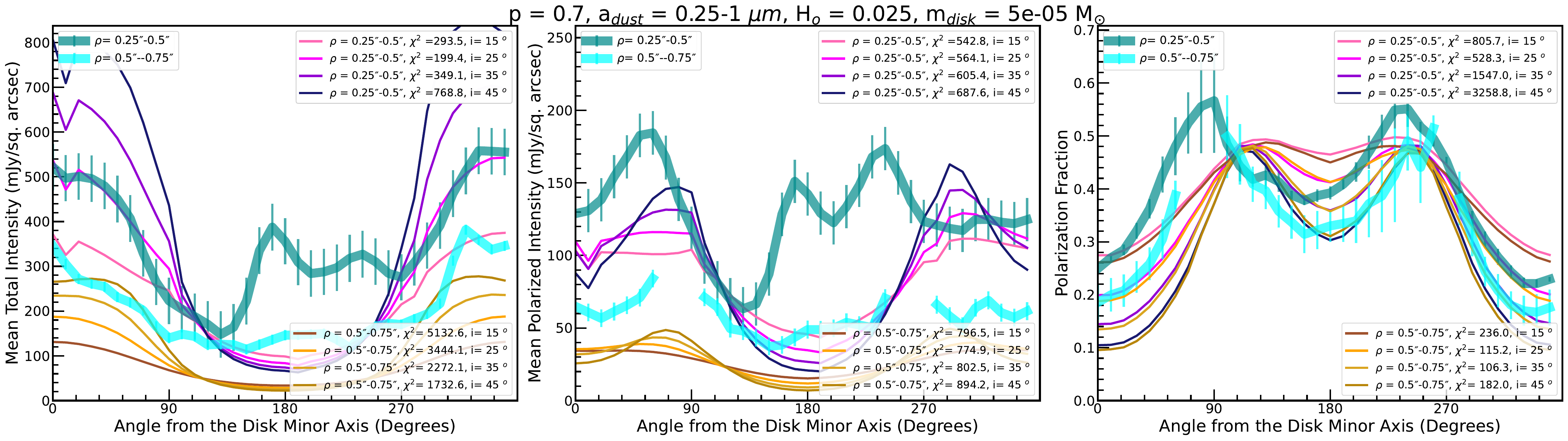}
      \vspace{-0.0in}
    \includegraphics[width=0.95\textwidth]{phasefunc_comp_07_75e-05_0025_025_chisquare.pdf}
      \vspace{-0.0in}
    \includegraphics[width=0.95\textwidth]{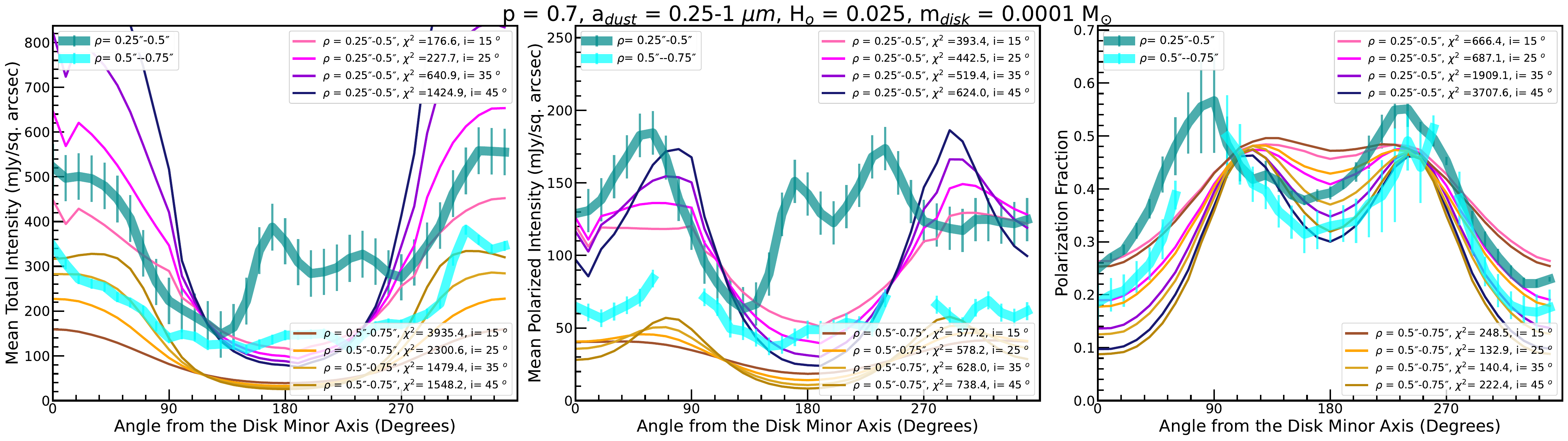}
    \vspace{-0.025in}
    \caption{Gallery of synthetic polarization fraction maps for different models at inclinations of 15--45$^{o}$, showing the effect of varying the disk mass from 10$^{-5}$ $M_{\rm \odot}$ to 10$^{-4}$ $M_{\rm \odot}$.}
    \label{fig:phasefuncgallery3}
\end{figure*}

Figures \ref{fig:phasefuncgallery}, \ref{fig:phasefuncgallery2}, and \ref{fig:phasefuncgallery3} show representative MCFOST models, illustrating how the azimuthal profile depends on parameters, as described in the main text.

\newpage

\begin{figure*}
    \includegraphics[width=.95\textwidth]{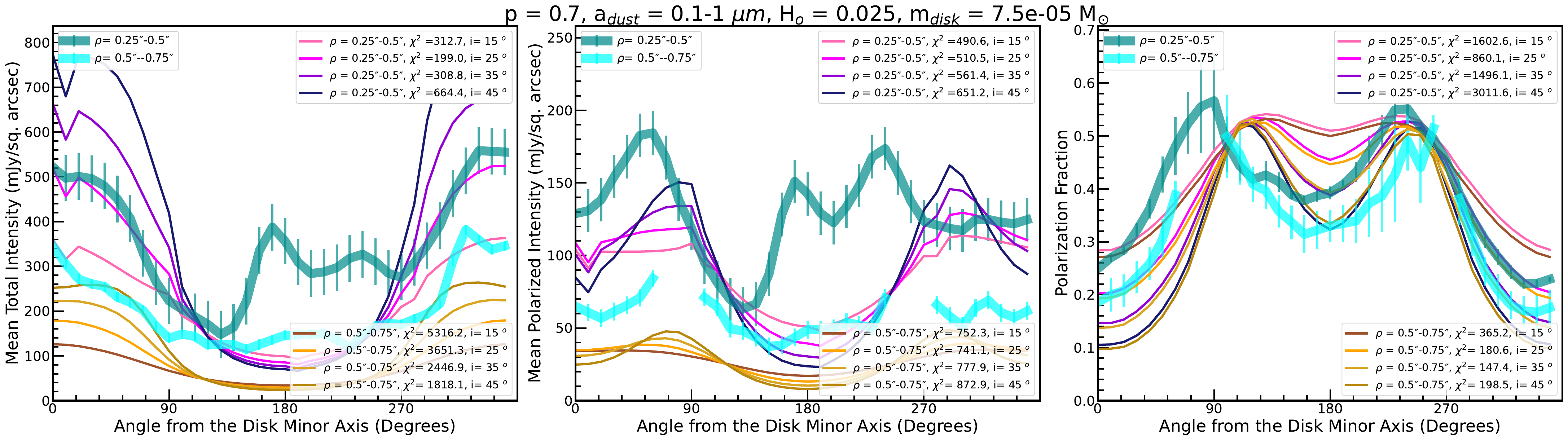}
      \vspace{-0.0in}
    \includegraphics[width=.95\textwidth]{phasefunc_comp_07_75e-05_0025_025_chisquare.pdf}
      \vspace{-0.0in}
    \includegraphics[width=0.95\textwidth]{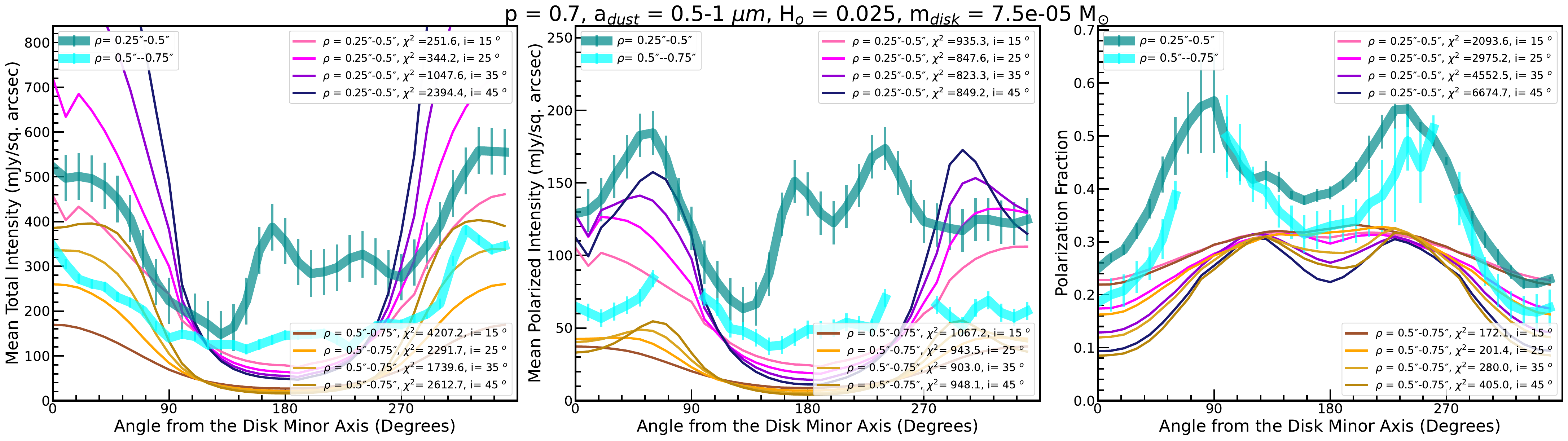}
    \vspace{-0.025in}
    \caption{Gallery of synthetic polarization fraction maps for different models at inclinations of 15--45$^{o}$, showing the effect of varying the minimum grain size from 0.1 to 0.5 $\mu m$.}
    \label{fig:phasefuncgallery4}
\end{figure*}

\section{MCFOST Model Chi-Square Distribution Dependendent on Different Input Parameters} 
\begin{figure*}
    \includegraphics[width=1\textwidth]{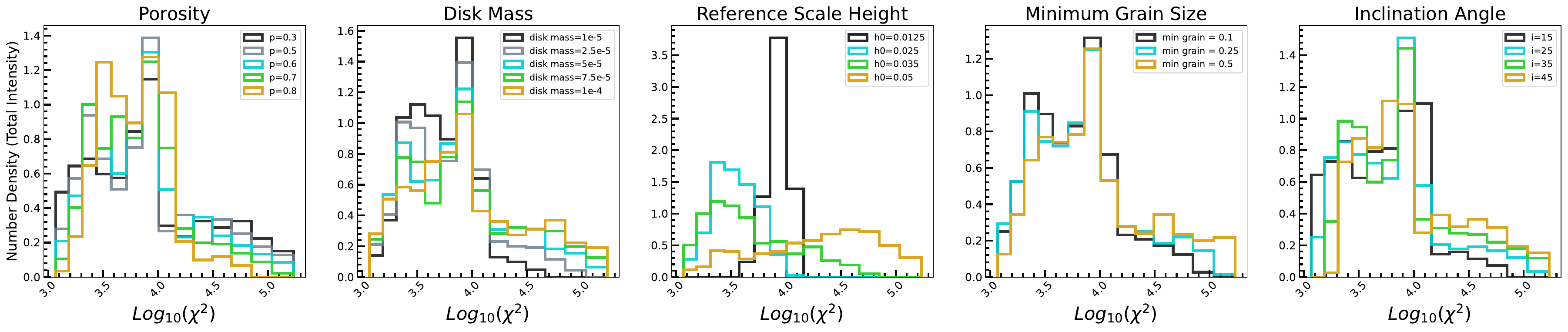}
    \includegraphics[width=1\textwidth]{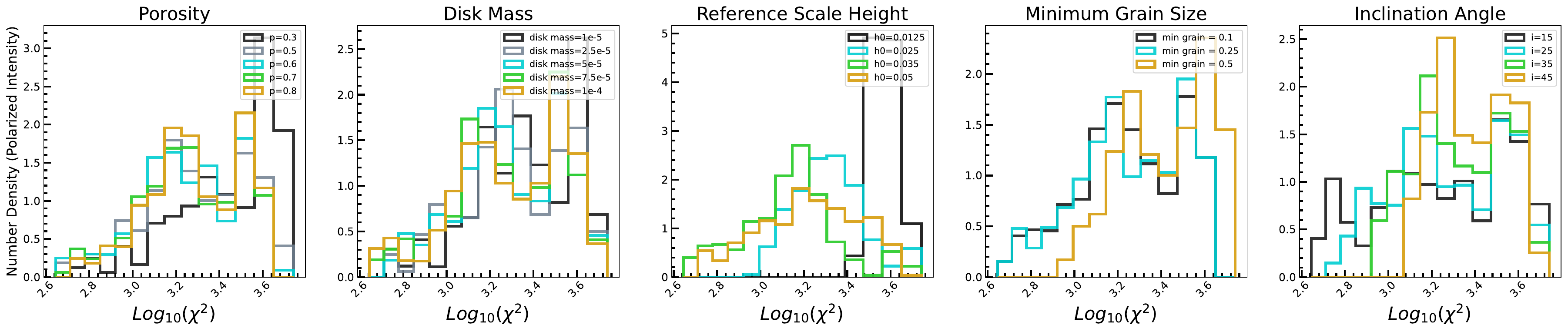}
    \includegraphics[width=1\textwidth]{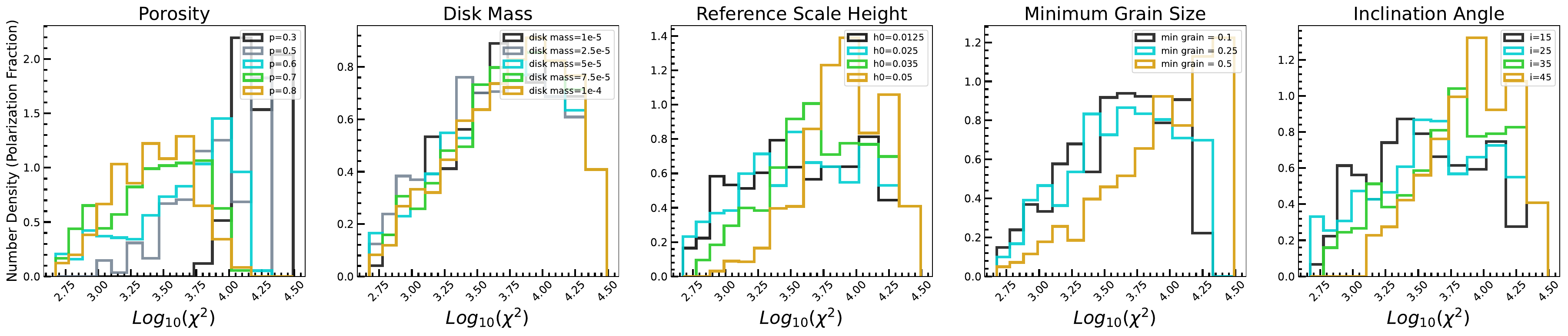}
    \includegraphics[width=1\textwidth]{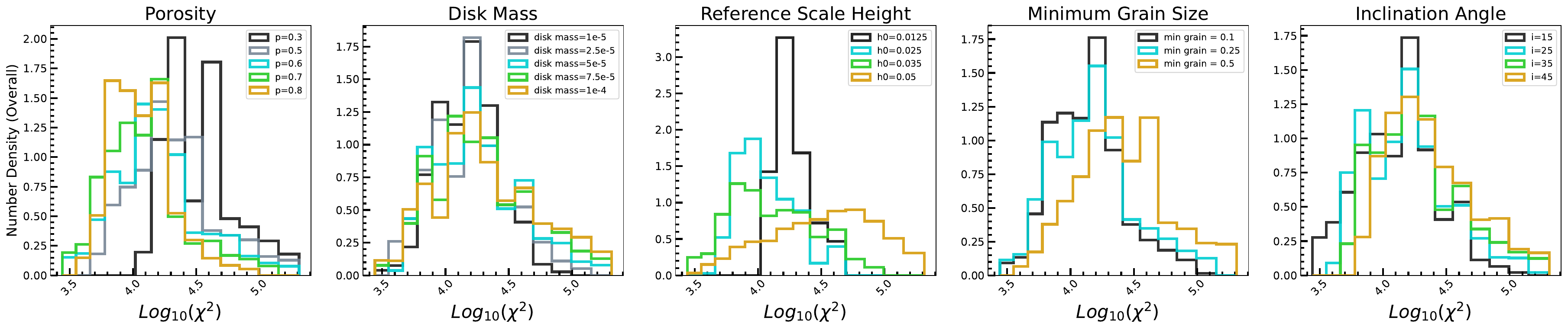}
    \caption{Gallery of chi-square distributions comparing the azimuthal profiles of our MCFOST models to our total intensity data (first row), polarized intensity data (second row), and polarization fraction (third row) for each of the free parameters. The cummulative chi-square distribution of all three data types is shown in the fourth row. The chi-square values are plotted on a logarimethic scale due to the large range.}
    \label{fig:chisquaregallery}
\end{figure*}

Figure \ref{fig:chisquaregallery} demonstrates how the chi-square values vary with over the range of each free parameter for total intensity, polarized intensity, and polarization fraction as well as the cummulative chi-square.

\end{document}